\documentclass[sigconf, natbib=false]{acmart}

\usepackage{amsmath}

\usepackage{float,relsize,dblfloatfix}
\usepackage{url}
\usepackage{graphicx}
\usepackage{epstopdf}
\usepackage[subrefformat=parens, labelformat=parens]{subfig}
\usepackage[font=footnotesize, labelfont=footnotesize]{caption}
\usepackage{diagbox, color, xcolor, soul}
\usepackage{lipsum}
\usepackage{enumitem}
\usepackage{listings}
\usepackage{hyperref}
\usepackage{multirow}
\usepackage{multicol}
\usepackage[style=ACM-Reference-Format,backend=bibtex,firstinits=true,citestyle=numeric-comp,sorting=none]{biblatex}
\makeatletter
\AtBeginDocument{%
  \let\c@figure\c@lstlisting
  
  \let\ftype@lstlisting\ftype@figure 
}
\makeatother

\usepackage[keeplastbox]{flushend}

\usepackage{siunitx} 
\DeclareSIUnit \MHz{ MHz }
\DeclareSIUnit \GHz{ GHz }
\DeclareSIUnit \dBm{ dBm }
\DeclareSIUnit \dB{ dB}

\newcommand\dB[1]{#1\thinspace{dB}}
\newcommand\dBm[1]{#1\thinspace{dBm}}
\newcommand\MHz[1]{#1\thinspace{MHz}}

\newcommand\commentsMK[1]{{\bf\color{red}$<$MK: #1$>$}}

\newcommand\addedMK[1]{{\color{black}#1}}
\newcommand\addedJD[1]{{\color{black}#1}}

\makeatletter
    \def\@authorfont{\normalsize}
\makeatother
\makeatletter
    \def\@affiliationfont{\normalsize}
\makeatother

\makeatletter
\def\blfootnote{\gdef\@thefnmark{}\@footnotetext}
\makeatother

\renewcommand\footnotetextcopyrightpermission[1]{}

\newif\ifcolumbia
\columbiafalse

\newif\ifarxiv
\arxivtrue

\begin{document}

\title{\addedMK{Outdoor-to-Indoor 28 GHz Wireless Measurements in Manhattan: Path Loss, Environmental Effects, and 90\% Coverage}}


\author{Manav Kohli$^1$, Abhishek Adhikari$^1$, Gulnur Avci$^{1}$, Sienna Brent$^{1}$, Aditya Dash$^1$, Jared Moser$^{2}$, Sabbir Hossain$^{3}$, Igor Kadota$^1$, \\ Carson Garland$^1$, Shivan Mukherjee$^1$, Rodolfo Feick$^4$, Dmitry Chizhik$^{5}$, Jinfeng Du$^{5}$, Reinaldo A. Valenzuela$^5$, Gil Zussman$^1$}
\affiliation{%
\institution{$^1$Columbia University, $^2$Stuyvesant H.S., $^3$City College of New York, $^4$Universidad T\'ecnica Federico Santa Maria, $^5$Nokia Bell Labs \\
\{mpk2138, aa4832, gza2102, scb2197, add2162, ik2496, ctg2137, sm5155, gz2136\}@columbia.edu, jmoser20@stuy.edu, \\ shossai009@citymail.cuny.edu, 
rodolfo.feick@usm.cl, \{dmitry.chizhik, jinfeng.du, reinaldo.valenzuela\}@nokia-bell-labs.com
\country{}
}
}

\renewcommand{\shortauthors}{M. Kohli et al.}
\renewcommand{\shorttitle}{Outdoor-to-Indoor Measurements of 28\thinspace{GHz} Wireless in a Dense Urban Environment}
\renewcommand*{\bibfont}{\normalfont\footnotesize}

\setcopyright{none}
\copyrightyear{2022}
\acmConference[]{ArXiv Report}{May 2022}{}
\acmYear{2022}



\begin{abstract}
\addedMK{Outdoor-to-indoor (OtI) signal propagation further challenges the already tight link budgets at millimeter-wave (mmWave).} To gain insight into \addedMK{OtI mmWave scenarios at 28\thinspace{GHz}, we conducted an extensive measurement campaign consisting of} over \addedMK{2,200} link measurements. \addedMK{In total, \addedMK{43} OtI scenarios were measured in West Harlem, New York City, covering seven highly diverse buildings.} 
\addedJD{The measured OtI path gain can vary by up to 40\thinspace{dB} for a given link distance, and the} empirical path gain model for \addedMK{all data shows} 
\addedJD{an average of 30\thinspace{dB} excess loss over free space at distances beyond 50\thinspace{m}, with an RMS fitting error of \thinspace{11.7 dB}. The} type of glass \addedJD{is found to be the single dominant feature for OtI loss, with} 20\thinspace{dB} observed  difference between \addedJD{empirical path gain models for scenarios with} low-loss and high-loss glass. \addedJD{The presence of scaffolding, tree foliage, or elevated subway tracks, as well as difference in floor height are each found to have an impact between 5--10\thinspace{dB}. }
\addedMK{We show that} \addedJD{for urban buildings with high-loss glass, OtI coverage can support} 500\thinspace{Mbps} for 90\% of indoor user equipment (UEs) with a base station (BS) antenna placed up to 49\thinspace{m} away. \addedJD{For buildings with low-loss glass, such as} \addedMK{our case study covering multiple} classrooms \addedMK{of a} public school\addedMK{, data rates} \addedJD{over 2.5/1.2 Gbps} are possible from a BS 68/175\thinspace{m} away \addedJD{from the school building, when a \addedMK{line-of-sight} path is available}. 
We expect these results to \addedMK{be useful for} the deployment of mmWave networks in dense urban environments as well as the development of relevant \addedMK{scheduling and beam management} algorithms.
\end{abstract}

\maketitle

\section{Introduction}
\label{sec:intro}

Millimeter-wave (mmWave) wireless is a key enabler of 5G-and-beyond networks. \addedMK{Its high-throughput potential makes it particularly viable in a variety of novel solutions, including fixed wireless access for providing Internet connectivity to public schools and public housing, which could help address the digital divide~\cite{fcc2021addressing, nyc2020masterplan}.} However, a major challenge in using mmWave links, particularly in dense urban environments, is their high path loss, which is often exacerbated in outdoor-to-indoor (OtI) scenarios. In order to \addedMK{guide the development of algorithms} (e.g., for beam management~\cite{polese2021deepbeam, fazliu2020graph, sanchez2022millimeter} or link scheduling~\cite{aggarwal2020libra, he2017link}) and to \addedMK{support} deployments (including for indoor coverage by fixed wireless access), there is a need for measurement-based models. 

However, while outdoor-to-outdoor (OtO) and indoor-to-indoor (ItI) propagation scenarios have been extensively measured~\cite{chizhik2020path, raghavan2018millimeter, du202128, jun2020penetration, zhao201328, aslam2020analysis, du2020suburban,  chen201928, xing2018propagation, ko2017millimeter, du2022outdoor, samimi2013angle, shkel2021configurable, xing2021millimeter, zhang2018improving, narayanan2020first, koymen2015indoor, nistlab}, existing OtI datasets are relatively small in size~\cite{diakhate2017millimeter, bas2018outdoor, larsson2014outdoor, du202128, zhao201328, jun2020penetration}. In this paper \emph{we present the results of an extensive OtI mmWave measurement campaign that we conducted in a dense urban environment}.

\begin{figure}[t]
  \centering
  \includegraphics[width=0.88\columnwidth]{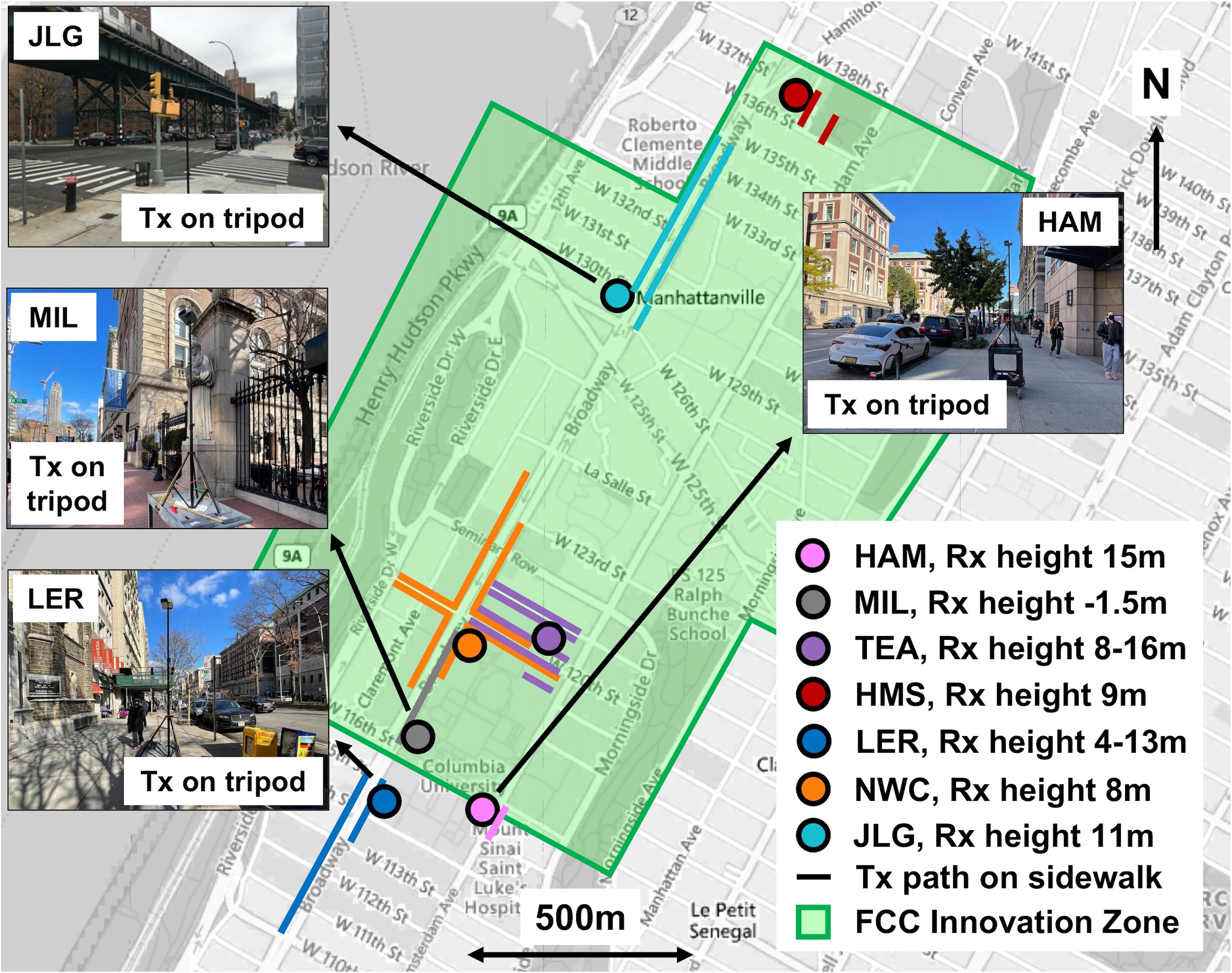}
    \vspace{-0.5\baselineskip}
  \caption{Buildings and corresponding sidewalks where over 2,000 link measurements were collected in and around the COSMOS FCC Innovation Zone in West Harlem, NYC (more details are in Tables~\ref{T:meas_locations} and~\ref{T:measurements}, and Figure~\ref{fig:meas_locations}).}
  \vspace{-1\baselineskip}
  \label{fig:measurementLocations_v2}
\end{figure}



\begin{table*}[t]
\caption{Overview of prior ItI, OtO, and OtI measurement studies in urban or suburban environments in various frequency ranges and various equipment designs.}
\vspace{-\baselineskip}
\footnotesize
\begin{tabular}{|c|c|c|c|c|c|c|c|}
\hline
\textbf{Ref.}     & \textbf{Type} & \textbf{Frequency}   & \textbf{Environment}    & \textbf{Tx Design}               & \textbf{Rx Design}           & \textbf{Bandwidth} & \textbf{\# Tx-Rx Links}  \\ \hline
\cite{chizhik2020path} & ItI & 28\thinspace{GHz} & Urban & Stationary Horn & Rotating Horn & Narrowband  & \textgreater{}1,500            \\ \hline
\cite{raghavan2018millimeter} &  ItI, OtO & 29 \& 60\thinspace{GHz}    & Urban \& Suburban & Rotating Horn         & Rotating Horn       & 200\thinspace{MHz}          & 785                  \\ \hline
\cite{du202128}         & ItI, OtO, OtI      & 28\thinspace{GHz}    & Suburban    & Stationary Horn        & Stationary Horn    & 2\thinspace{GHz}          & 153                  \\ \hline
\cite{jun2020penetration}         &  ItI, OtI      & 60\thinspace{GHz}    & Urban    & 8x1 MIMO Array        & 8x2 MIMO Array    & 4\thinspace{GHz}          & 150                  \\ \hline
\cite{zhao201328}           & ItI, OtI & 28\thinspace{GHz}       & Urban    & Gimbal-mounted Horn     & Gimbal-mounted Horn   & 400\thinspace{Mcps}          & 18                  \\ \hline
\cite{aslam2020analysis}      & OtO   & 60\thinspace{GHz}       & Urban    & 36x8 Phased Array       & 36x8 Phased Array   & 2.16\thinspace{GHz}         & 15                  \\ \hline
\cite{du2020suburban}          & OtO   & 28\thinspace{GHz}      & Suburban & Stationary Horn         & Rotating Horn       & Narrowband      & \textgreater{}2,000            \\ \hline
\cite{chen201928}          & OtO   & 28\thinspace{GHz}      & Urban & Omnidirectional         & Rotating Horn       & Narrowband      & \textgreater{}1,500            \\ \hline
\cite{diakhate2017millimeter}  & OtI  & 60\thinspace{GHz}       & Urban    & Stationary Horn         & Stationary Horn     & 125\thinspace{MHz}          & 76                  \\ \hline
\cite{bas2018outdoor}          & OtI  & 28\thinspace{GHz}       & Urban    & 8x2 Phased Array        & 8x2 Phased Array    & 400\thinspace{MHz}          & 29                  \\ \hline
\cite{larsson2014outdoor}          & OtI  & 28\thinspace{GHz}       & Suburban    & Stationary Slot Array        & Stationary Parabolic Dish    & 50\thinspace{MHz}          & 43                  \\ \hline
\textit{This work}                   & OtI    & 28\thinspace{GHz}    & Urban    & Omnidirectional         & Rotating Horn       & Narrowband                & \textgreater{}2,200               \\ \hline
\end{tabular}
\vspace{-0.5\baselineskip}
\label{T:work_comparison}
\end{table*}

\noindent \textbf{Measurements:} As illustrated in Figure~\ref{fig:measurementLocations_v2}, we conducted a large-scale measurement campaign in and around the COSMOS FCC Innovation Zone in West Harlem, New York City (NYC)~\cite{fcc2021innovation, raychaudhuri2020challenge}. Using a 28\thinspace{GHz} channel sounder \cite{du2020suburban}, we collected over \addedMK{2,200} OtI measurements (comprising over 32 million individual power measurements) \addedMK{across 43 OtI scenarios in seven very diverse buildings, covering a variety of construction materials and building utility.}



\noindent\textbf{Models:} \addedMK{We develop path gain models for each OtI scenario using a single-slope exponent fit to the measured data as a function of distance, and record the CDF of the measured azimuth beamforming (BF) gain. We also develop clustered models covering scenarios at each building, and aggregate models to  study specific effects, including (i) the type of glass used for the windows (low- or high-loss glass), (ii) base station (BS) antenna placement in front of or behind an elevated subway track, (iii) user equipment (UE) placement on upper/lower floors of the building, and (iv) the angle of incidence (AoI) of the mmWave signal into the building. Additionally, we measure the impact of scaffolding and tree foliage on the path loss and azimuth beamforming gain models. Using these clusters, we show, among other things: (i) a $20$\thinspace{dB} additional loss for the high-loss glass when compared to low-loss glass, (ii) a 10\thinspace{dB} difference in path gain for BSes blocked by elevated subway tracks or UEs on different floors, and (iii) a 5--6\thinspace{dB} total impairment on link budget caused by scaffolding or tree foliage.}

\noindent\textbf{Case Study - Public School:} \addedMK{We consider the Hamilton Grange public school in West Harlem as a case study and provide an in-depth discussion of the OtI scenarios in that school. The low path loss that we observed in this building along with its location in an area with below-average Internet access make it of particular interest for mmWave OtI coverage via fixed wireless access. We find that the path gain models associated with different classrooms are within 5\thinspace{dB} across an 80\thinspace{m} span of BS placements, suggesting that uniform OtI coverage can be achieved.}

\noindent\addedMK{\textbf{Inter-user Interference (IUI):} We evaluate IUI with an OtI scenario in a classroom building using a typical street intersection BS placement. For indoor users located far from the BS, we find that IUI can be significant, with a median correlation coefficient of 0.75 between the directions of received power at the BS. This could hamper the BS' ability to serve multiple users with multiple beams.}

\noindent\textbf{Coverage:} \addedMK{We calculate achievable data rates for an indoor UE using the path gain models for low-loss and high-loss glass. Our analysis shows that data rates in excess of \addedMK{2.5}\thinspace{Gbps} are possible in low-loss glass OtI scenarios for up to 90\% of users with the BS up to 68\thinspace{m} away, \addedMK{and over 1.2\thinspace{Gbps} up to 175\thinspace{m} away}. For high-loss glass OtI scenarios, we find achievable data rates in excess of 500\thinspace{Mbps} for BS placements up to 49\thinspace{m} away, demonstrating the significant impact of the glass material.}


To the best of our knowledge, this is the first paper to present an extensive OtI mmWave measurement campaign and accompanying path gain models which are then used to study OtI coverage. We anticipate that these results will \addedMK{be useful for} the deployment of mmWave BSes capable of providing OtI coverage in dense urban environments as well as the development of relevant algorithms.

\addedMK{The rest of the paper is organized as follows.} In Section~\ref{sec:related}, we discuss related work. In Section~\ref{sec:measurements}, we describe the measurement campaign, including equipment, locations, and method. In Section~\ref{sec:results}, we develop path gain models from the measurement data. In Section~\ref{sec:school}, we focus on \addedMK{the} case study of a public school. In Section~\ref{sec:reverse_tc} we discuss the potential of multi-user support in OtI scenarios, and in Section~\ref{sec:OtI-coverage} we derive achievable data rates. Finally, we conclude and discuss future work in Section~\ref{sec:conclusion}.

\section{Related Work}
\label{sec:related}

\begin{table*}[t!]
\caption{Measurement locations considered, as shown in Figures~\ref{fig:measurementLocations_v2} and \ref{fig:meas_locations}. Corresponding OtI scenarios are in Table~\ref{T:measurements}.}
\vspace{-\baselineskip}
\footnotesize
\begin{tabular}{|l|c|l|c|c|c|}
\hline
\multicolumn{1}{|l|}{\textbf{Building Name}} & \textbf{Abbreviation}  & \multicolumn{1}{|l|}{\textbf{Purpose}}        & \textbf{Year}  & \textbf{Construction}          & \textbf{Glass Type}  \\ \hline
Hamilton Hall & \textbf{HAM} & Classroom Building & 1907  & Brick \& concrete     & Low-e       \\ \hline
Miller Theatre & \textbf{MIL} & Theater & 1918  & Brick \& concrete     & Low-e       \\ \hline
Teachers' College & \textbf{TEA} & Classroom and Office Building & 1924  & Brick \& concrete     & Low-e       \\ \hline
M209 Hamilton Grange Middle School & \textbf{HMS} & Public School & 1928  & Brick \& concrete     & Traditional \\ \hline
Lerner Hall & \textbf{LER} & Student Center & 1999  & Brick \& concrete     & Low-e       \\ \hline
Northwest Corner Building & \textbf{NWC} & Laboratory Building & 2008  & Glass, metal, stone   & Low-e       \\ \hline
Jerome L. Greene Science Center & \textbf{JLG} & Laboratory Building & 2017  & Glass \& metal        & Low-e       \\ \hline
\end{tabular}
\label{T:meas_locations}
\end{table*}

\ifcolumbia
\else

\begin{figure*}[t]
\subfloat[Hamilton Hall (HAM)]{
\includegraphics[width=0.15\linewidth]{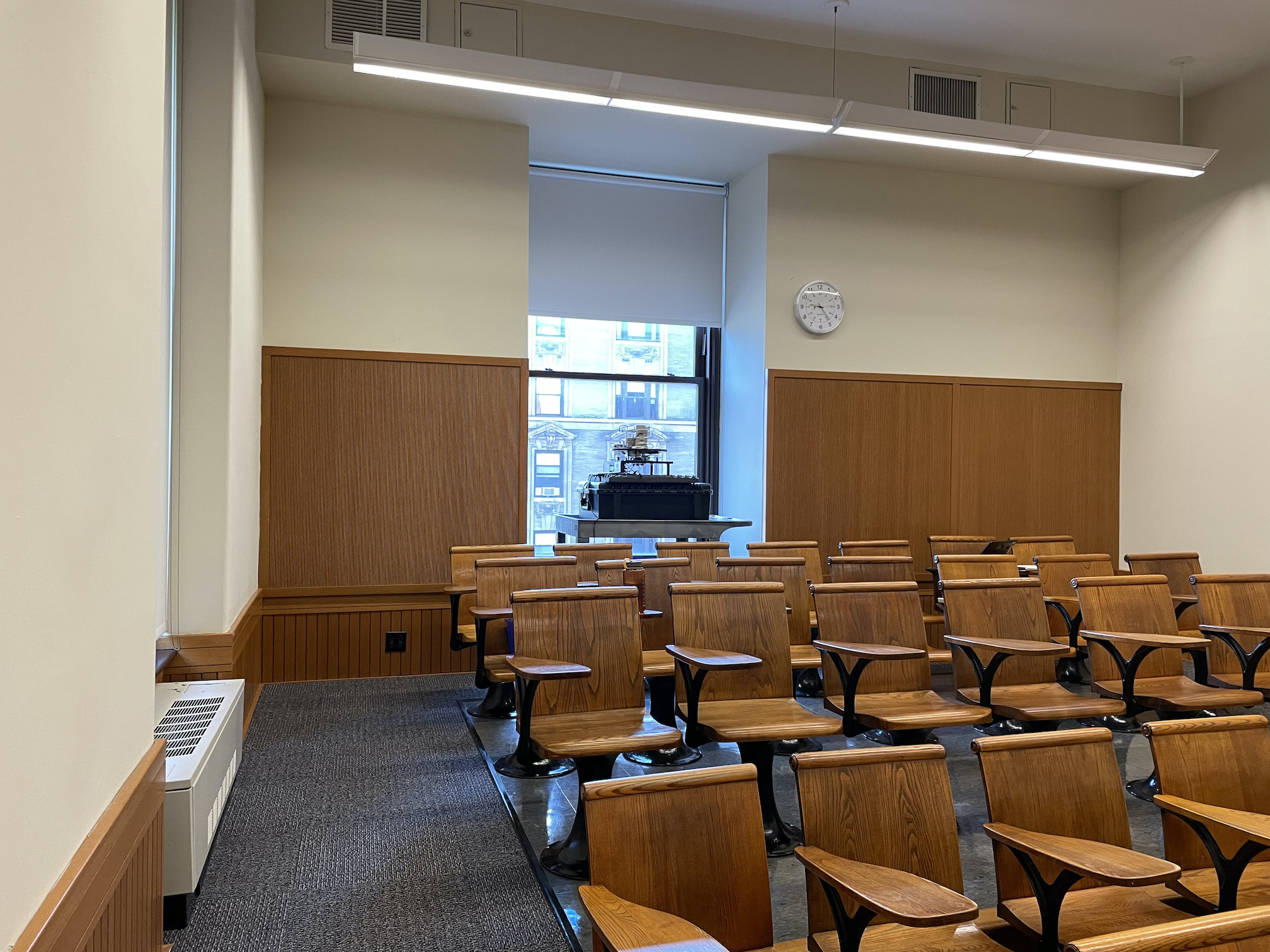}
\hspace{0.5pt}
\includegraphics[width=0.15\linewidth]{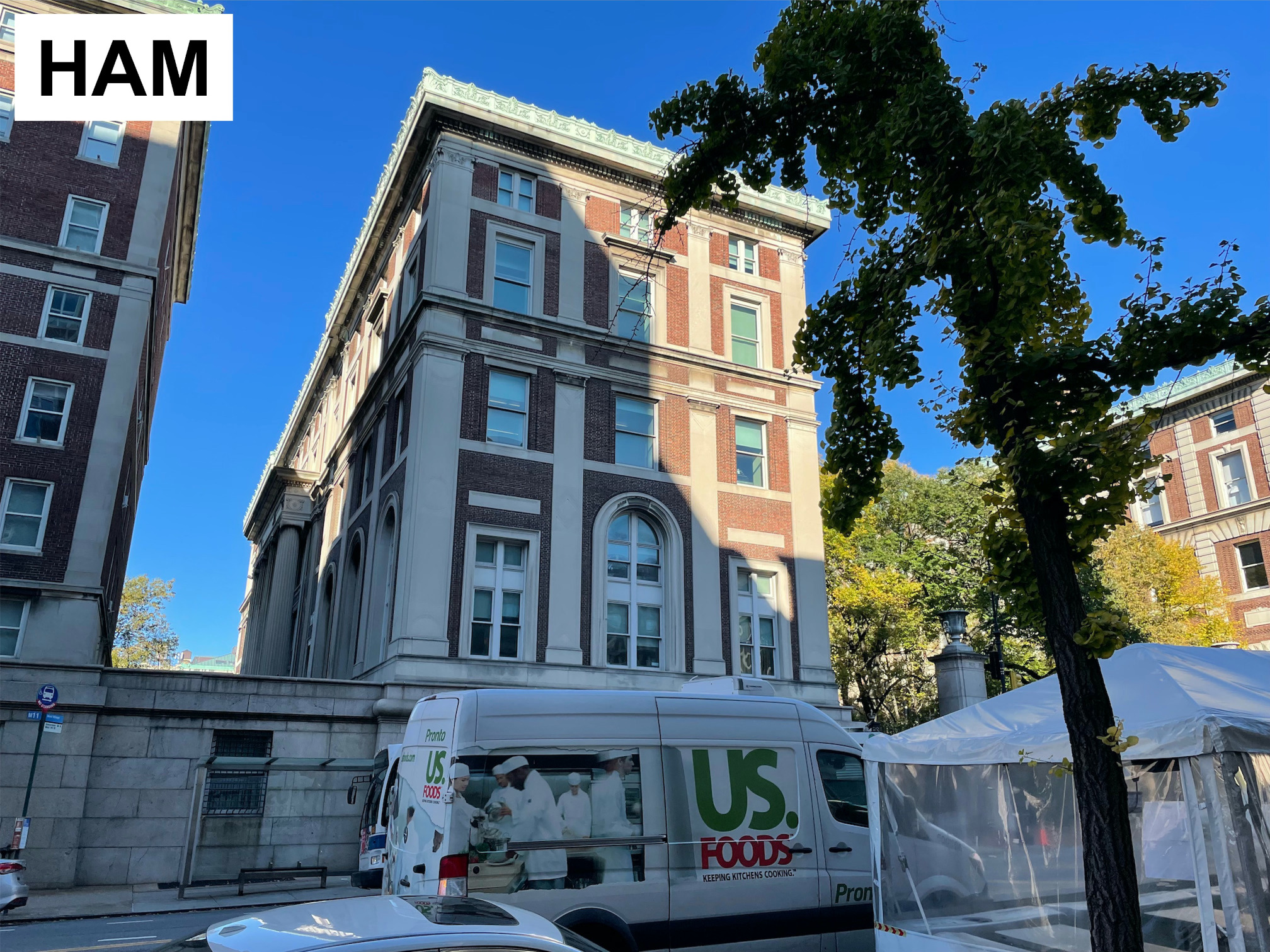}
\label{fig:rx-ham}}
\hspace{1pt}
\subfloat[Miller Theatre (MIL)]{
\includegraphics[width=0.15\linewidth]{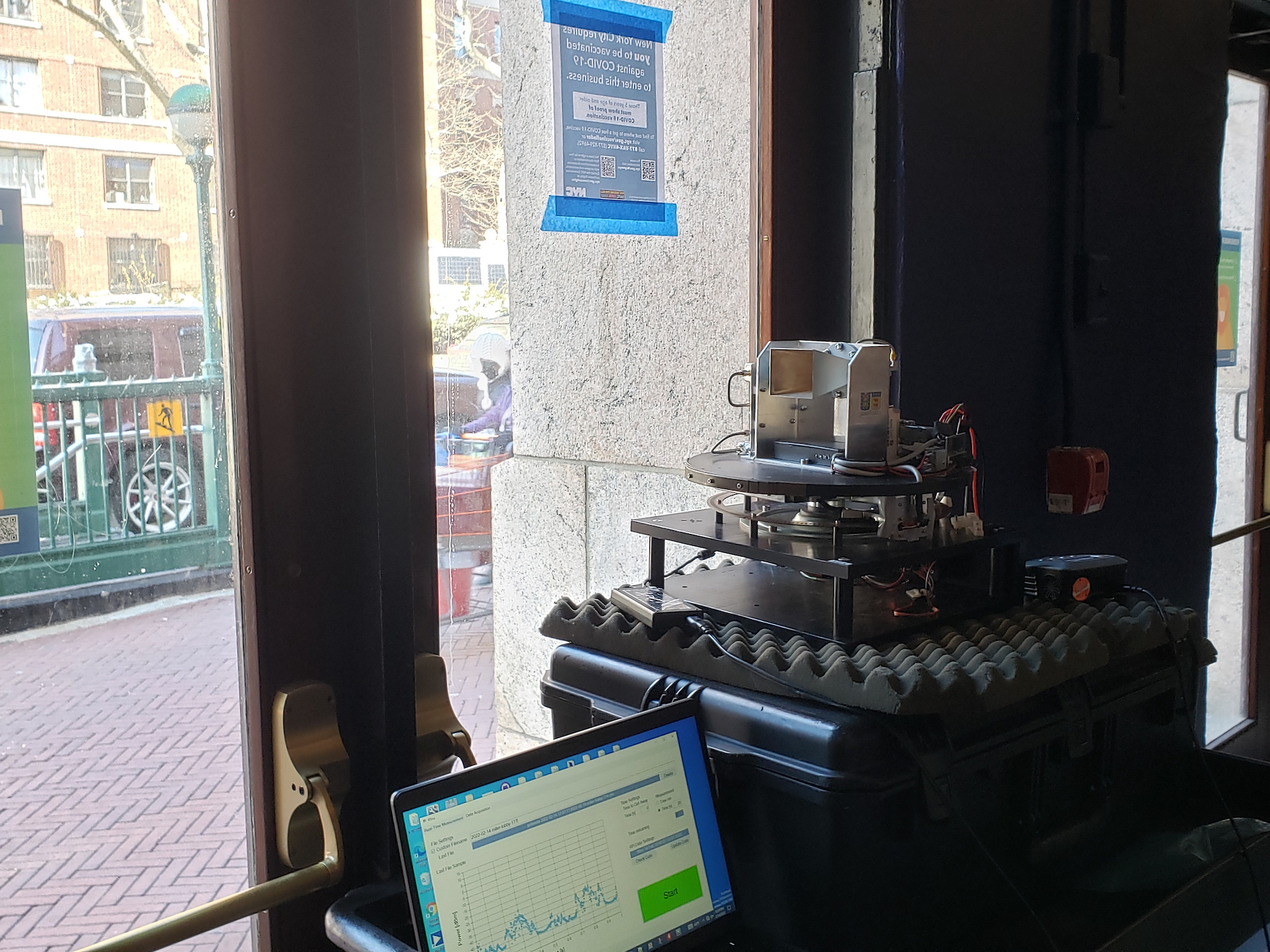}
\hspace{0.5pt}
\includegraphics[width=0.15\linewidth]{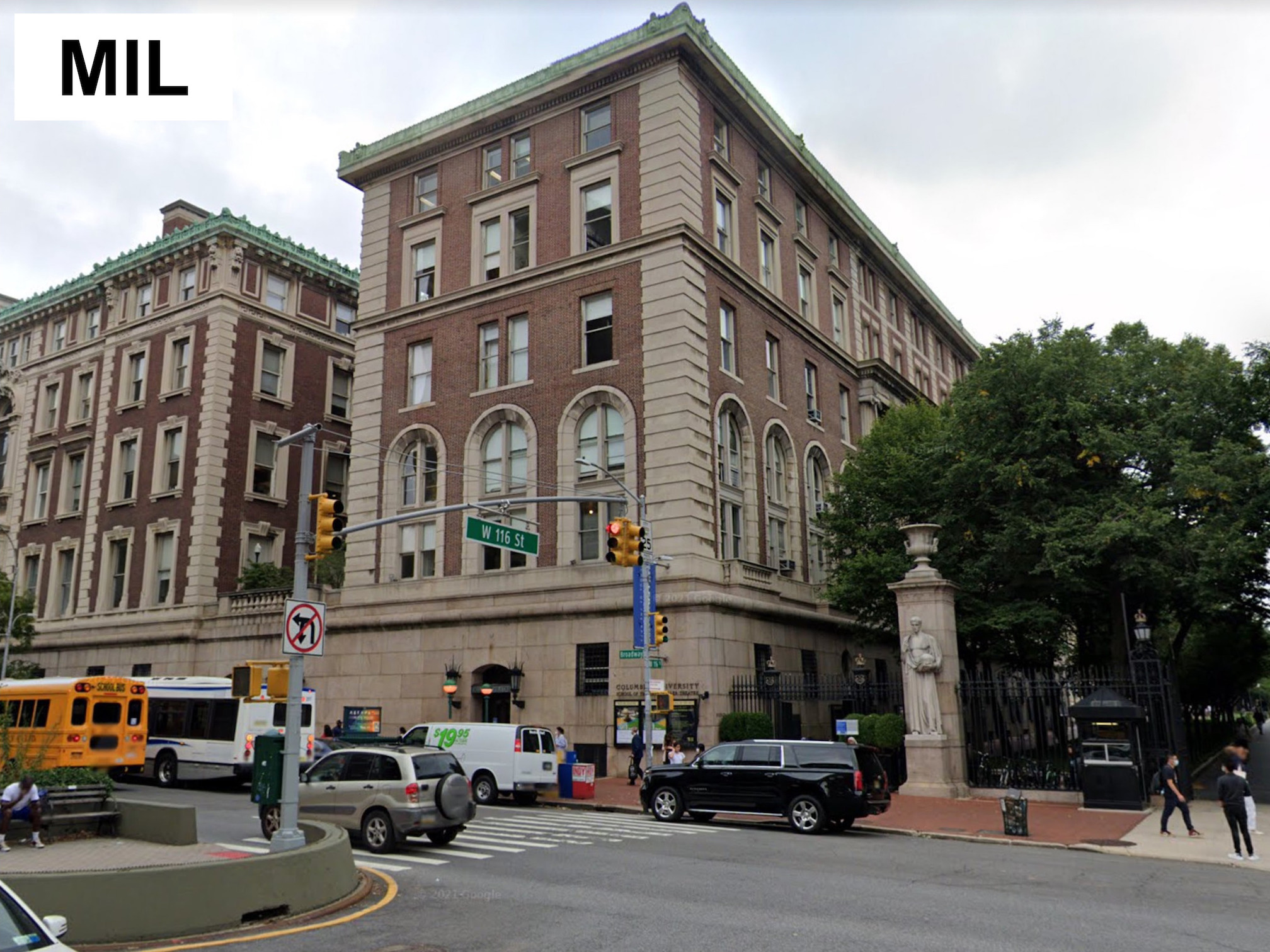}
\label{fig:rx-mil}}
\hspace{1pt}
\subfloat[Teachers' College (TEA)]{
\includegraphics[width=0.15\linewidth]{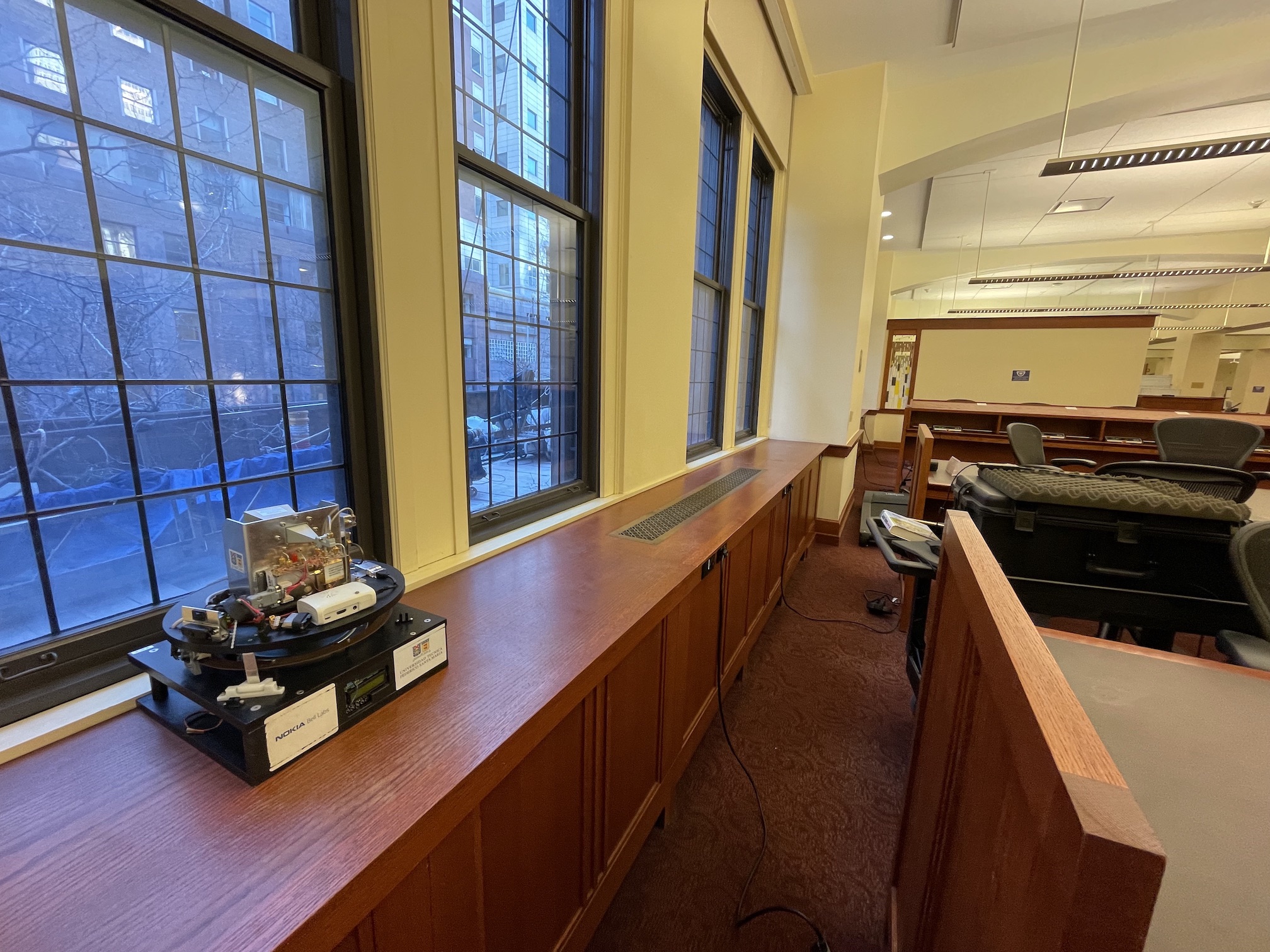}
\hspace{0.5pt}
\includegraphics[width=0.15\linewidth]{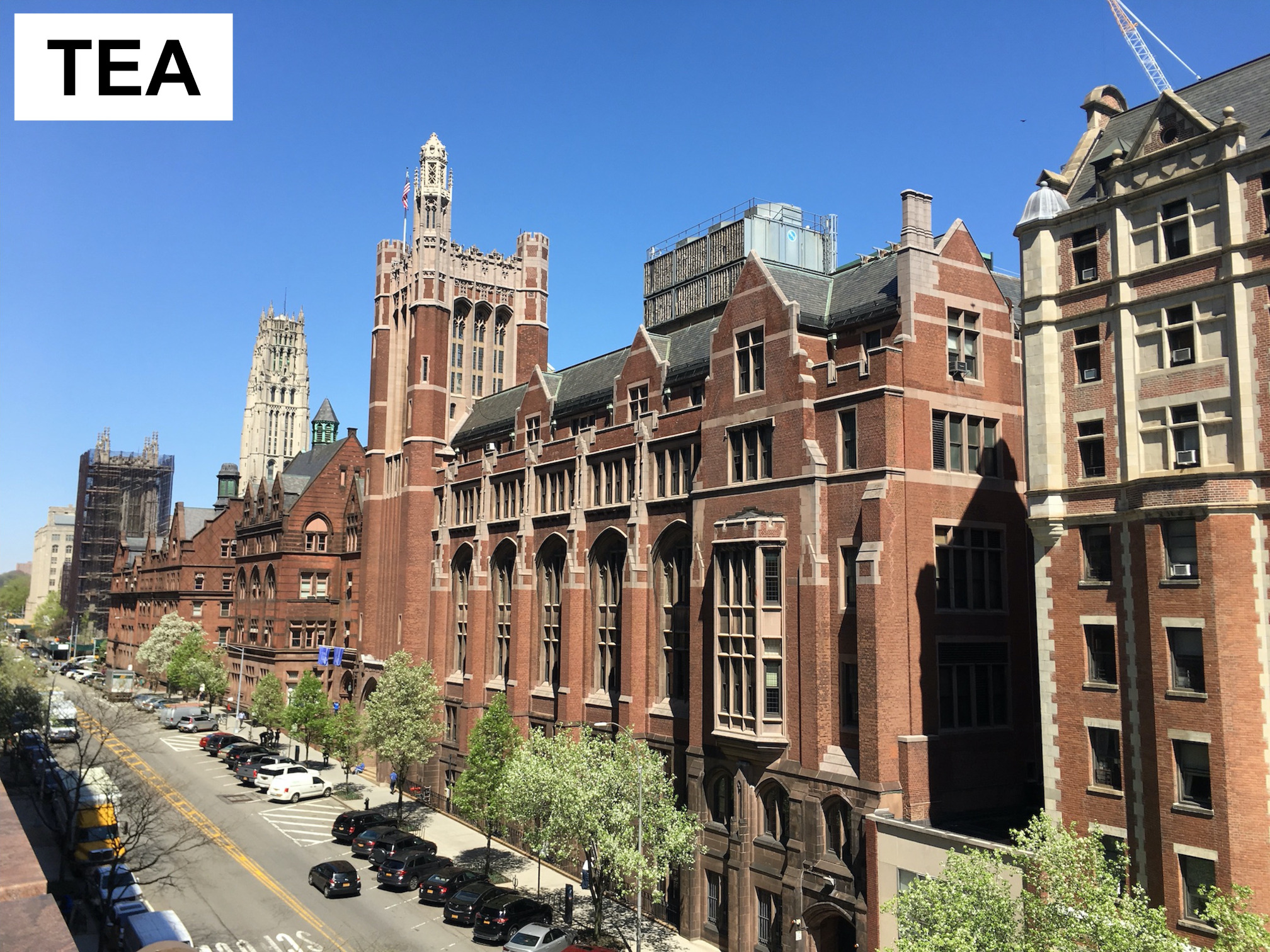}
\label{fig:rx-tc}}
\\
\subfloat[Hamilton Grange Middle School (HMS)]{
\includegraphics[width=0.15\linewidth]{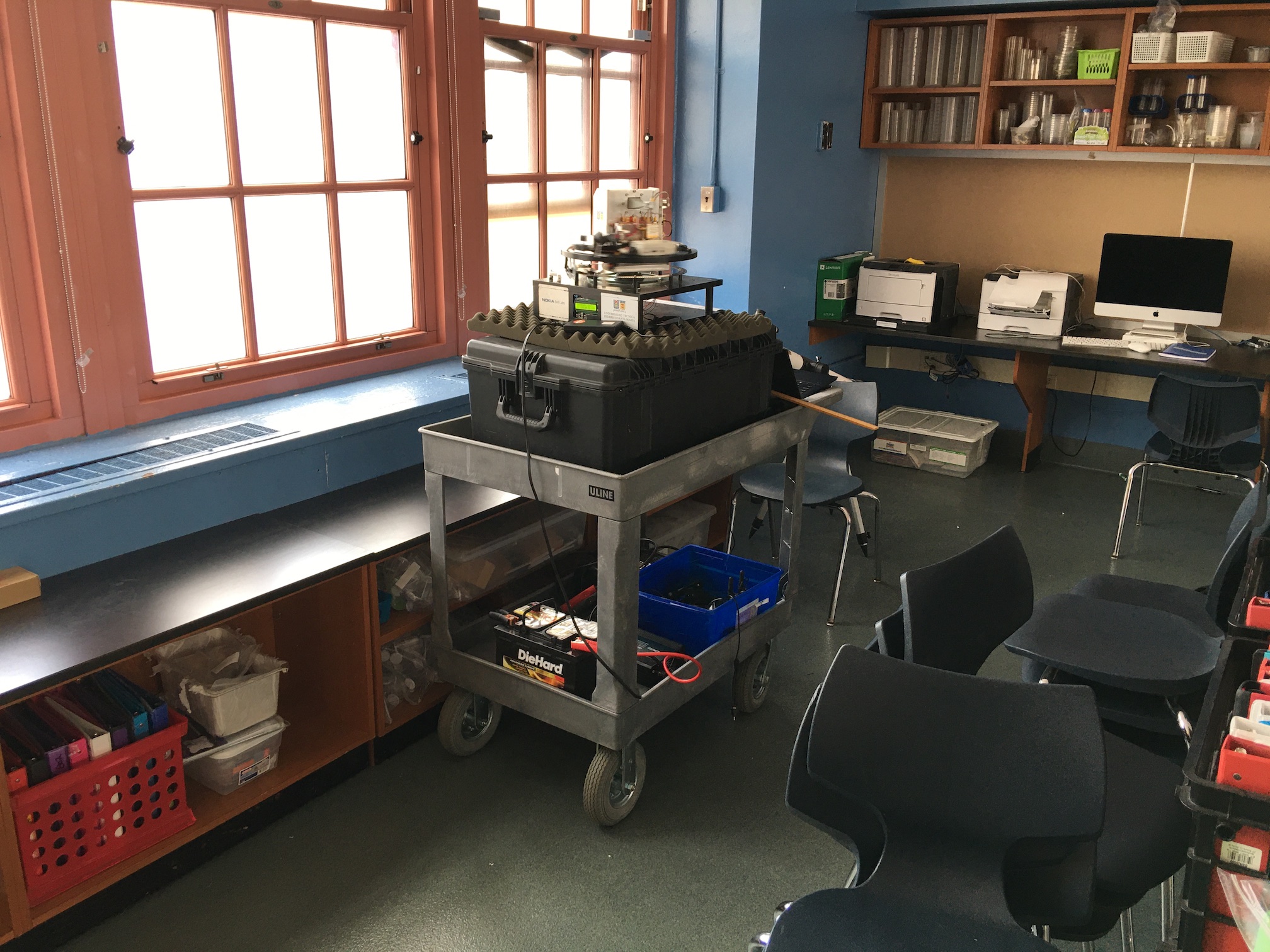}
\hspace{0.5pt}
\includegraphics[width=0.15\linewidth]{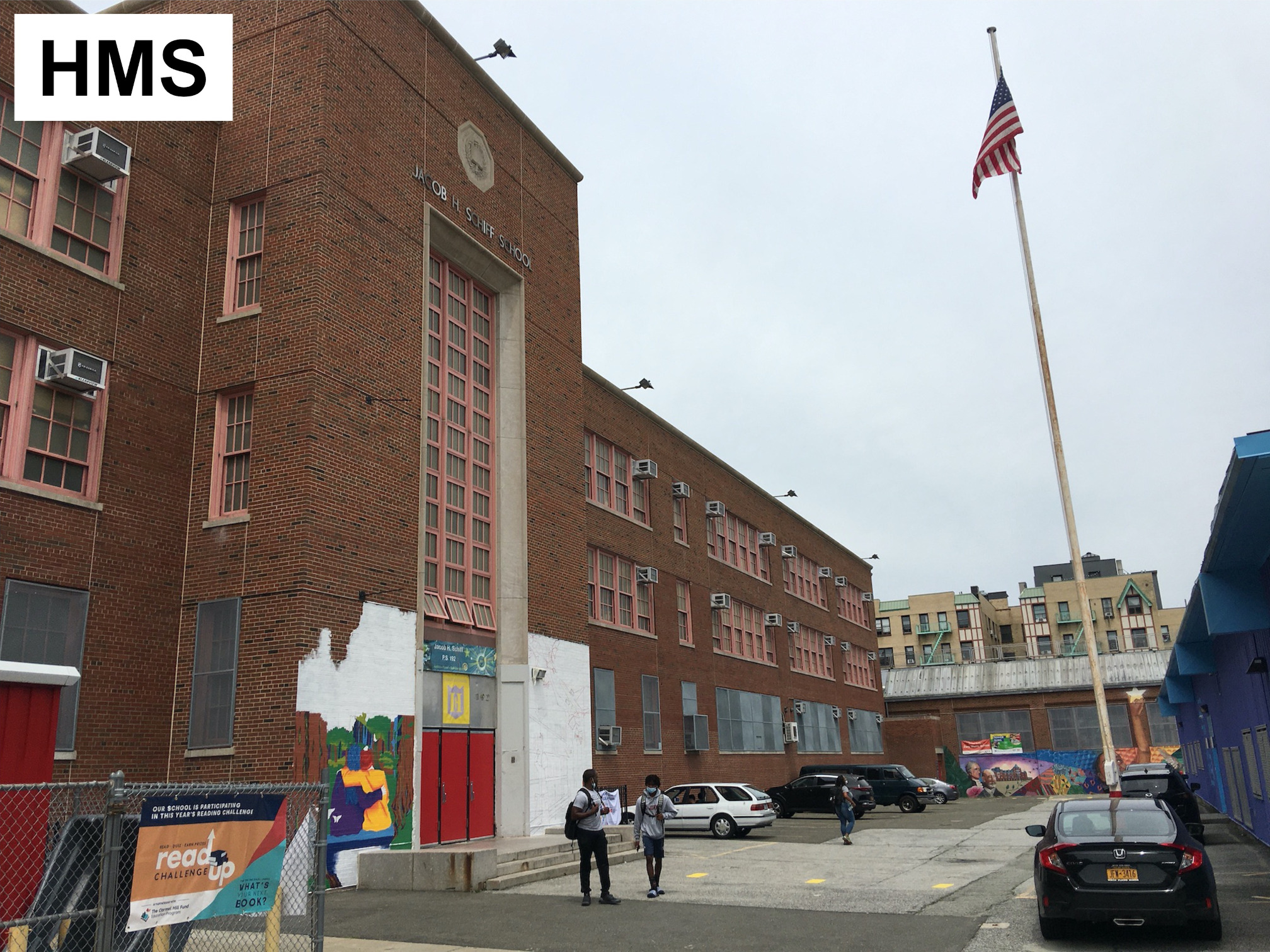}
\label{fig:rx-hms}}
\hspace{1pt}
\subfloat[Lerner Hall (LER)]{
\includegraphics[width=0.15\linewidth]{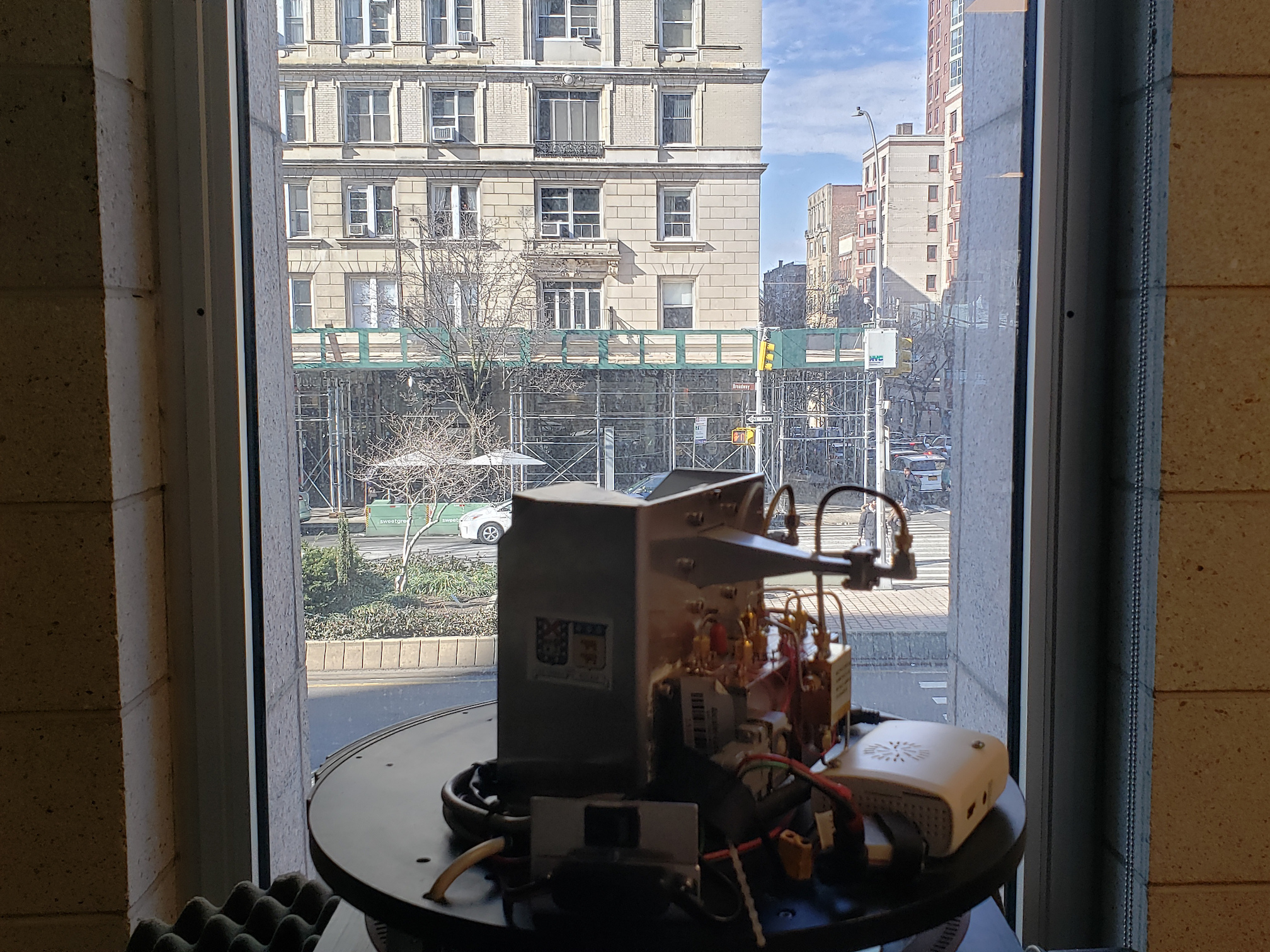}
\hspace{0.5pt}
\includegraphics[width=0.15\linewidth]{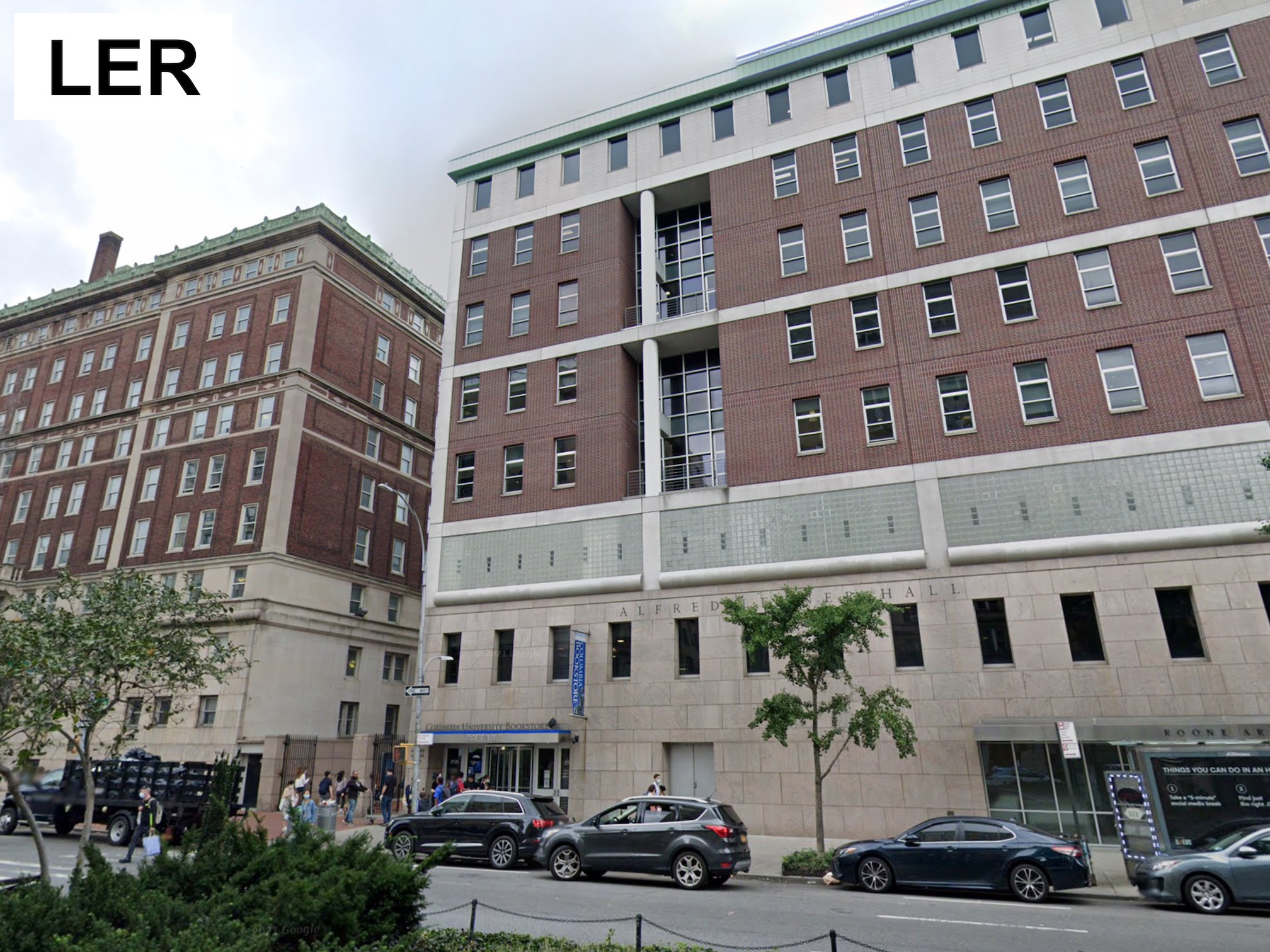}
\label{fig:rx-ler}}
\hspace{1pt}
\subfloat[Northwest Corner Building (NWC)]{
\includegraphics[width=0.15\linewidth]{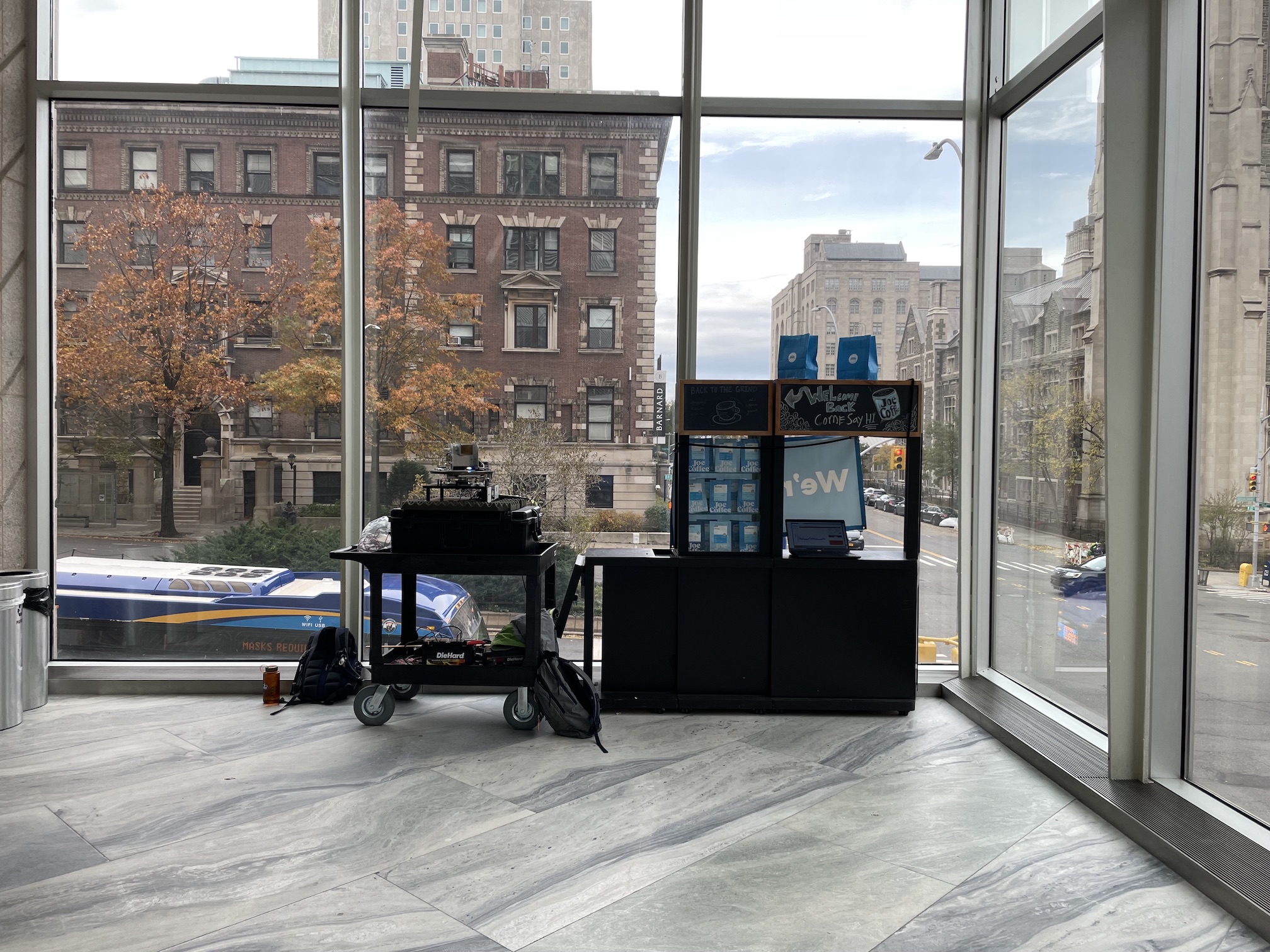}
\hspace{0.5pt}
\includegraphics[width=0.15\linewidth]{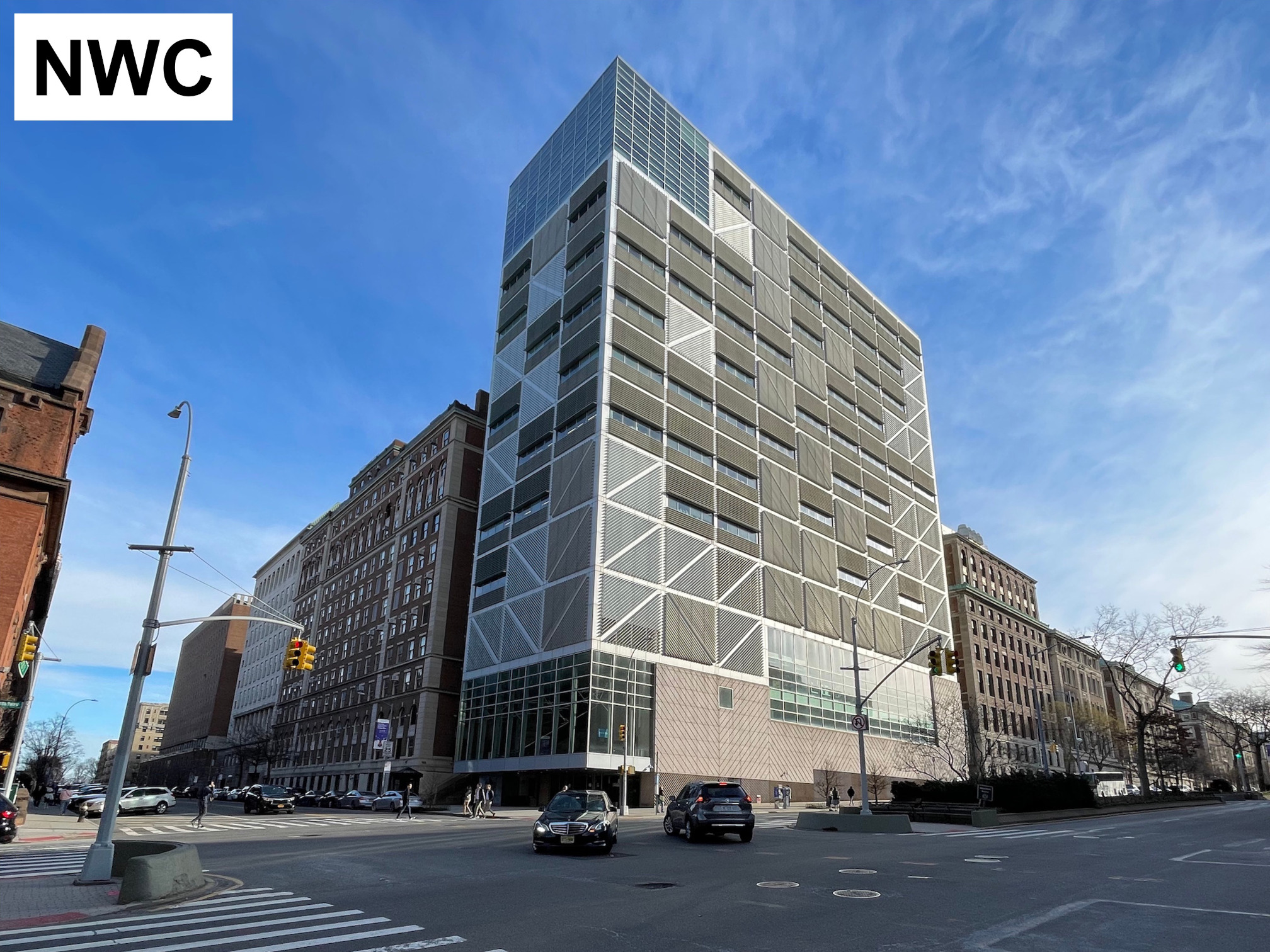}
\label{fig:rx-nwc}}
\vspace{-1\baselineskip}
\caption{Representative interior and exterior views of six of the seven locations (shown in Figure~\ref{fig:measurementLocations_v2} and Table~\ref{T:meas_locations}).}
\vspace{-0.5\baselineskip}
\label{fig:meas_locations}
\end{figure*}

\fi

Table~\ref{T:work_comparison} provides an overview of a subset of prior efforts. As seen in the table, mmWave measurement studies typically require the use of specialized channel sounders and may be further categorized based on the type: OtO~\cite{raghavan2018millimeter, chen201928, ko2017millimeter, samimi2013angle, du2020suburban, xing2021millimeter, du202128, du2022outdoor, aslam2020analysis, shkel2021configurable, zhang2018improving, narayanan2020first, nistlab}, ItI~\cite{chizhik2020path, raghavan2018millimeter, ko2017millimeter, jun2020penetration, zhao201328, xing2018propagation, du202128, shkel2021configurable, koymen2015indoor, nistlab}, and OtI~\cite{diakhate2017millimeter, bas2018outdoor, larsson2014outdoor, zhao201328, jun2020penetration, du202128}, as well as the frequency range and urban or suburban environment. Datasets that include outcomes of some of these studies are available in~\cite{nistlab} and a review of several efforts at 60\thinspace{GHz} for a specific type of sounder is available in~\cite{shkel2021configurable}.

OtO measurements have focused on a variety of environments, including urban~\cite{chen201928, du202128},  suburban~\cite{du2020suburban, du202128}, and rural~\cite{schmieder2018measurement} mmWave deployment scenarios. Conversely, ItI measurements have primarily focused on office buildings~\cite{chizhik2020path, zhao201328, jun2020penetration, koymen2015indoor}. While such indoor environments represent a significant use case for mmWave wireless, especially with the recent approval of the 802.11ay standard~\cite{ieee202180211ay}, they represent only one building type. 

Previous OtI measurements include those at a regional airport~\cite{du202128} and measurements of office space using a \addedMK{receiver (Rx)} mounted on a robot and a stationary \addedMK{transmitter (Tx)}~\cite{jun2020penetration}. Other forms of Tx/Rx mounting have been used, such as a Tx mounted on a van with indoors Rx~\cite{diakhate2017millimeter}. Phased array antennas have also been used at the Tx and Rx~\cite{bas2018outdoor}, with 90$^\circ$ beamsteering capability and 5$^\circ$ resolution. Longer-term measurements have also been studied, including a four-day measurement with the indoor Rx and outdoor Tx both kept stationary~\cite{larsson2014outdoor}. Finally, 28\thinspace{GHz} OtI measurements have been collected in NYC using a fixed Tx and Rx~\cite{zhao201328}.


While some OtI measurements are available, to the best of our knowledge (and as can be seen in Table~\ref{T:work_comparison}), this paper is the \emph{first large-scale, measurement-driven study into the mmWave channel for OtI scenarios in a dense urban environment, leading to \addedMK{reliable} statistical models for the path loss and beamforming gain degradation as well as quantitative insights into realistic data rates.}


\section{Measurement Campaign}
\label{sec:measurements}
In this section, we describe the measurement locations, equipment, and scenarios in the OtI measurement campaign. 

\noindent\textbf{Locations:} Figure~\ref{fig:measurementLocations_v2} and Table~\ref{T:meas_locations} show seven buildings where measurements were conducted. These buildings are located in and around the FCC Innovation Zone~\cite{fcc2021innovation} associated with the NSF PAWR COSMOS testbed~\cite{raychaudhuri2020challenge} in West Harlem, NYC. In Figure~\ref{fig:measurementLocations_v2}, the locations of these buildings are shown along with the corresponding outdoor sites (sidewalks, parking lot, and basketball court). Photos of these buildings are shown as insets in Figure~\ref{fig:measurementLocations_v2} and in Figure~\ref{fig:meas_locations}. Table~\ref{T:measurements} lists the OtI scenarios for the seven buildings. \ifarxiv \addedMK{Each building is described in detail below, including their location in terms of NYC street intersections, and the type of glass used, which is discussed in further detail in Section~\ref{sec:results-glass}.} \fi

\ifarxiv

\noindent\addedMK{\textit{\textbf{HAM}: \textbf{Ham}ilton Hall}. \textbf{HAM} is a fifth-floor classroom at Hamilton Hall, located at the intersection of W. 116\textsuperscript{th} Street and Amsterdam Avenue. This building was completed in 1907 with a brick-and-concrete construction shown in Figure~\ref{fig:meas_locations}\subref{fig:rx-ham}. The windows have been renovated with modern glass in recent years.}
\\
\noindent\addedMK{\textit{\textbf{MIL}: \textbf{Mil}ler Theatre}. \textbf{MIL} is the first-floor entrance to the Miller Theatre, located by the intersection of W. 116\textsuperscript{th} Street and Broadway. This brick-and-concrete building was originally completed in 1918 and renovated in 1988. The windows are glazed with modern glass panels. The exterior construction is very similar to \textbf{HAM}, as seen in Figure~\ref{fig:meas_locations}\subref{fig:rx-mil}.}
\\
\noindent\addedMK{\textit{\textbf{TEA}: \textbf{Tea}chers College}. \textbf{TEA} covers a first-floor cafeteria and the first, second, and third floors of a library located within Russell Hall at Teachers' College. This building was completed in 1924 and has a complex facade constructed with brick and concrete seen in Figure~\ref{fig:meas_locations}\subref{fig:rx-tc}. The library and cafeteria overlook 120\textsuperscript{th} and 121\textsuperscript{st} Street between Broadway and Amsterdam Avenue, respectively. The windows at \textbf{TEA} were renovated in 2001 with modern glass.}
\\
\noindent\addedMK{\textit{\textbf{HMS}: \textbf{H}amilton Grange \textbf{M}iddle \textbf{S}chool}. \textbf{HMS} covers a set of third-floor classrooms at M209 Hamilton Grange Middle School (HGMS) located in West Harlem, NYC, at Broadway and 135\textsuperscript{th} Street. Each measured classroom contains a length of older single-glazed windows spanning one-third of the exterior wall, shown in Figure~\ref{fig:meas_locations}\subref{fig:rx-hms}. Each classroom is an open space, with no pillars, but has a relatively low ceiling due to the older building construction. The building is primarily constructed out of brick and concrete.}
\\
\noindent\addedMK{\textit{\textbf{LER}: \textbf{Ler}ner Hall}. \textbf{LER} is a second- and fifth-floor student recreation building located at the intersection of 115\textsuperscript{th} Street and Broadway. The building was completed in 1999 and the exterior facing Broadway has a brick face. The windows on the second and fifth floors are recessed 2\thinspace{m} into the building face, as seen in Figure~\ref{fig:meas_locations}\subref{fig:rx-ler}.}
\\
\noindent\addedMK{\textit{\textbf{NWC}: \textbf{Northwest} \textbf{C}orner Building}. \textbf{NWC} is a coffee shop within the second floor of the Northwest Corner Building at the intersection of 120\textsuperscript{th} Street and Broadway. This building was completed in 2008 with an exterior primarily made of glass and metal. The coffee shop is an open space with a high ceiling and floor-to-ceiling modern glass windows overlooking the intersection seen in Figure~\ref{fig:meas_locations}\subref{fig:rx-nwc}. Stone walls encircle the rest of the coffee shop not facing the street.}
\\
\noindent\addedMK{\textit{\textbf{JLG}: \textbf{J}erome \textbf{L}. \textbf{G}reene Science Center}. \textbf{JLG} is a third-floor corner office and common area at the Jerome L. Greene Science Center, located on the northwest corner of 129\textsuperscript{th} Street and Broadway. \textbf{JLG} was completed in 2017 and has an exterior construction primarily made of glass and metal. In particular, the common area and corner office are encircled by floor-to-ceiling windows overlooking Broadway, including a raised portion of the 1 line of the NYC Subway. The building and subway track are visible in the inset in Figure~\ref{fig:measurementLocations_v2}.}

\fi

\begin{table*}[t!]
\centering
\caption{\addedMK{40} OtI measurement scenarios with computed path gain model and median azimuth beamforming gain.}
\vspace{-\baselineskip}
\footnotesize
\begin{tabular}{| c | c | c | c | c | c | c | c | c | c | c |}
\hline
\textbf{Name} & \textbf{Color} & \textbf{Group} & \textbf{Range \addedMK{(m)}} & \textbf{Step (m)} & \textbf{\# Links} & \textbf{Slope (dB)} & \textbf{Intercept (dB)} & \textbf{RMS (dB)} & \textbf{Median $G_{az}$ (dBi)} \\
\hline
HAM-S-E &  Pink  & HAM  & 61 & 1 & 62 & -6.61 & -23.7 & 3.5 & 11.1 \\ \hline
MIL-N-E &  Gray  & MIL  & 155 & 2.5 & 76 & -3.53 & -59.1 & 2.8 & 11.0 \\ \hline
TEA-S-N-1-Sc & Purple & TEA  & 230 & 6/8 & 35 & -2.56 & -95.3 & 5.6 & 11.0 \\ \hline
TEA-S-S-1-Sc & Purple & TEA  & 228 & 4/8 & 45 & -3.49 & -75.1 & 4.8 & 10.9 \\ \hline
TEA-S-S-2 & Purple & TEA  & 155 & 3 & 52 & -5.52 & -40.5 & 2.6 & 7.7 \\ \hline
TEA-S-S-3 & Purple & TEA  & 232 & 3 & 77 & -5.13 & -36.1 & 3.3 & 8.8 \\ \hline
TEA-S-Bal-1 & Purple & TEA  & 85 & 3 & 29 & -1.61 & -107.9 & 4.7 & 9.7 \\ \hline
TEA-S-Bal-2 & Purple & TEA  & 85 & 3 & 29 & -0.69 & -111.3 & 4.2 & 7.8 \\ \hline
TEA-S-Bal-3 & Purple & TEA  & 37 & 3 & 13 & -5.20 & -33.6. & 4.3 & 10.0 \\ \hline
TEA-N-N & Purple & TEA  & 243 & 3 & 68 & -4.45 & -53.0 & 4.1 & 10.8 \\ \hline
TEA-N-S & Purple & TEA  & 243 & 3 & 81 & -4.80 & -41.0 & 4.1 & 10.1 \\ \hline
HMS-Lot-307 & Maroon & HMS  & 62 & 1 & 63 & -3.22 & -60.4 & 1.6 & 10.4 \\ \hline
HMS-Lot-317 & Maroon & HMS  & 62 & 1 & 63 & -3.48 & -52.0 & 3.4 & 11.5 \\ \hline
HMS-Lot-321 & Maroon & HMS  & 62 & 1 & 63 & -4.12 & -44.1 & 3.4 & 11.8 \\\hline
HMS-Lot-323 & Maroon & HMS  & 62 & 1 & 63 & -4.10 & -47.2 & 2.5 & 9.9 \\\hline
HMS-Lot-325 & Maroon & HMS  & 62 & 3 & 22 & -3.40 & -54.8 & 2.5 & 10.8 \\ \hline
HMS-Court-307 & Maroon & HMS  & 42 & 1 & 43 & -5.47 & -3.9 & 2.9 & 13.3 \\ \hline
HMS-Court-317 & Maroon & HMS  & 39 & 1 & 40 & -6.48 & 11.2 & 3.2 & 12.0 \\\hline
HMS-Court-321 & Maroon & HMS  & 57 & 1 & 58 & -8.50 & 51.1 & 3.1 & 11.0 \\ \hline
HMS-Court-323 & Maroon & HMS  & 57 & 1 & 58 & -8.13 & 43.6 & 1.6 & 9.8 \\ \hline
HMS-Court-325 & Maroon & HMS  & 58 & 1 & 59 & -1.88 & -84.3 & 2.2 & 10.2 \\ \hline
LER-S-W-5 &  Blue  & LER  & 298 & 3 & 96 & -5.29 & -19.6 & 3.0 & 10.8 \\ \hline
LER-S-W-2 &  Blue  & LER  & 110 & 8 & 14 & -6.72 & -22.8 & 4.2 & 9.4 \\ \hline
LER-S-E-2 &  Blue  & LER  & 95 & 6 & 23 & -3.97 & -75.2 & 3.8 & 9.4 \\ \hline
NWC-N-W-Sc-NLe &  Orange  & NWC & 227 & 3/6 & 70  & -3.26 & -76.0 & 3.5 & 11.1 \\ \hline
NWC-N-W-NSc-NLe &  Orange  & NWC & 197 & 3/6 & 65  & -3.03 & -76.9 & 4.7 & 12.8 \\ \hline
NWC-N-E-NSc-Le &  Orange  & NWC & 201 & 3 & 60 & -3.52 & -73.0 & 1.9 & 11.2 \\ \hline
NWC-N-E-NSc-NLe &  Orange  & NWC & 174 & 3 & 51 & -3.38 & -71.5 & 2.5 & 12.1 \\ \hline
NWC-E-N-Sc-NLe &  Orange  & NWC  & 131 & 3 & 44 & -4.62 & -56.4 & 2.0 & 8.8 \\ \hline
NWC-E-N-NSc-NLe &  Orange  & NWC  & 131 & 3 & 44 & -4.83 & -48.7 & 2.9 & 11.1 \\ \hline
NWC-E-S-NSc-Le &  Orange  & NWC  & 242 & 3 & 78 & -3.08 & -83.2 & 2.8 & 10.8 \\ \hline
NWC-S-E-NSc-NLe &  Orange  & NWC  & 105 & 1 & 106 & -3.30 & -86.7 & 4.9 & 9.8 \\ \hline
NWC-S-W-NSc-NLe &  Orange  & NWC  & 180 & 2/3/6 & 72 & -3.36 & -74.9 & 4.5 & 10.9 \\ \hline
NWC-W-S-NSc-Le &  Orange  & NWC  & 153 & 3 & 46 & -4.36 & -55.4 & 4.2 & 12.1 \\ \hline
NWC-W-S-NSc-NLe &  Orange  & NWC  & 135 & 3 & 40 & -4.85 & -42.3 & 3.2 & 13.7 \\ \hline
NWC-W-N-NSc-Le &  Orange  & NWC  & 146 & 3 & 47 & -2.70 & -92.2 & 2.2 & 8.3 \\ \hline
NWC-W-N-NSc-NLe &  Orange  & NWC  & 173 & 3 & 56 & -2.02 & -102.4 & 3.2 & 10.3 \\ \hline
JLG-N-W &  Cyan  & JLG & 291  & 3/6 & 75 & -2.94  & -72.5 & 2.5 & 10.8 \\ \hline
JLG-N-E &  Cyan  & JLG  & 224 & 3 & 68 & -3.20 & -77.7 & 2.3 & 8.9 \\ \hline
JLG-E-E &  Cyan  & JLG & 49 & 3 & 17 & 11.61 & -355.6 & 2.9 & 13.1 \\ \hline
\end{tabular}
\label{T:measurements}
\end{table*}


\begin{table*}[t!]
\centering
\caption{\addedMK{3 additional OtI measurement scenarios with different Tx and Rx locations}}
\vspace{-\baselineskip}
\footnotesize
\begin{tabular}{| c | c | c | c | c | c | c | c | c | c | c |}
\hline
\textbf{Name} & \textbf{Group} & \textbf{Range (m)} & \textbf{\# Links} & \textbf{Tx \& Rx placement} & \textbf{Purpose} \\
\hline
TEA-S-Bal-1-Reverse & TEA & 230 & 17 & Rx stationary outdoors, Tx moved inside & Evaluating multi-user coverage potential (\S\ref{sec:reverse_tc}) \\ \hline
HMS-Lot-Hallway & HMS & 57 & 58 & Tx stationary outdoors, Rx moved inside & Studying signal loss and propagation further indoors (\S\ref{sec:school-hallway}) \\ \hline
HMS-Court-Hallway & HMS & 57 & 58 & Tx stationary outdoors, Rx moved inside & Studying signal loss and propagation further indoors (\S\ref{sec:school-hallway}) \\ \hline
\end{tabular}
\label{T:measurements_alt}
\end{table*}

\noindent\textbf{Equipment:} We utilize a 28\thinspace{GHz} channel sounder consisting of a separate Tx and Rx, which is described in detail in~\cite{du2020suburban,chen201928}. The Tx is equipped with an omnidirectional antenna with 0\thinspace{dBi} gain, and transmits a +22\thinspace{dBm} continuous-wave tone. The Rx is fed by a 24\thinspace{dBi} rotating horn antenna  (14.5\thinspace{dBi} in azimuth) with 10$^{\circ}$ 3\thinspace{dB} beamwidth. The antenna feeds a mixer which downconverts the received signal to 10\thinspace{MHz} intermediate frequency (IF). The IF signal is then passed through two switchable low-noise amplifiers (LNAs) and a bandpass filter. Finally, the IF signal's power is recorded by a power meter with 20\thinspace{kHz} bandwidth and 5\thinspace{dB} noise figure.

\noindent\textbf{Scenarios:} 
\addedMK{For the scenarios in Table~\ref{T:measurements}}, we placed the rotating Rx indoors (emulating a UE) and the omnidirectional Tx outdoors (emulating a BS). The Tx was moved along a linear path, such as a sidewalk, at a height of 3.4\thinspace{m}. This emulates lightpole deployments of mmWave BSes along streets, which are slated for widespread use in NYC and other urban areas~\cite{nycdoitt5g, samsungmacro}. 

\addedMK{A total of 43 scenarios are listed in Tables~\ref{T:measurements} and~\ref{T:measurements_alt}. For the measurements in Table~\ref{T:measurements},} an OtI scenario is defined by the indoor Rx placement within a given building and the outdoors Tx path. \addedMK{In each scenario}, we placed the Tx at set intervals along the path whose length in defined by the ``Range'' column in Table~\ref{T:measurements}. At each such location (namely, every interval), we measured a link to the indoor Rx. The number of links for each scenario is listed in Table~\ref{T:measurements}. For each single link measurement, the rotating Rx measured the channel for 20 seconds, corresponding to 40 full rotations at 120\thinspace{RPM}. A power reading was taken 740 times per second, providing at least 14,800 power readings per link measurement. 

\addedMK{Using the same equipment but in different setups, three additional OtI scenarios were studied, which are listed in Table~\ref{T:measurements_alt}. The first, detailed in Section~\ref{sec:reverse_tc}, investigates the potential of supporting multiple users, a consideration for multi-user MIMO systems. The Rx was kept stationary outdoors and the Tx moved indoors. The second and third, detailed in Section~\ref{sec:school-hallway}, investigate the path loss and signal propagation within an interior hallway in a public school building. The Tx was kept stationary outdoors and the Rx moved indoors.} \emph{In total, we took over \addedMK{2,200} Tx-Rx link measurements representing over \addedMK{32} million individual power measurements}.

\section{Measurement Results}
\label{sec:results}
In this section, we use the data obtained from the measurement campaign to develop path gain models for the 40 OtI scenarios covered in Figure~\ref{fig:measurementLocations_v2} and Table~\ref{T:measurements}. Each scenario name in Table~\ref{T:measurements} is structured as \textbf{LOC-X-Y-\#}, where \textbf{LOC} is a location in Figure~\ref{fig:measurementLocations_v2}, \textbf{X} is the cardinal direction of the Tx relative to the Rx, \textbf{Y} is the sidewalk along which the Tx was moved, and \textbf{\#} is the floor of the building in which the Rx was placed, if applicable. In some OtI scenarios at \textbf{TEA}, the Tx was moved along an outdoors balcony on the opposite side of the street instead of a sidewalk, indicated by ``Bal''. The measurements at \textbf{HMS} use a different naming scheme. The first value refers to  the Tx path that was used (along a parking lot or along a basketball court) and the number refers to the classroom in which the Rx was placed. \addedMK{Lastly, the measurements for \textbf{NWC} have additional descriptors which mark if the measurement has (no) scaffold ((N)Sc) or (no) tree leaves ((N)Le). Two measurements from \textbf{TEA} are also marked with the scaffolding descriptor.}

\begin{figure*}[t]
\centering
\begin{minipage}[t]{.59\linewidth}
    \subfloat[]{
    \includegraphics[width=\textwidth]{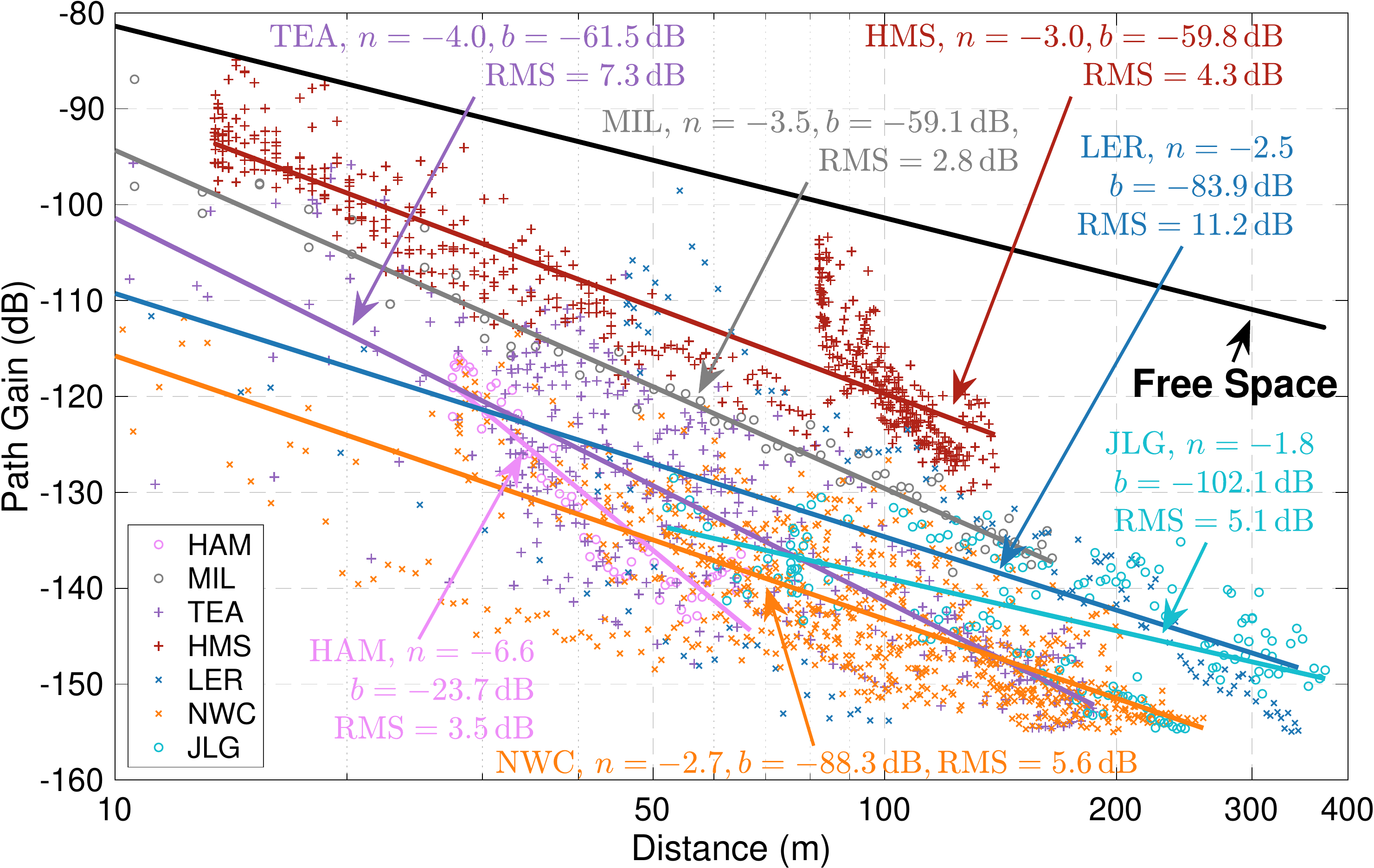}
    \label{fig:indoor_locs_pg}}
\end{minipage}
\hspace{1pt}
\begin{minipage}[t]{.36\linewidth}
    \subfloat[]{
    \includegraphics[width=\textwidth]{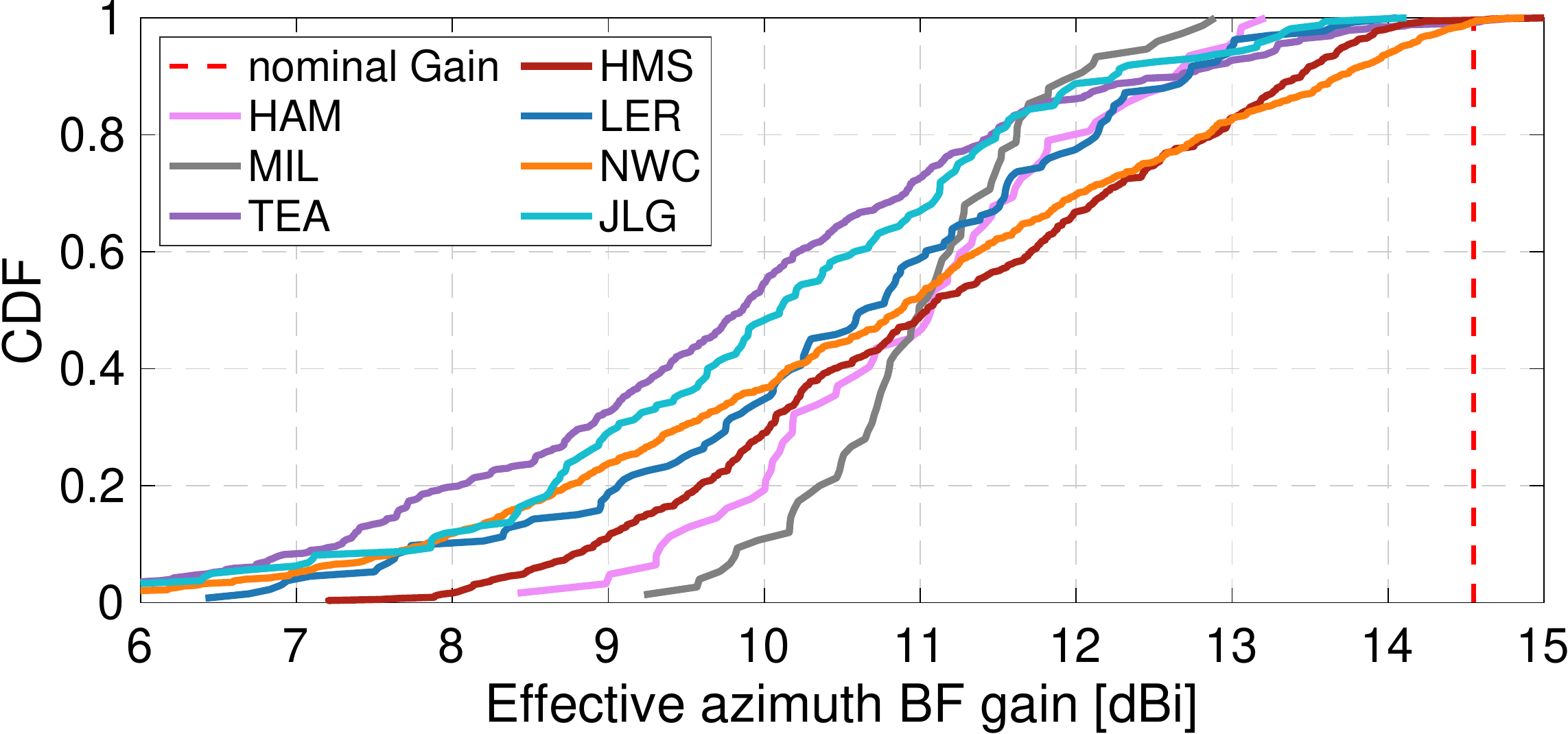}
    \label{fig:indoor_locs_bfg}}
    \vspace{-0.5\baselineskip}
    \\
    \subfloat[]{
    \includegraphics[width=\textwidth]{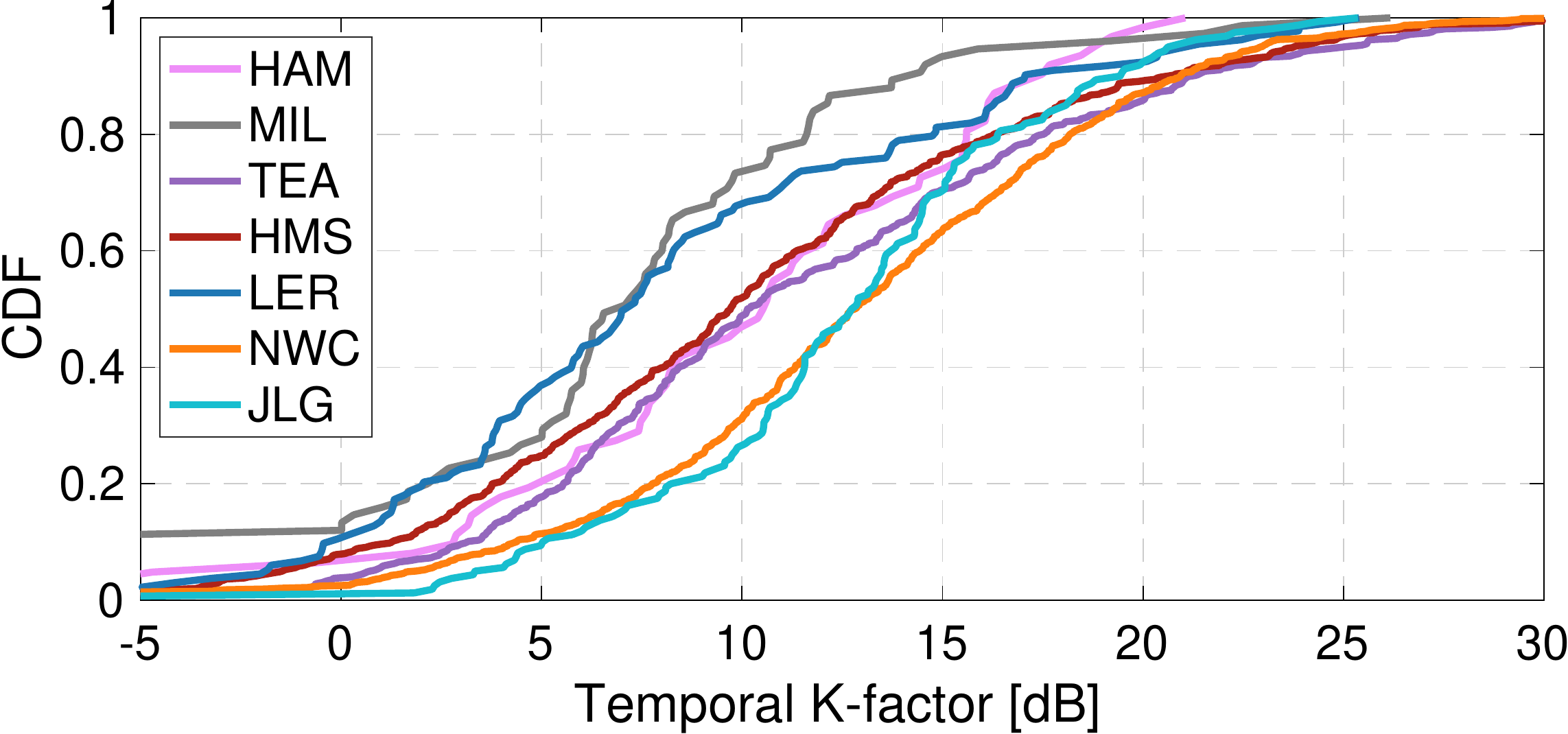}
    \label{fig:indoor_locs_kfac}}
\end{minipage}
\vspace{-1\baselineskip}
\caption{\addedMK{Measurement results clustered by building: (a) average path gain as a function of the 3-dimensional Tx-Rx link distance, with models for each building plotted and noted, (b) CDF of azimuth beamforming gain, (c) CDF of temporal $k$-factor.}}
\vspace{-1\baselineskip}
\label{fig:indoor_locs}
\end{figure*}

The path gain models in Table~\ref{T:measurements} show large differences even between OtI scenarios at the same building, for example the measurements \textbf{JLG}-E-E\footnote{The very high positive slope is due to a small measurement range at a comparatively large offset from the building.} and \textbf{JLG}-N-E. This means very few conclusions can be drawn from these models on an individual basis. Therefore, we develop insights by clustering OtI scenarios in certain ways. We first cluster the \addedMK{OtI scenarios by building}, as seen in Figure~\ref{fig:indoor_locs}. We compute a path gain model for each building, along with distributions of the azimuth beamforming gain and temporal $k$-factor. Significant differences between buildings with outwardly similar appearances are found. 

\begin{figure*}[t]
\subfloat[]{
\includegraphics[width=0.48\linewidth]{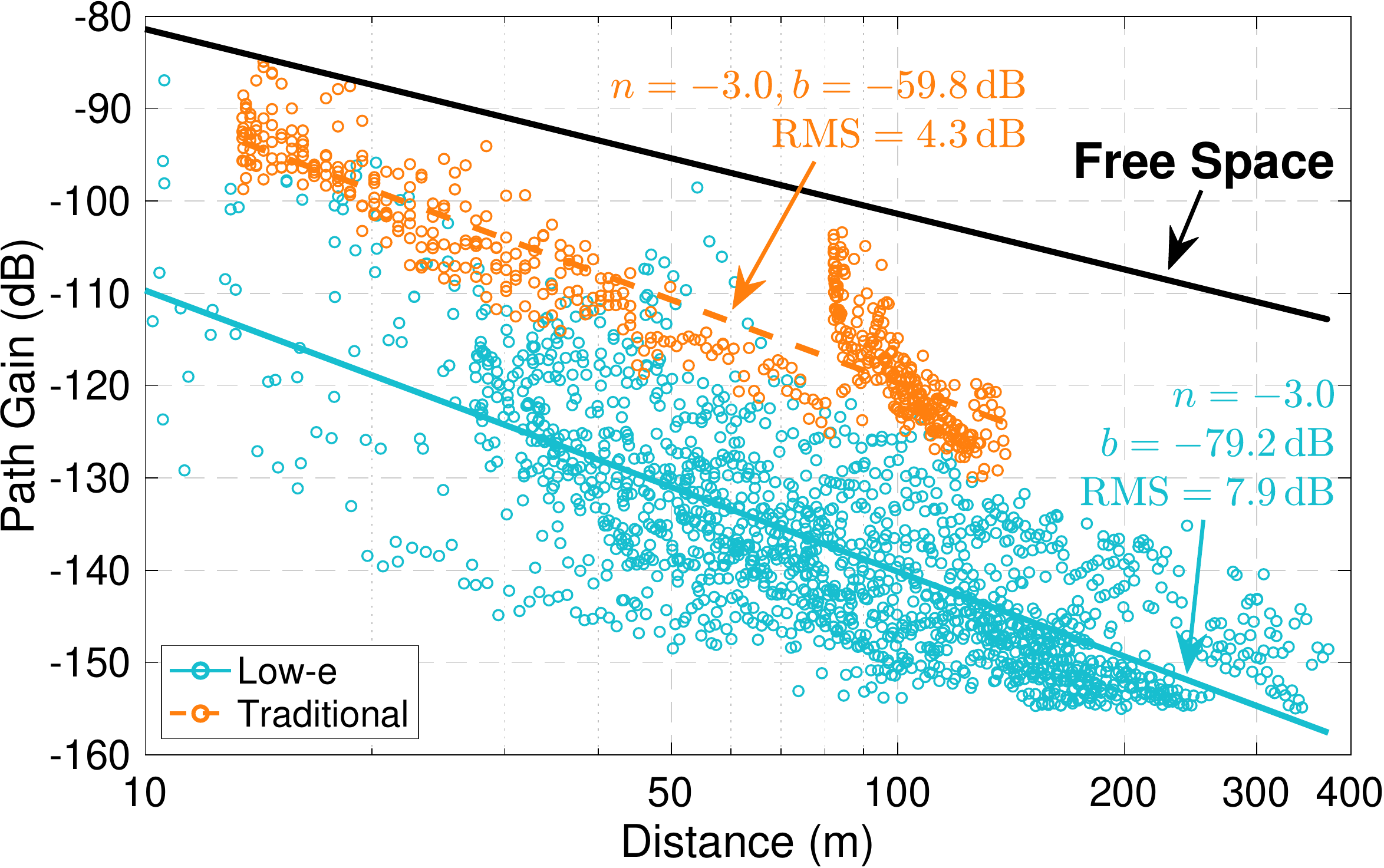}
\vspace{-1\baselineskip}
\label{fig:glass_pg}}
\vspace{-0\baselineskip}
\hspace{0pt}
\subfloat[]{
\includegraphics[width=0.48\linewidth]{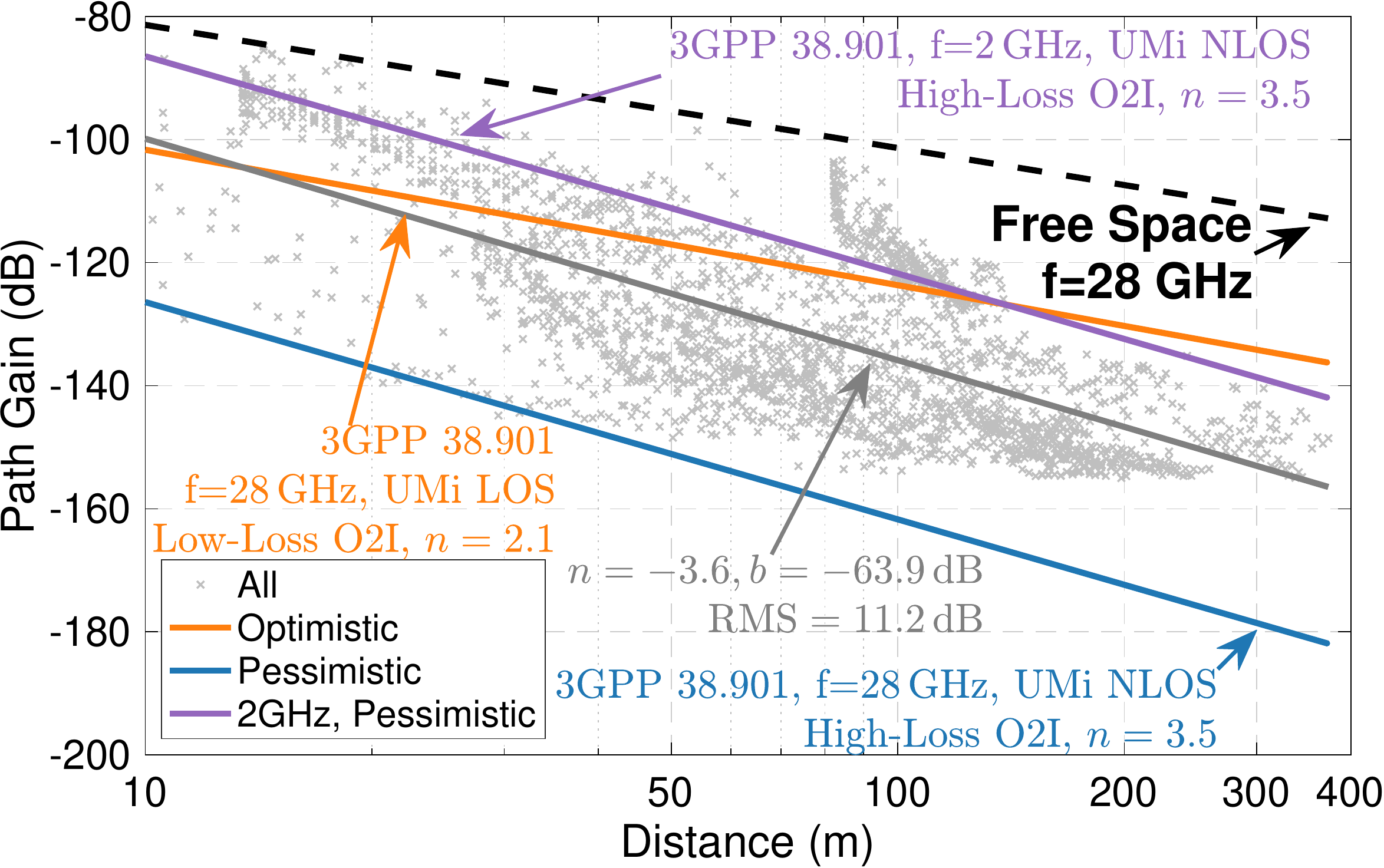}
\vspace{-1\baselineskip}
\label{fig:glass_model_compare}}
\vspace{-1\baselineskip}
\caption{\addedMK{Models for glass: (a) models for ``traditional'' and Low-e glass calculated from OtI scenario clusters, (b) comparison of the path gain model for the cluster of all OtI scenarios to optimistic and pessimistic models developed from 3GPP UMi path loss predictions for different types of glass.}}
\label{fig:glass_compare}
\end{figure*}

\begin{figure}[t]
\vspace{-1\baselineskip}
\subfloat[]{
\includegraphics[width=0.46\linewidth]{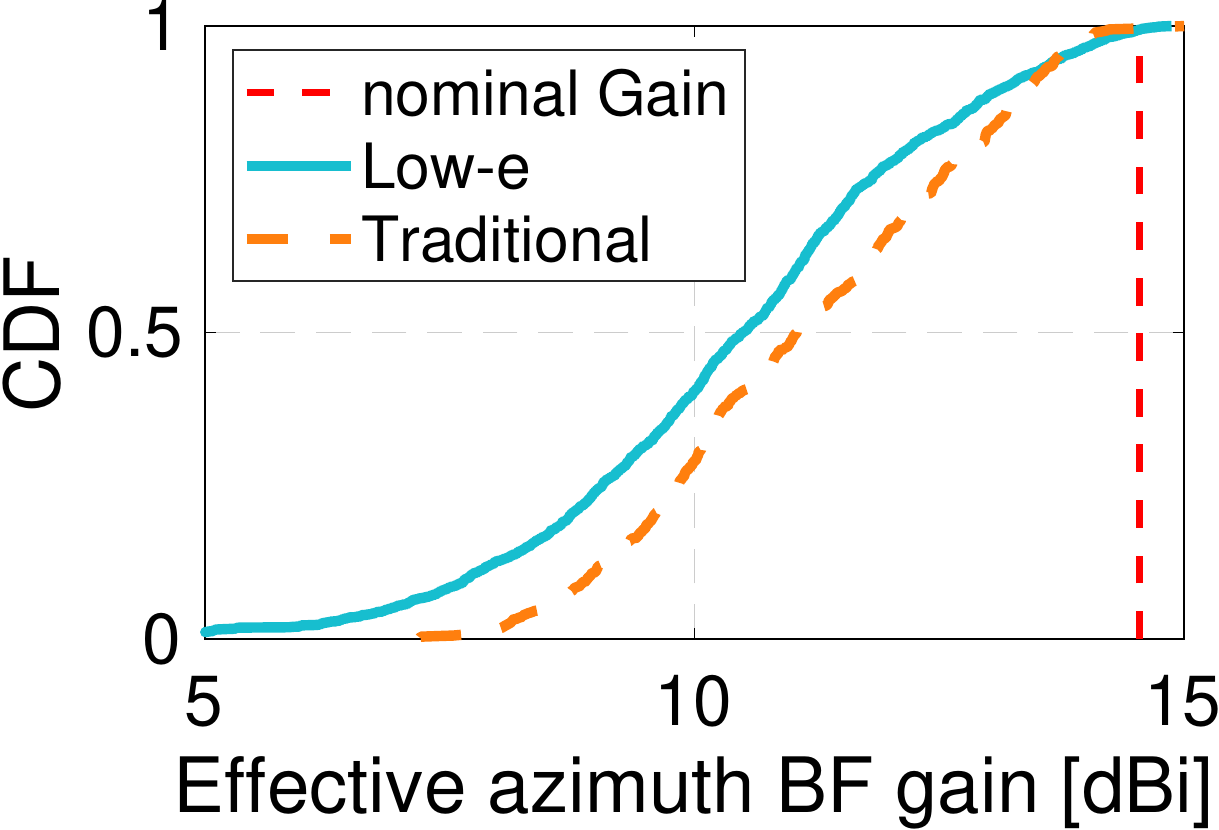}
\vspace{-1\baselineskip}
\label{fig:glass_bfg}}
\hspace{1pt}
\subfloat[]{
\includegraphics[width=0.46\linewidth]{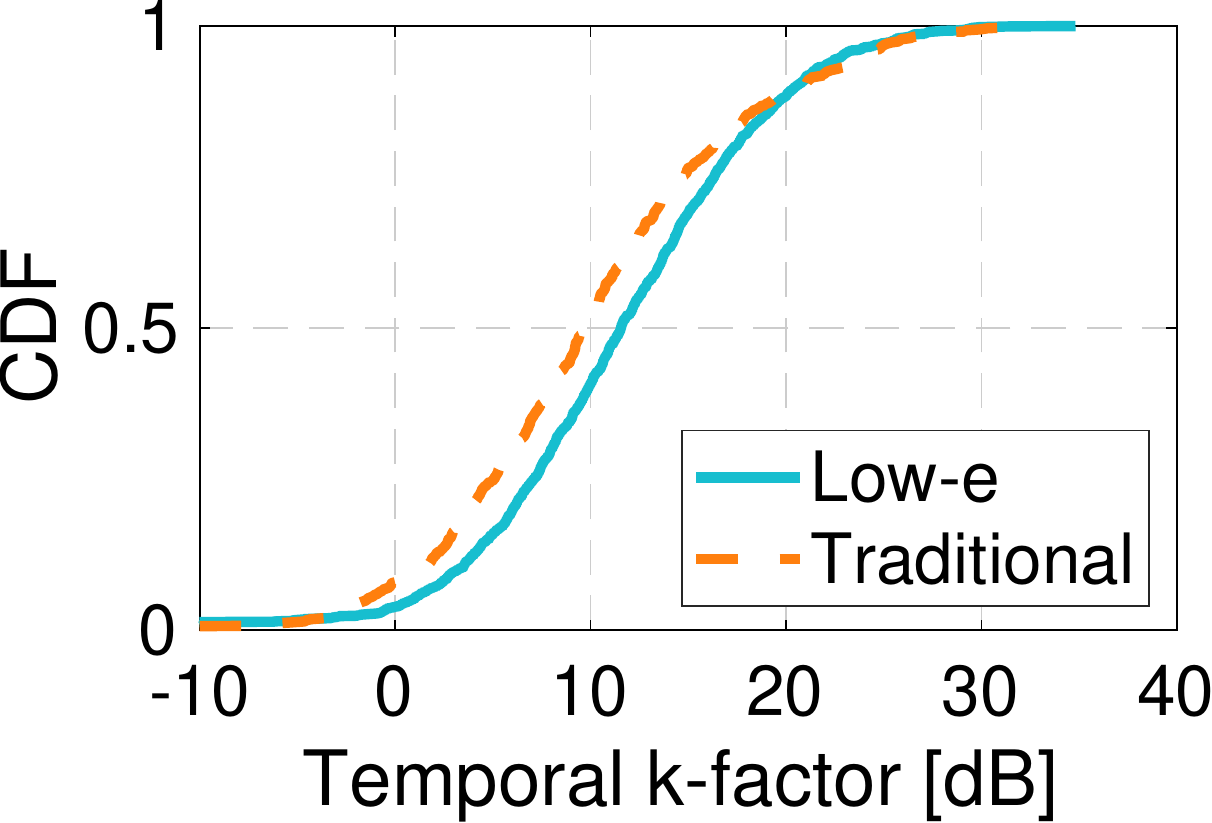}
\vspace{-1\baselineskip}
\label{fig:glass_kfac}}
\vspace{-1\baselineskip}
\caption{\addedMK{Measurement results categorized by the type of glass used. CDFs of (a) effective azimuth beamforming gain and (b) temporal $k$-factor}}
\label{fig:glass_bfgkfac}
\vspace{-1\baselineskip}
\end{figure}

\begin{figure*}[t]
\subfloat[]{
\includegraphics[width=0.3\linewidth]{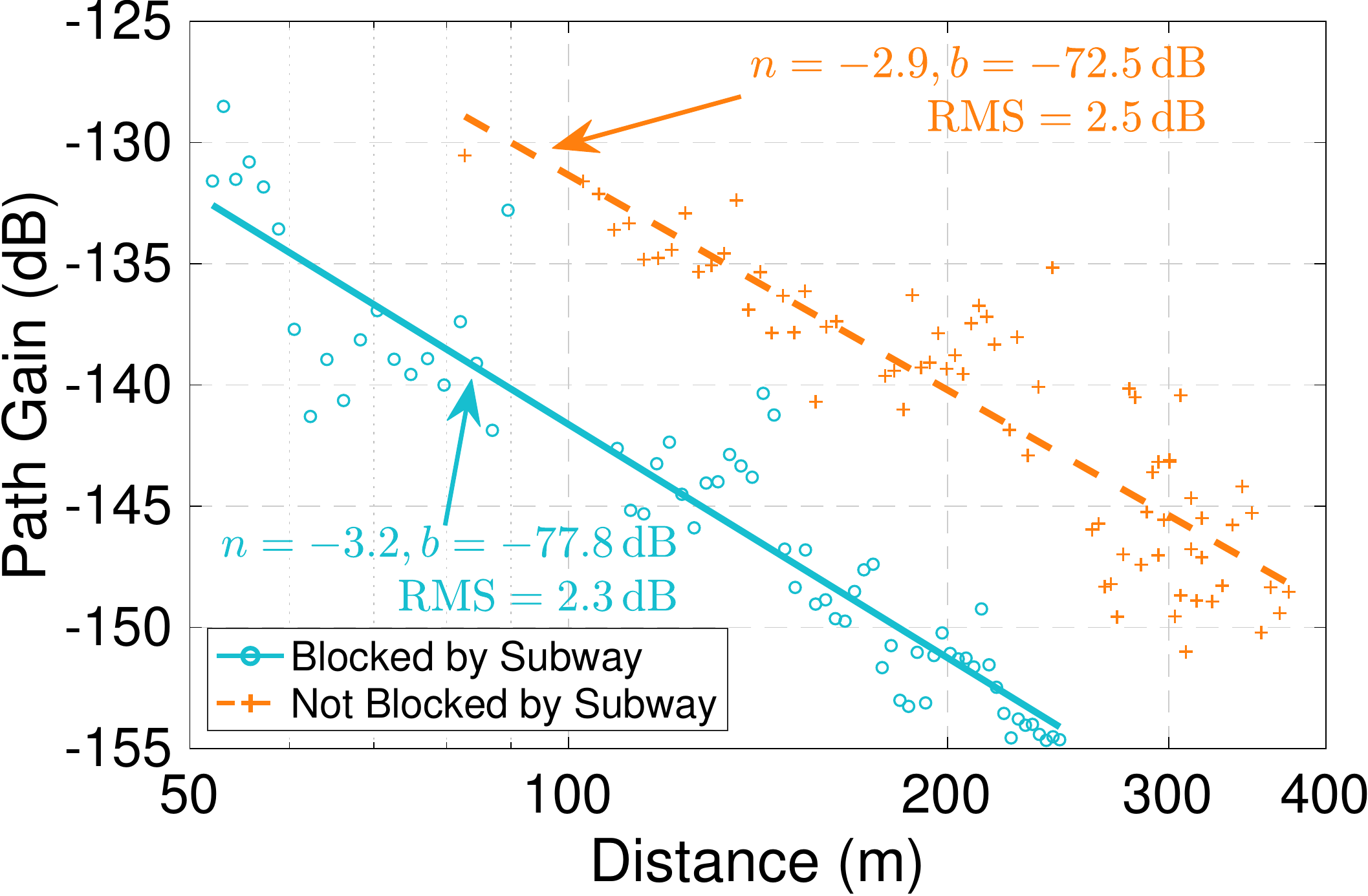}
\vspace{-1\baselineskip}
\label{fig:new1-street-compare-pg}}
\hspace{10pt}
\setcounter{subfigure}{2}
\subfloat[]{
\includegraphics[width=0.3\linewidth]{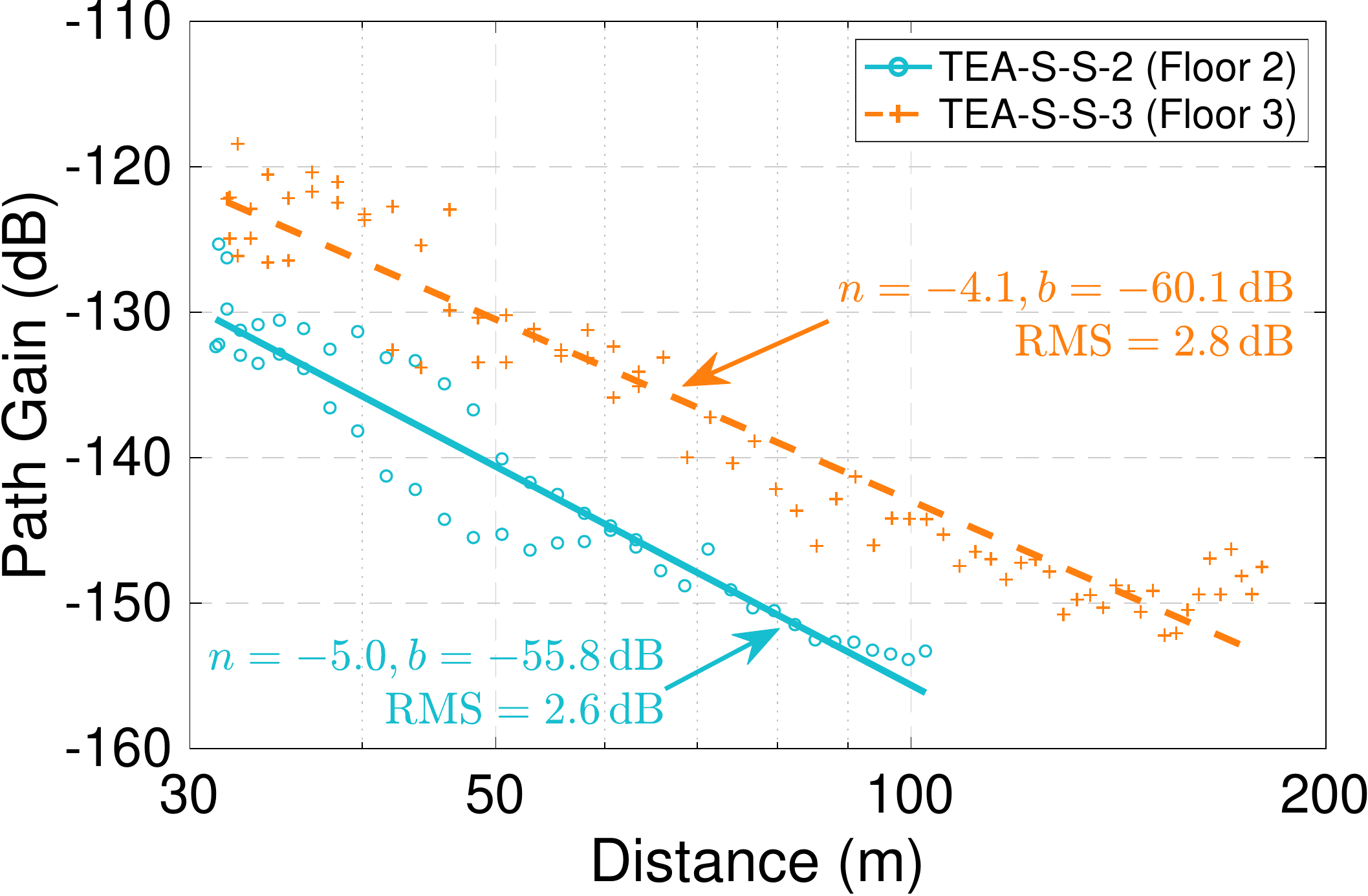}
\vspace{-1\baselineskip}
\label{fig:old2-height-compare-pg}}
\hspace{10pt}
\setcounter{subfigure}{4}
\subfloat[]{
\includegraphics[width=0.3\linewidth]{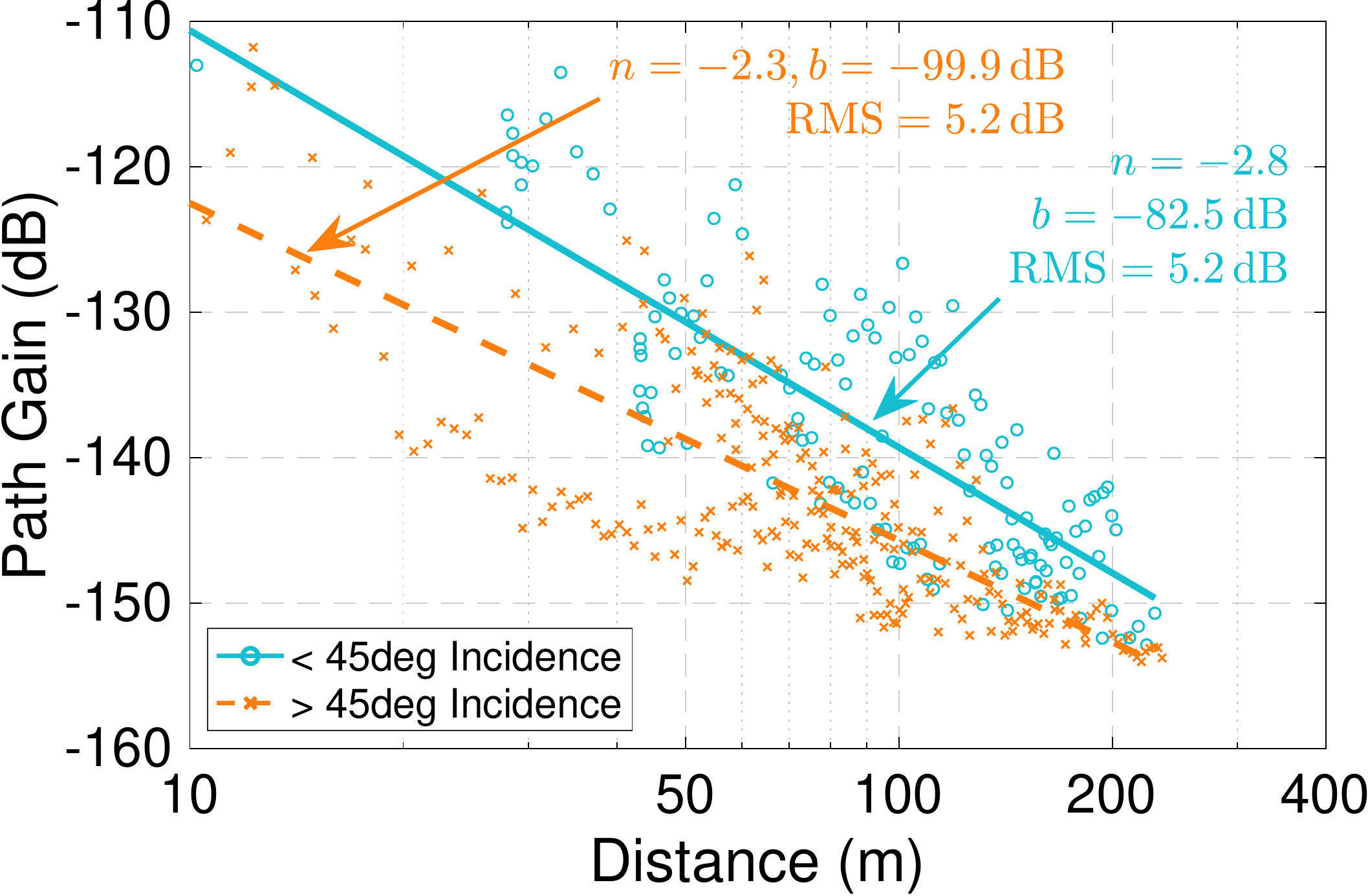}
\vspace{-1\baselineskip}
\label{fig:oblique-compare-pg}}
\vspace{-1\baselineskip}
\setcounter{subfigure}{1}
\subfloat[]{
\includegraphics[width=0.3\linewidth]{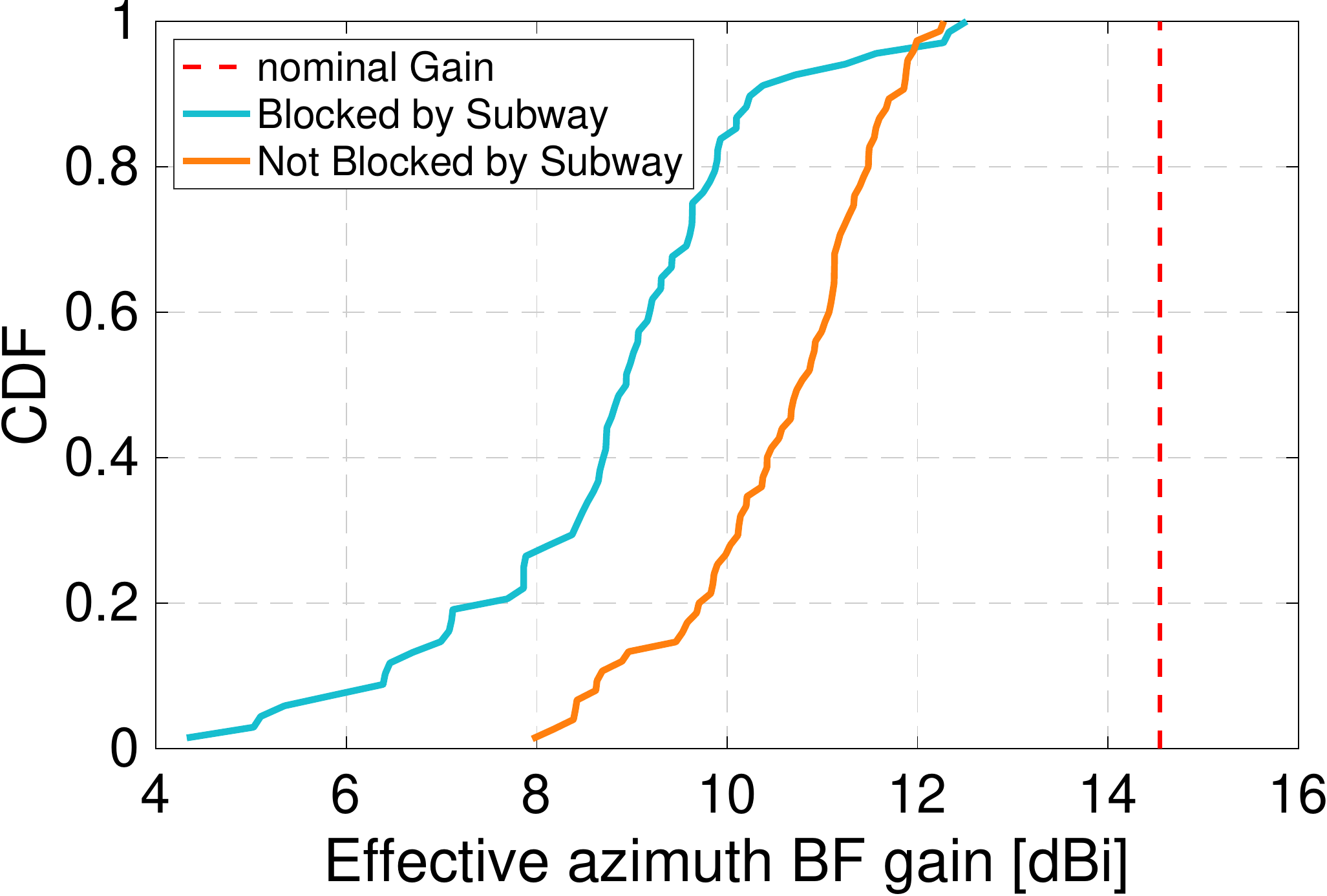}
\vspace{-1\baselineskip}
\label{fig:new1-street-compare-bfg}}
\hspace{10pt}
\setcounter{subfigure}{3}
\subfloat[]{
\includegraphics[width=0.3\linewidth]{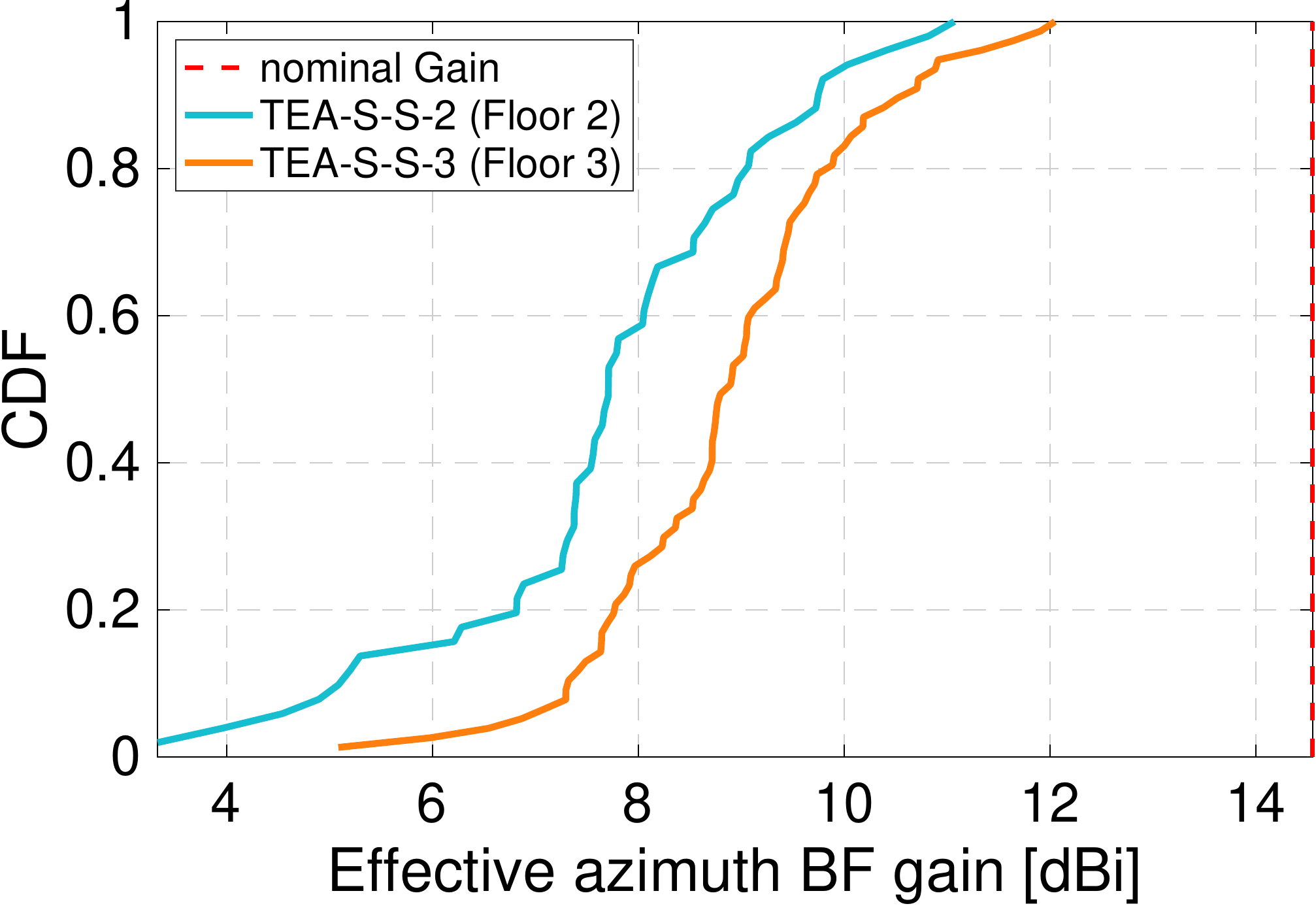}
\vspace{-1\baselineskip}
\label{fig:old2-height-compare-bfg}}
\hspace{10pt}
\setcounter{subfigure}{5}
\subfloat[]{
\includegraphics[width=0.3\linewidth]{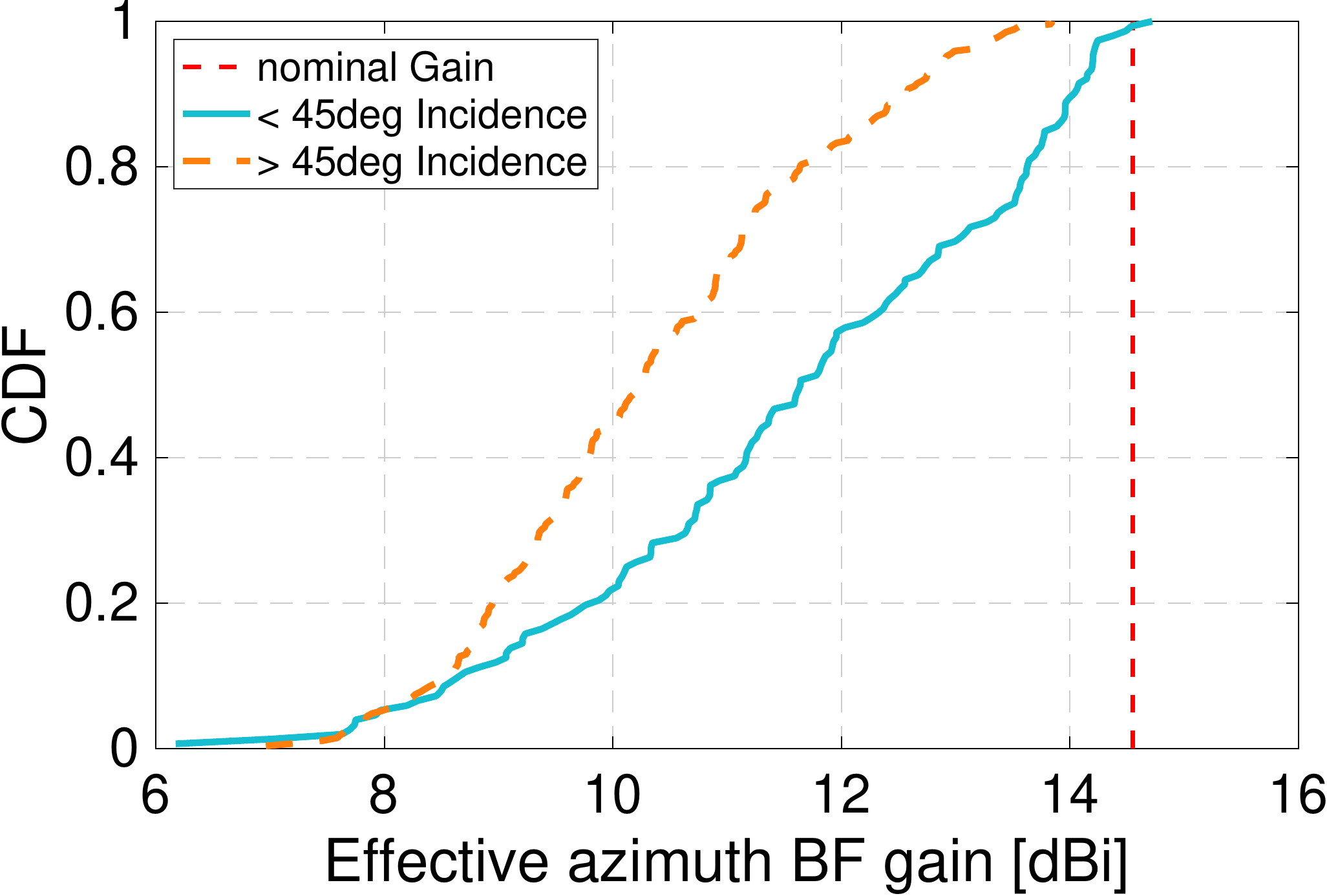}
\vspace{-1\baselineskip}
\label{fig:oblique-compare-bfg}}
\vspace{-1\baselineskip}
\caption{\addedMK{Path gain and azimuth beamforming gain measurements for different placements of Tx and Rx: (a,b) Tx placed on different sides of the same street measured from \textbf{JLG}, (c,d) Rx placed on different floors of \textbf{TEA}, and (e,f) AoI above or below 45$^\circ$ at \textbf{NWC}.}}
\label{fig:placement}
\end{figure*}

\addedMK{We then cluster OtI scenarios based on the type of window glass used, considering ``traditional'' and low-emissivity (Low-e) glass, the latter measured as shown in Figure~\ref{fig:aoi-test}. The results are presented in Figures~\ref{fig:glass_compare}\subref{fig:glass_pg} and ~\ref{fig:glass_bfgkfac}. We also compare the measurement data to OtI models derived from 3GPP UMi models for path loss and building penetration loss in Figure~\ref{fig:glass_compare}\subref{fig:glass_model_compare}. We then consider a variety of Tx-Rx placements: (i) Tx behind/in front of an elevated subway track, (ii) different floors of the same building, and (iii) \addedMK{angle of incidence (AoI) less than or greater than 45$^\circ$} into the window near the Rx with results in Figure~\ref{fig:placement}. Lastly, Figure~\ref{fig:environment} shows an analysis on the impact of leaves on trees lining sidewalks and scaffolding surrounding the building containing the Rx.}

\subsection{Measurable Parameters}
\label{sec:meas-parameters}
Four parameters are calculated from the data: (i) path gain, $G_{path}(d)$, (ii) power angular spectra, $\bar{S}(d, \phi)$, which describes the received power from all azimuth directions, (iii) effective azimuth beamforming gain $G_{az}(d)$, which represents the effect of angular spread, and (iv) the temporal $k$-factor $K(d)$, which represents the time-varying component of the channel. 

\addedMK{To compute $G_{path}(d)$, we note that averaging the received power $P_{horn}(d)$ over all directions $\phi$ gives the equivalent power that would be received by an omnidirectional antenna~\cite{du2020suburban}:
\[
    P_{omni}(d) = \frac{1}{2\pi}\int^{2\pi}_0P_{horn}(d, \phi)d\phi
\]

\noindent By taking the average $\bar{P}_{omni}(d)$ over all turns, we can compute the path gain $G_{path}(d)$ as
\[
    G_{path}(d) = \frac{\bar{P}_{omni}(d)}{(P_{Tx}\cdot G_{el})}
\]

\noindent $P_{Tx}$ is the transmit power, and $G_{el}(d)$ is the elevation gain, a value calculated from the antenna patterns of the Tx and Rx measured in an anechoic chamber. $G_{el}(d)$ is used to correct for the misalignment of the Rx horn as it spins in the azimuthal plane. The power angular spectra is computed by averaging $P_{horn}(d, \phi)$ for every integer azimuth value between $0^\circ$ and $359^\circ$:
\[
    \bar{S}(d, \phi) = \frac{1}{N}\sum_{i=1}^NP^{(i)}_{horn}(d, \phi)
\]

\noindent Where $P^{(i)}_{horn}(d, \phi)$ represents the power recorded at angle $\phi$ on the $i$\textsuperscript{th} turn. $\bar{S}(d, \phi)$ can then be directly used to compute $G_{az}(d)$:
\[
    G_{az}(d) = \frac{\max_\phi\{\bar{S}(d, \phi)\}}{\bar{P}_{omni}(d)}
\]

\begin{figure}[t!]
\vspace{-0.5\baselineskip}
\subfloat[]{
\includegraphics[width=0.32\columnwidth]{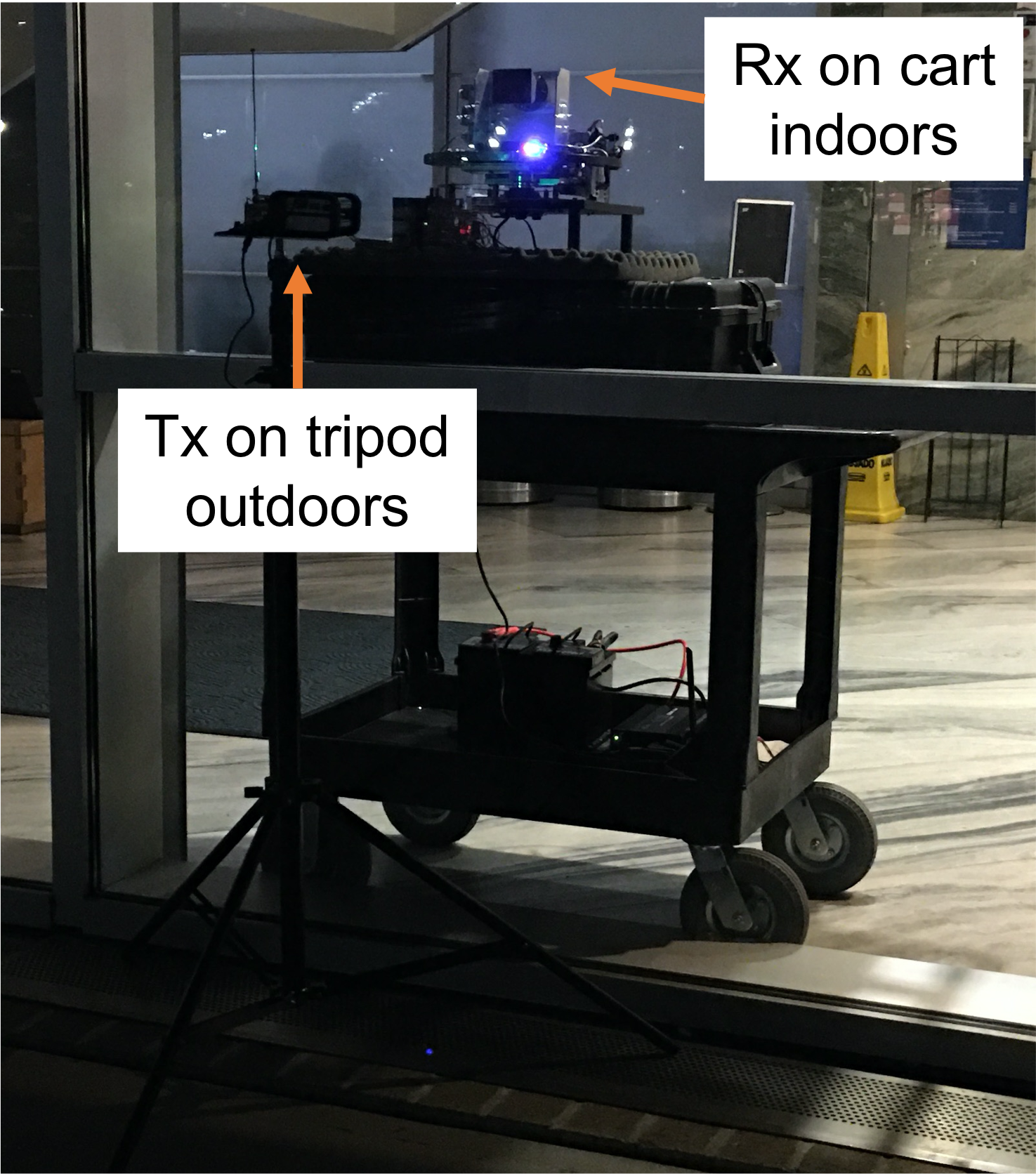}
\vspace{-1\baselineskip}
\label{fig:aoi-test-photo}}
\vspace{-0\baselineskip}
\hspace{5pt}
\subfloat[]{
\includegraphics[width=0.5\columnwidth]{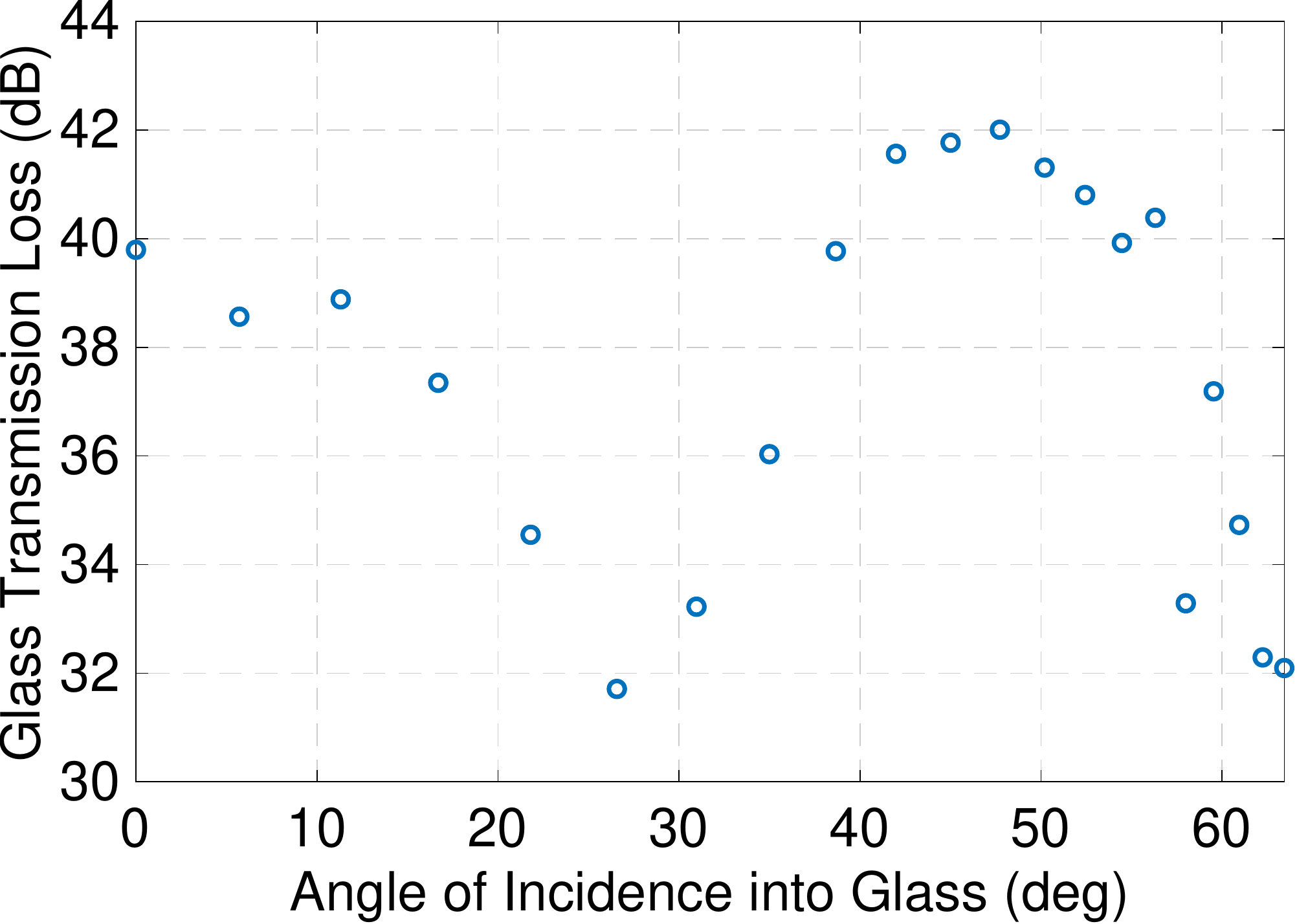}
\label{fig:aoi-test-loss}}
\vspace{-1\baselineskip}
\caption{Measurement to study impact of AoI on path loss. (a) Equipment setup and (b) measured glass transmission loss as a function of AoI.}
\vspace{-2\baselineskip}
\label{fig:aoi-test}
\end{figure}

\noindent Lastly, $K(d)$ represents the level of time variation in the wireless channel, and is computed using the method of moments~\cite{greenstein1999moment}. 

Any $G_{az}(d)$ below the nominal value of 14.5\thinspace{dBi} indicates beamforming gain degradation, which will result from environmental scattering of the 28\thinspace{GHz} signal. $K(d)$ captures time-varying changes in the urban environment, such as the movement of cars, pedestrians, foliage etc. as a fraction of the signal power. For example, a measured $K(d)$ of 20\thinspace{dB} indicates that the time-varying component of the signal accounts for $\frac{1}{100}$ of the total power.}

\hspace{-1pt}

\begin{figure}[t]
  \centering
  \includegraphics[width=0.7\columnwidth]{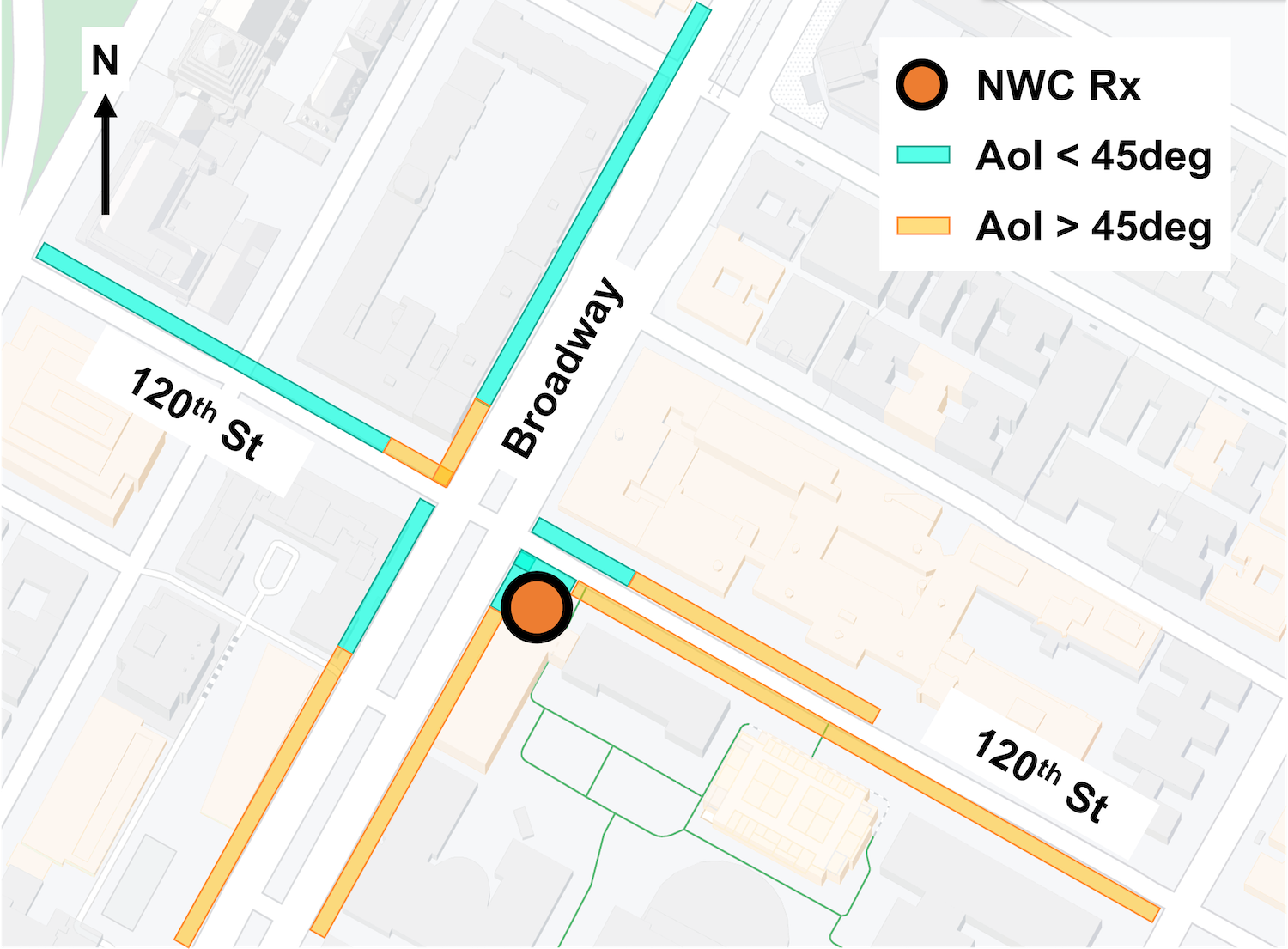}
    \vspace{-0.5\baselineskip}
  \caption{\addedMK{Overview of AoI measurement clustering for six sidewalks at NWC. Tx locations on sidewalks marked in cyan have less than 45$^\circ$ degrees incidence into the glass, those in orange have greater than 45$^\circ$ degrees incidence.}}
  \vspace{-1\baselineskip}
  \label{fig:aoi_measurement_map}
\end{figure}

\subsection{\addedMK{Different Buildings}}
\label{sec:results-buildings}

Figure~\ref{fig:indoor_locs} shows the measured path gain, azimuth beamforming gain, and temporal $k$-factor for all building clusters. The best-fit path gain models for each building are also displayed in Figure~\ref{fig:indoor_locs}\subref{fig:indoor_locs_pg}.

Most notably, Figure~\ref{fig:indoor_locs}\subref{fig:indoor_locs_pg} shows that \textbf{HMS} experiences path gain 10--25dB higher than other buildings at 50\thinspace{m} \addedMK{three-dimensional Euclidean} distance between Tx and Rx. Figure~\ref{fig:indoor_locs}\subref{fig:indoor_locs_bfg} shows that the median azimuth beamforming gain for all buildings is within around 1.2\thinspace{dB}, which is overall an inconsequential difference, though we do note that buildings with larger windows (\textbf{TEA}, \textbf{NWC}, and \textbf{JLG}) tend to have lower azimuth beamforming gain than others. Figure~\ref{fig:indoor_locs}\subref{fig:indoor_locs_kfac} shows that the median temporal $k$-factor can vary by around $8$\thinspace{dB} between locations, though the temporal $k$-factor is a characteristic of \addedMK{the time-varying propagation environment rather than the static one determined by factors such as building construction.}

The physical appearance of a building is thus not a good indicator of the expected loss. As seen in Figure~\ref{fig:meas_locations}, \textbf{HMS} has a relatively similar brick construction to \textbf{HAM} or \textbf{MIL} yet experiences a 10--25\thinspace{dB} lower path loss at 50\thinspace{m}. Even \textbf{HAM} and \textbf{MIL}, with almost identical exteriors, have a 15\thinspace{dB} difference in path loss at 50\thinspace{m}.

\subsection{Low-e and Traditional Glass}
\label{sec:results-glass}
\subsubsection{Measurements}
In order to understand the specific factors that may impact the path loss for a building, we first group the measurements based on the \addedMK{type} of glass. ``Traditional'' glass, often used in buildings predating the availability of float glass in the 1960s, typically has less than 1\thinspace{dB} loss at 28\thinspace{GHz}~\cite{du2021subterahertz}. \addedMK{Modern Low-e glass can have losses in excess of 25\thinspace{dB}~\cite{hyunjin2021transmission}; Figure~\ref{fig:aoi-test} shows a measured normal incidence loss of 40\thinspace{dB} from Low-e glass at \textbf{NWC}. Loss as high as 50~\thinspace{dB} through concrete walls at 28\thinspace{GHz}~\cite{vargas2017measurements} implies that the majority of the mmWave signal will be received via windows, suggesting them to be a significant factor impacting path loss.}

\textbf{HMS} uses ``traditional" glass, while the other six locations use Low-e glass in their construction; the windows at older buildings have been reglazed in recent years. The results of this analysis are presented in Figures~\ref{fig:glass_compare} and \ref{fig:glass_bfgkfac}. The path gain models for both categories are shown in  Figure~\ref{fig:glass_compare}\subref{fig:glass_pg}. We observe that the models have identical slopes, with the difference being a uniform 20\thinspace{dB} additional loss experienced by the buildings with Low-e glass. 

The results for the azimuth beamforming gain and temporal $k$-factor are shown in Figures~\ref{fig:glass_bfgkfac}\subref{fig:glass_bfg} and \ref{fig:glass_bfgkfac}\subref{fig:glass_kfac}. These two quantities are very similar, with an azimuth beamforming gain degradation of 3.5--4.5\thinspace{dB} and median $k$-factor value of 10--12\thinspace{dB}. This is to be expected, as these values are influenced primarily by the overall measurement environment rather than by the type of glass. The results indicate a moderate level of beamforming gain degradation and reasonable channel stability over time. The median $k$-factor value of around 10\thinspace{dB} indicates that the \addedMK{varying component of the received signal is $\frac{1}{10}$ the total power.}

\subsubsection{Comparison to 3GPP Predictive Models}
Figure~\ref{fig:glass_compare}\subref{fig:glass_model_compare} shows a model aggregated over all OtI scenarios \addedMK{in Table~\ref{T:measurements}} compared to pessimistic and optimistic models \addedMK{at 28\thinspace{GHz}} developed from 3GPP TR 38.901~\cite{3gppmodels}. The pessimistic model is defined as $PL_{pess}(d) = PL_{OtO,NLOS}(d) + PL_{OtI,High}$, which is the sum of the non-line-of-sight (NLOS) urban street canyon model (USCM) and building transmission loss with Low-e glass. \addedMK{We use the NLOS model for two reasons. First is that beyond 52\thinspace{m}, the 3GPP NLOS probability will exceed 50\%~\cite{3gppmodels}, and the majority of our measurement data is at distances larger than 52\thinspace{m} and thus prone to occlusion by trees and other sidewalk clutter. Second is this model can provide an upper bound for the expected path loss. Similarly, to give a lower bound on the expected path loss,} the optimistic model is defined as $PL_{opt}(d) = PL_{OtO,LOS}(d) + PL_{OtI,Low}$, using the LOS USCM and building transmission loss with ``traditional'' multi-pane glass. \addedMK{In addition, a pessimistic model at 2\thinspace{GHz} is included in the figure. In all models we set the} BS height to 10\thinspace{m} and the UE height to 3.5\thinspace{m}. 

\addedMK{The 40 OtI scenario measurements predominantly fall between the  two 28\thinspace{GHz} models.} There are a number of points which lie above the optimistic line; these are mostly from \textbf{HMS}. This is largely due to the single-pane ``traditional'' glass windows, which should produce even less building transmission loss than predicted by $PL_{OtI,Low}$. The tendency for the measured path gain to be either in between pessimistic and optimistic models, or greater than an optimistic one, was previously observed in OtO measurements~\cite{chen201928}. \addedMK{Lastly, we observe that the pessimistic model at 2\thinspace{GHz} predicts lower path loss than even optimistic model (and most of the measurement data) at 28\thinspace{GHz}.}


\begin{figure*}[t]
\subfloat[]{
\includegraphics[width=0.3\linewidth]{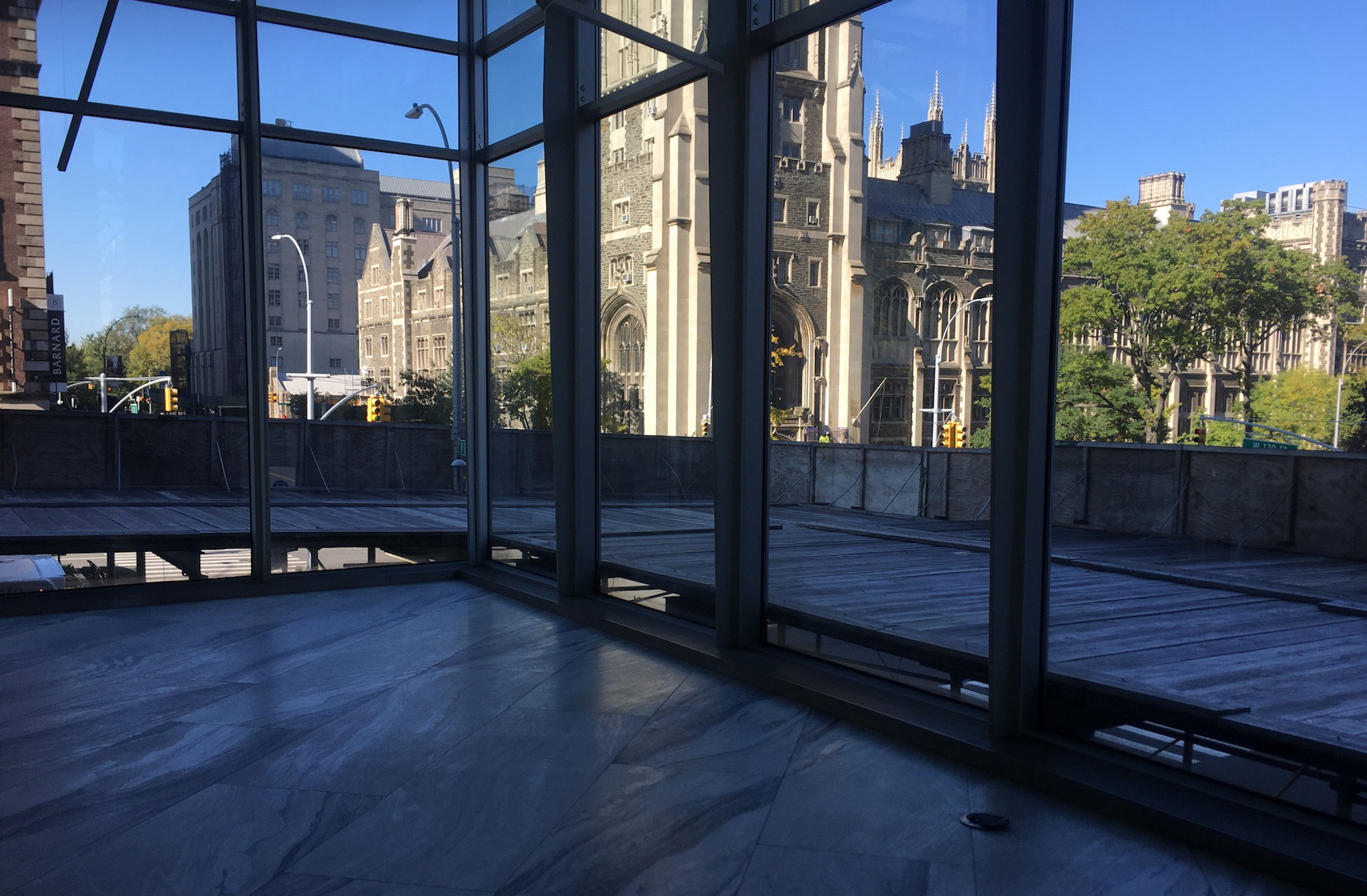}
\vspace{-1\baselineskip}
\label{fig:nwc-scaffold}}
\hspace{10pt}
\setcounter{subfigure}{2}
\subfloat[]{
\includegraphics[width=0.3\linewidth]{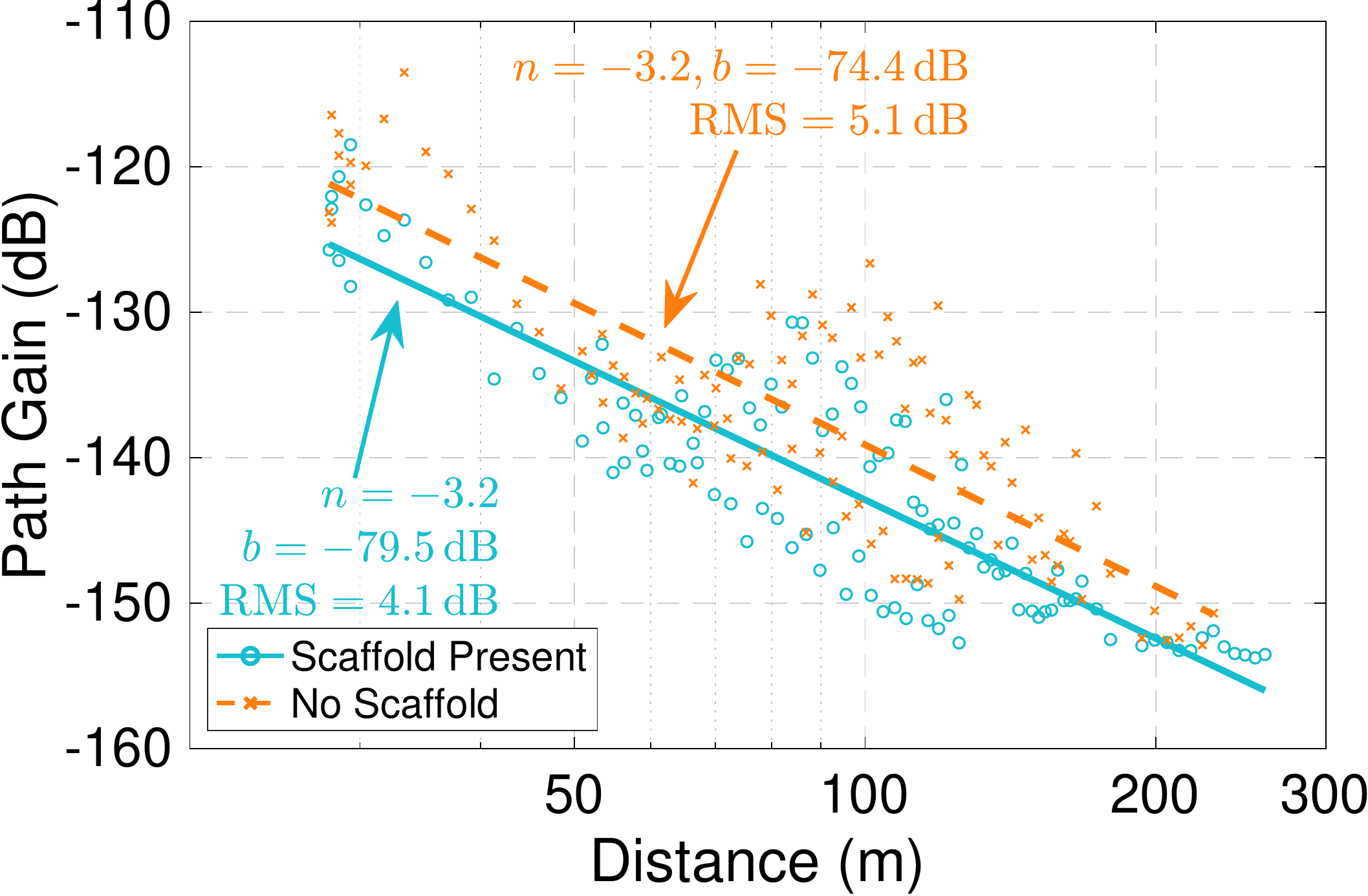}
\vspace{-1\baselineskip}
\label{fig:scaffold-compare-pg}}
\hspace{10pt}
\setcounter{subfigure}{4}
\subfloat[]{
\includegraphics[width=0.3\linewidth]{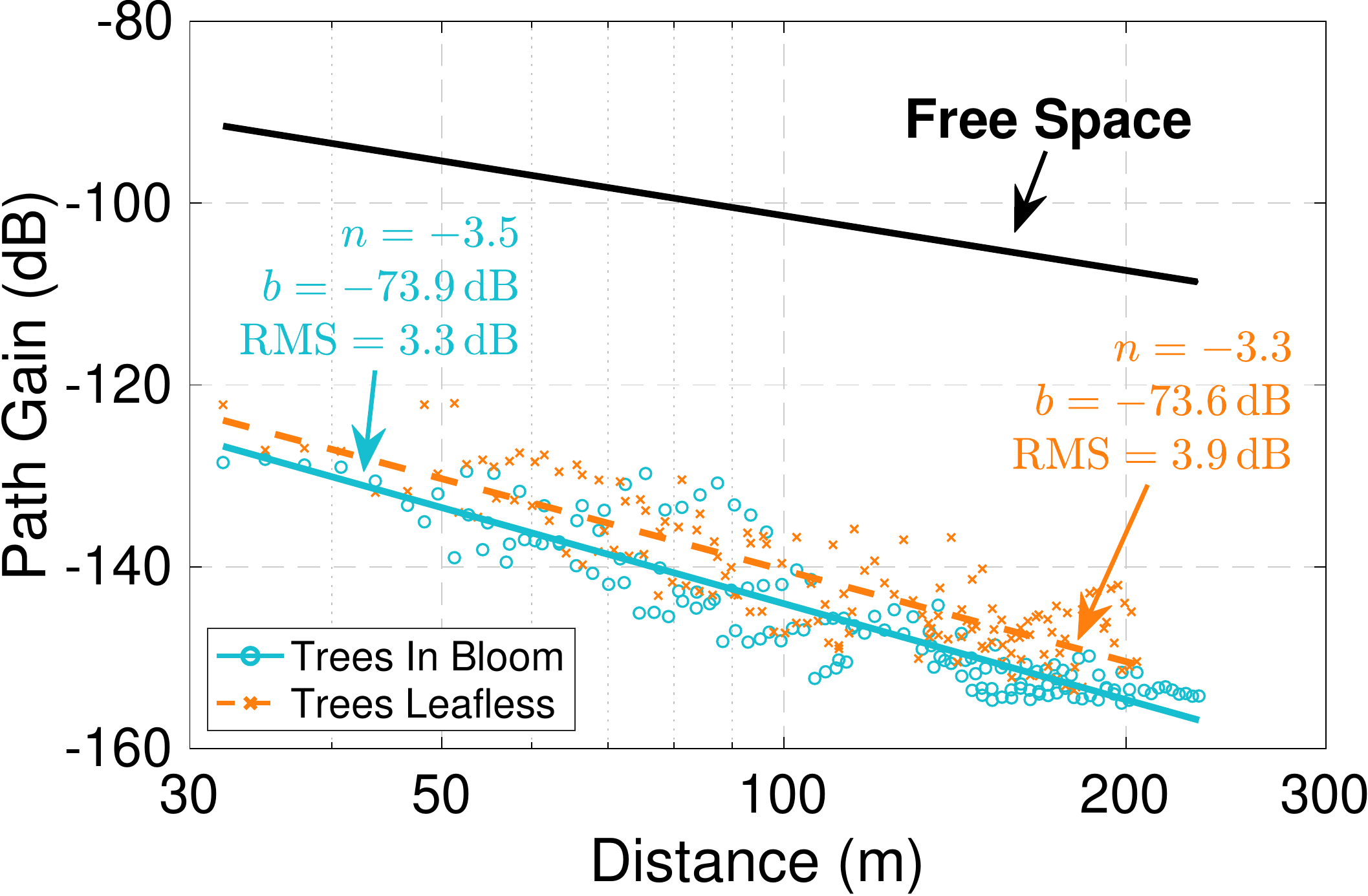}
\vspace{-1\baselineskip}
\label{fig:foliage-compare-pg}}
\vspace{-1\baselineskip}
\setcounter{subfigure}{1}
\subfloat[]{
\includegraphics[width=0.3\linewidth]{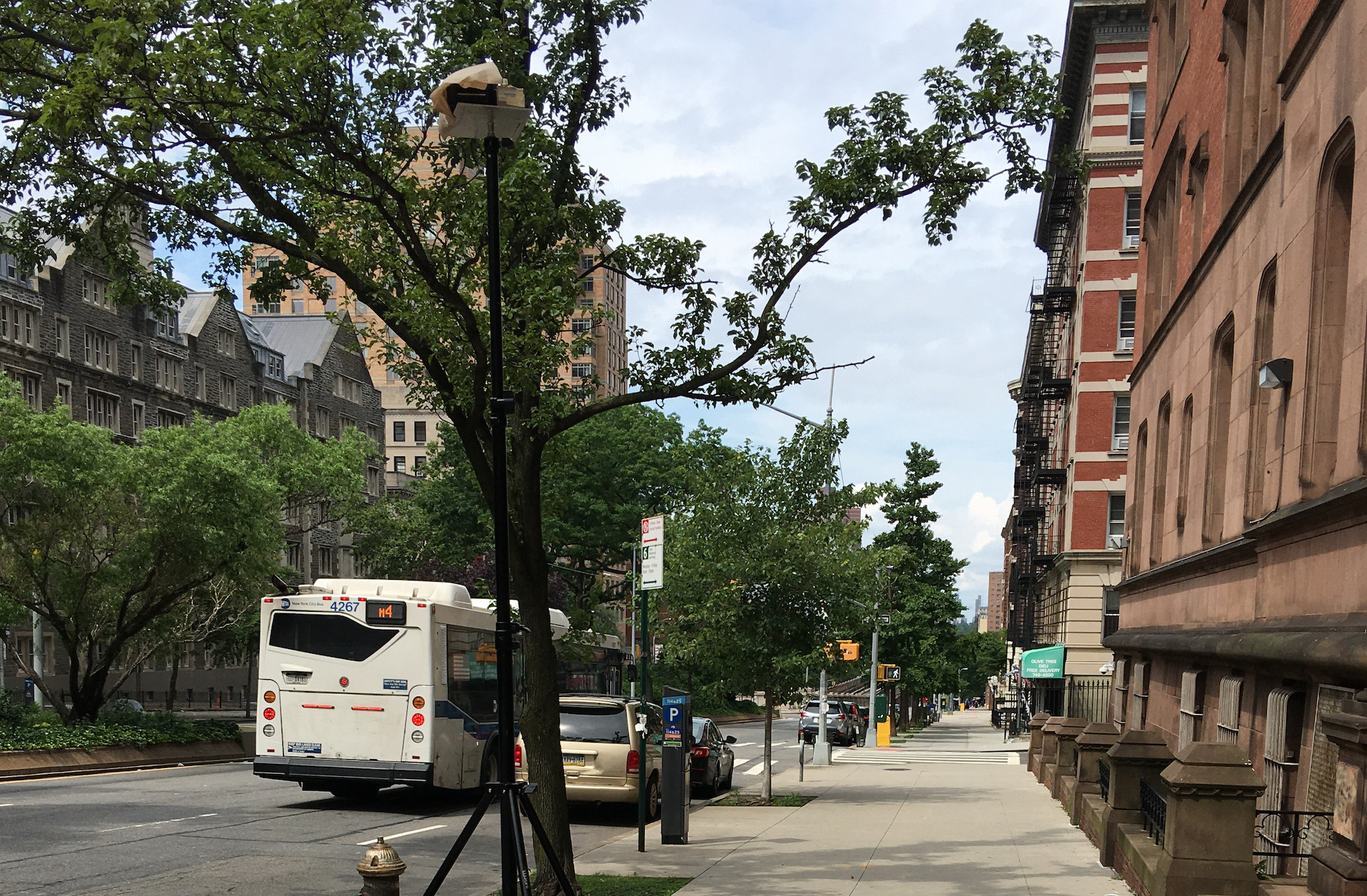}
\vspace{-1\baselineskip}
\label{fig:nwc-foliage}}
\hspace{10pt}
\setcounter{subfigure}{3}
\subfloat[]{
\includegraphics[width=0.3\linewidth]{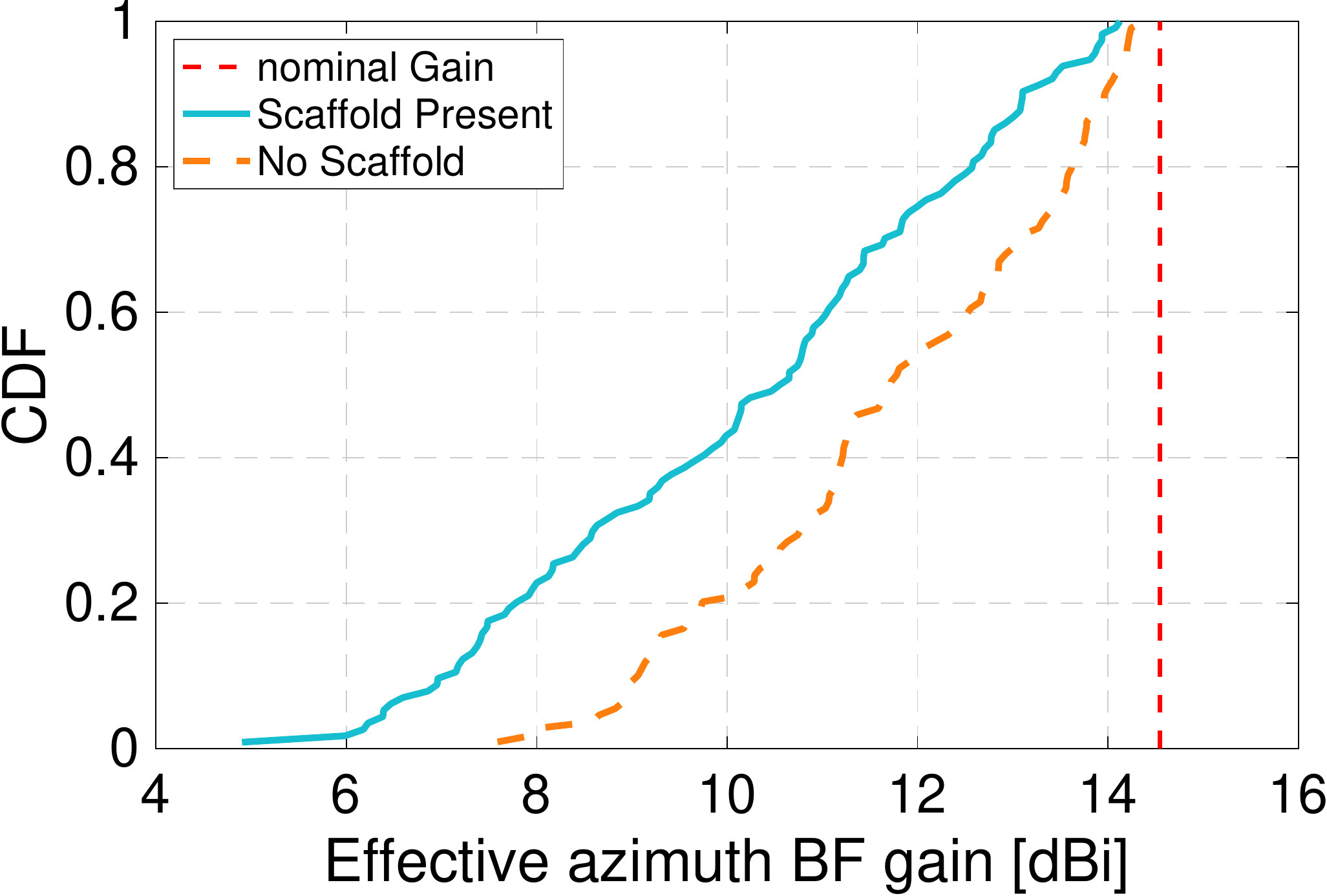}
\vspace{-1\baselineskip}
\label{fig:scaffold-compare-bfg}}
\hspace{10pt}
\setcounter{subfigure}{5}
\subfloat[]{
\includegraphics[width=0.3\linewidth]{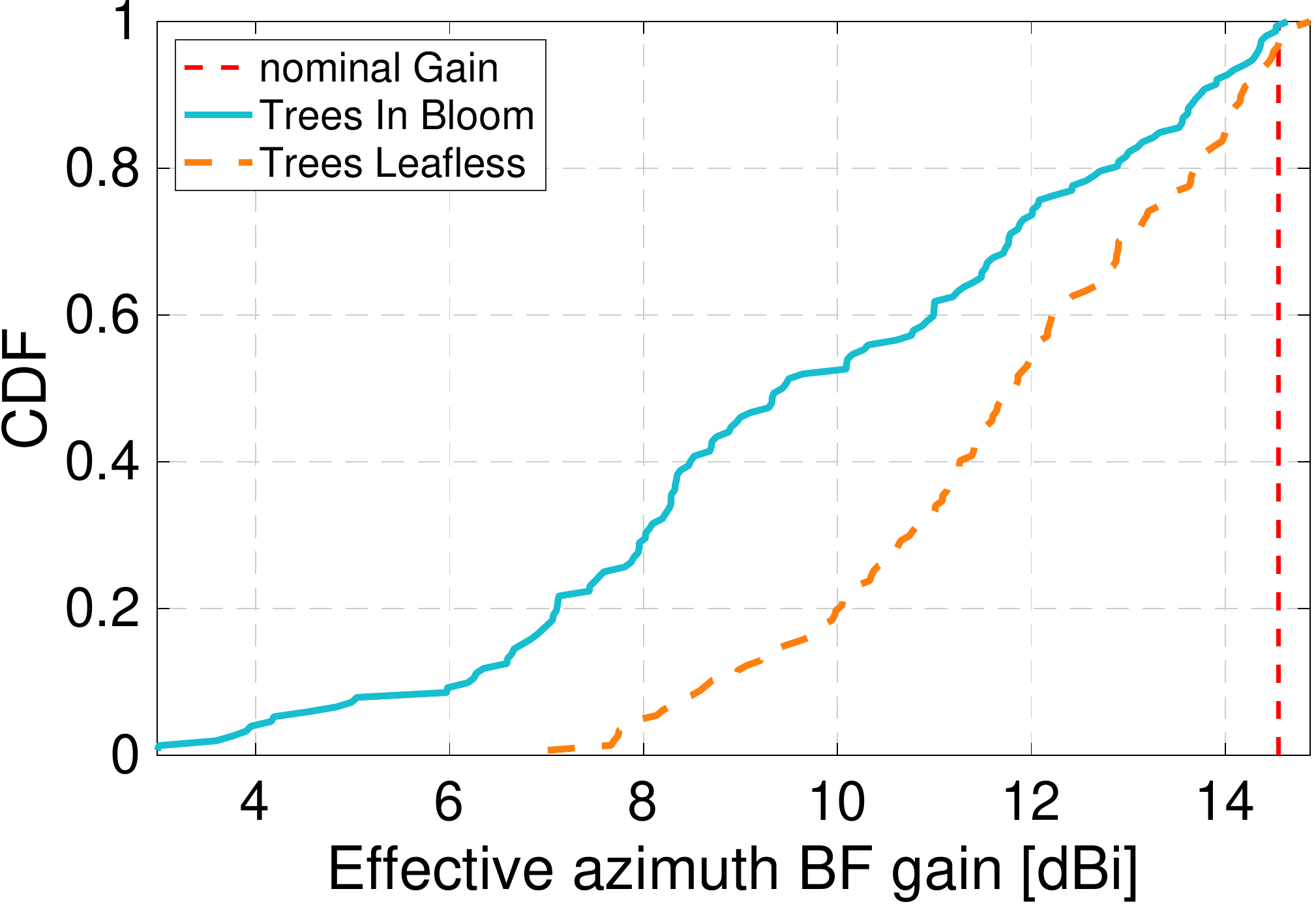}
\vspace{-1\baselineskip}
\label{fig:foliage-compare-bfg}}
\vspace{-1\baselineskip}
\caption{\addedMK{Path gain and azimuth beamforming gain measurements for different environmental factors. (a) scaffolding present at \textbf{NWC}, (b) typical in-bloom foliage viewed from the sidewalk used for scenario NWC-N-E, (c,d) scaffold/no scaffold measurements at \textbf{NWC} and (e,f) foliage/no foliage measurements at \textbf{NWC}.}}
\label{fig:environment}
\end{figure*}

\subsection{Impact of Tx and Rx Placement}
\label{sec:results-environment}
\addedMK{Having many OtI scenarios allows us to develop a sense of the ``average'' wireless channel by considering many data points as a single ensemble. However, having multiple locations means that we can also isolate specific features of the Tx and Rx placements from our OtI scenarios to understand their potential impact.}

\subsubsection{\addedMK{Different Sides of an Elevated Subway Track}}
\label{sec:results-environment-sidewalk}
We observe differences not only between measurement locations, but also between individual sidewalks measured at a single location. \addedMK{A notable example of this is shown for the JLG-N-W and JLG-N-E scenarios in Figures~\ref{fig:placement}\subref{fig:new1-street-compare-pg} and \ref{fig:placement}\subref{fig:new1-street-compare-bfg}. An elevated subway track bisects the two sides of the street; the receiver was placed at \textbf{JLG} directly in-line with JLG-N-W; JLG-N-E is the sidewalk on the far side of the street which has significant blockage from the subway track.}

Figure~\ref{fig:placement}\subref{fig:new1-street-compare-pg} shows a consistent 10\thinspace{dB} \addedMK{higher path loss for JLG-N-E} over the distances measured. Furthermore, Figure~\ref{fig:placement}\subref{fig:new1-street-compare-bfg} shows the median azimuth beamforming gain for JLG-E-E is degraded by a further 1.8\thinspace{dB}, for a total \addedMK{median} beamforming gain loss of almost 6\thinspace{dB}. This result indicates that elevated subway tracks \addedMK{or similar structures} add a significant amount of \addedMK{path loss and environmental scattering}, and more generally demonstrate how an OtI scenario is still heavily dependent on the outdoor propagation environment. 

\subsubsection{\addedMK{Different Floors of the Same Building}}
\label{sec:results-environment-floors}
As a typical building occupies more than one floor, it is useful to understand what effect, if any, the height of a user has on the wireless channel. We use the measurements from \textbf{TEA} where the Rx was placed on the second and third floors, such that the Rx is at the same distance along the street, only higher or lower in elevation. \addedMK{The indoor layout of the second and third floors where the Rx is placed is largely identical, meaning any observed difference will be due to the outdoors propagation environment.} The Tx was then placed along identical locations on the street sidewalks. The results of this comparison can be seen in Figures~\ref{fig:placement}\subref{fig:old2-height-compare-pg} and \ref{fig:placement}\subref{fig:old2-height-compare-bfg}, which show that the third floor placement of the Rx experiences an 8--10\thinspace{dB} \addedMK{lower path loss} than the second floor placement. 


We observe that the street sidewalks along \textbf{TEA} have trees planted at regular intervals. Therefore, a plausible explanation for this result is that the higher floor has a \addedMK{view of} the Tx which experiences less blockage due to foliage. We also note that the azimuth beamforming gain degradation is around 1\thinspace{dB} lower for the third floor. A lower blockage from foliage would also explain this effect, as foliage can create significant scattering~\cite{du2020suburban, yang2019impact}. 

\ifcolumbia
\else
\subsubsection{\addedMK{Angle of Incidence}}
The measurement presented in Figure~\ref{fig:aoi-test} shows that the AoI into the window can have an over 10
\thinspace{dB} impact on the amount of loss experienced by the 28\thinspace{GHz} signal. \addedMK{Therefore, we may observe a widespread impact of the AoI into the glass on the measured path loss.} In each OtI scenario, the Tx is moved perpendicular/parallel to the window by the Rx, leading to a normal/oblique AoI into the window. The Tx was moved in both ways during the measurements at \textbf{NWC}\addedMK{. As seen in Figure~\ref{fig:aoi_measurement_map}, we cluster the NWC OtI scenarios according to the measurable street geometry by considering what AoI the straight line between the Tx and Rx has on the window glass. We generate two clusters, one where AoI $<45^\circ$, and the second where AoI $\geq45^\circ$ For cases where LOS from Tx to Rx is blocked, the real AoI for the mmWave signal is difficult to determine. Hence we do not include NWC-N-E or NWC-W-S as they lose LOS to \textbf{NWC} as the Tx moves farther away.}

We observe a \addedMK{9\thinspace{dB}} difference between the two clusters at 50\thinspace{m} in Figure ~\ref{fig:placement}\subref{fig:oblique-compare-pg}, \addedMK{close to the observed 10\thinspace{dB} range of glass transmission loss in Figure~\ref{fig:aoi-test}. We also observe that the difference between the two clusters becomes smaller at greater distances; this is an expected result as the path loss is prone to impacts from other effects at larger link distances.} 
The azimuth beamforming plot in Figure~\ref{fig:placement}\subref{fig:oblique-compare-bfg} shows that the median beamforming gain is around \addedMK{1\thinspace{dB} lower for the higher AoI group, implying that OtI scenarios with a larger AoI experience not only a greater path loss but also a larger degree of environmental scattering.}
\fi

\ifarxiv
\subsection{\addedMK{Impact of Environmental Effects}}
\addedMK{To study specific environmental factors, we consider repeated OtI scenarios on identical sidewalks, with the only difference being a single controlled environmental variable. As the measurement campaign was spread out over a long time on account of the COVID-19 pandemic, we were able to measure across different seasons, giving control of two variables: the presence of scaffolding and the presence of tree foliage.}

\subsubsection{\addedMK{Scaffolding}} 
\addedMK{A common feature seen on NYC streets is scaffolding which typically encloses the sidewalk in front of a building. BSes located on street lightpoles may therefore lose direct LOS to UEs in lower floors, which may cause an increased path loss. In NYC, scaffold is commonly deployed in winter as a means to protect pedestrians from falling ice; the scaffolding considered in our measurements visible in Figure~\ref{fig:environment}\subref{fig:nwc-scaffold} was placed for this purpose. However, in Winter 2021, this scaffolding was not placed, allowing us to  measure NWC-N-W and NWC-E-N one year apart, with and without scaffold. Other variables which could impact the measured path loss, such the location of the Tx and Rx and presence of foliage, remained constant such that the only change in the static environment is the scaffold. TEA-S-N-1 and TEA-S-S-1 were also taken with scaffold outside the window, but we were not able to measure them without scaffold during the measurement campaign.} 

\addedMK{Figure~\ref{fig:environment}\subref{fig:scaffold-compare-pg} shows that the presence of scaffolding leads to a uniform 5\thinspace{dB} additional loss. This corresponds well to the 4-6\thinspace{dB} measurable penetration loss typical of pressed wood~\cite{vargas2017measurements}, which we observed to make up the majority of the scaffold construction. Figure~\ref{fig:environment}\subref{fig:scaffold-compare-bfg} shows an additional ~1\thinspace{dB} beamforming degradation with scaffolding present, indicating that scaffolding \addedMK{introduces more} environmental scattering. These results show that the common occurrence of scaffolding located directly outside of an indoors location can have a meaningful impact on the OtI path loss.} 

\subsubsection{\addedMK{Presence of Tree Leaves}}
\addedMK{City streets are typically lined with trees. As foliage can have a significant impact on path loss for a mmWave signal~\cite{yang2019impact}, we may observe seasonal differences in measured path loss for a given sidewalk. We can study potential effects by considering the  NWC-N-E, NWC-W-N, and NWC-W-S scenarios twice; once in summer with tree leaves present, and once in winter without tree leaves. As with the scaffolding comparison, all other variables in the static environment were controlled to the greatest degree possible to ensure the only difference is the presence of leaves on deciduous trees and shrubbery.}

\addedMK{Figure~\ref{fig:environment}\subref{fig:foliage-compare-pg} shows a 2--3\thinspace{dB} increase in path loss when tree leaves are present. This result is expected; while there are trees at regular intervals on the sidewalks, they are typically not densely packed enough to significantly impact even visual LOS. We also note that the sidewalk trees present in the testbed area in Figure~\ref{fig:meas_locations} are young and not very large, seen in Figure~\ref{fig:environment}\subref{fig:nwc-foliage}. Furthermore, the  the 35-45\thinspace{dB} gap additional loss above free space  Figure~\ref{fig:environment}\subref{fig:foliage-compare-pg} is mostly accounted for by the glass loss measured in Figure~\ref{fig:aoi-test}. With these considerations, it expected that the presence of leaves on trees would have only a minor impact on the path loss for the sidewalks measured. Figure~\ref{fig:environment}\subref{fig:foliage-compare-bfg} demonstrates a 2--3\thinspace{dB} further degradation in median azimuth beamforming gain when foliage is present, which is larger than the increased degradation with scaffold present.}

\addedMK{Altogether, by combining the measured increased path loss and beamforming gain degradation, we observe that the presence of scaffolding or tree leaves can reduce the mmWave link budget by 4--6\thinspace{dB}. This result is uniform over the five OtI scenarios considered from \textbf{NWC}. Due to the preponderance of these environmental factors in urban environments, we claim that their consideration is important in the discussion of mmWave OtI coverage.}
\fi


\begin{figure*}[t]
\subfloat[]{
\includegraphics[width=0.28\linewidth]{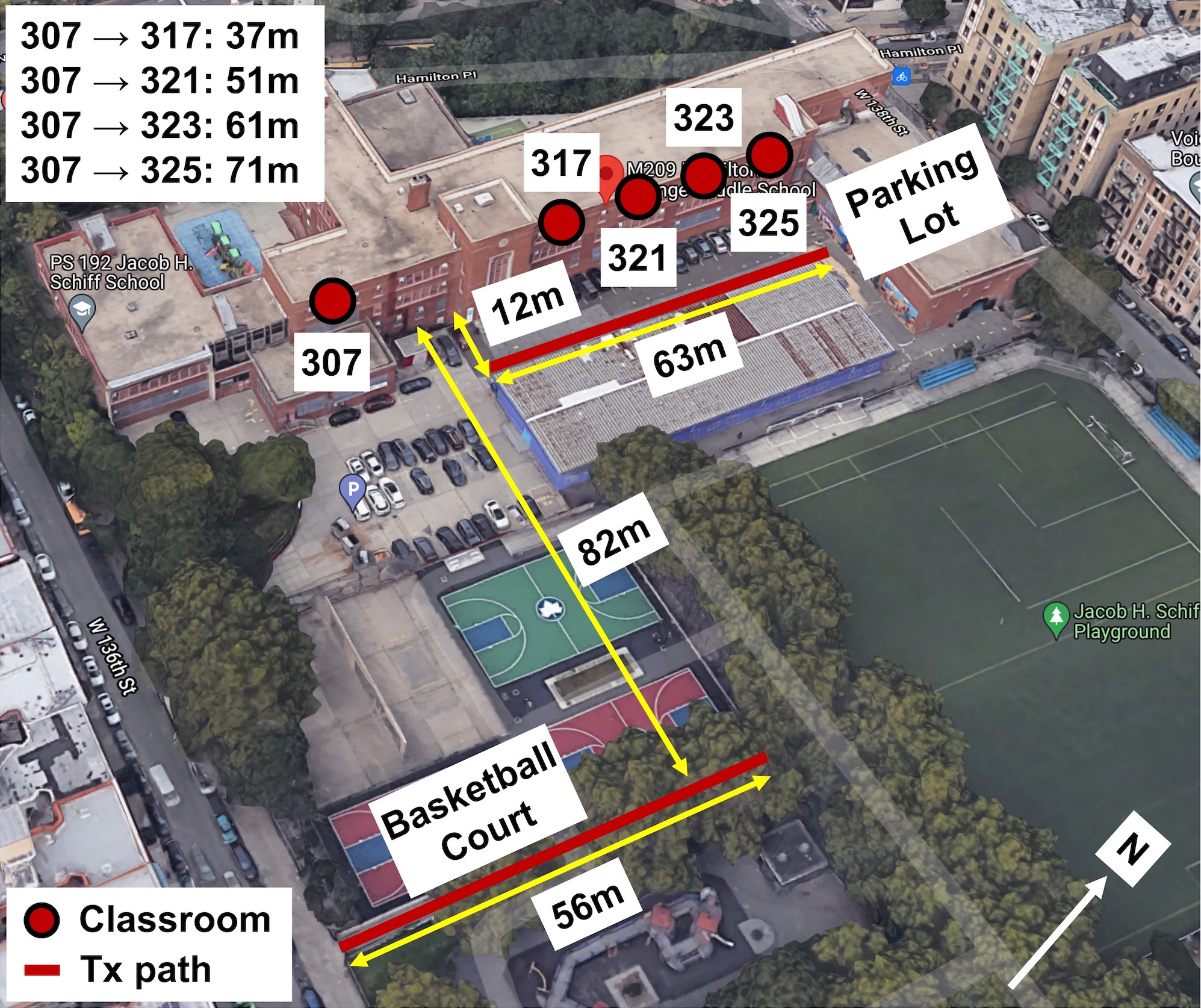}
\label{fig:hgms-map}}
\hspace{5pt}
\subfloat[]{
\includegraphics[width=0.33\linewidth]{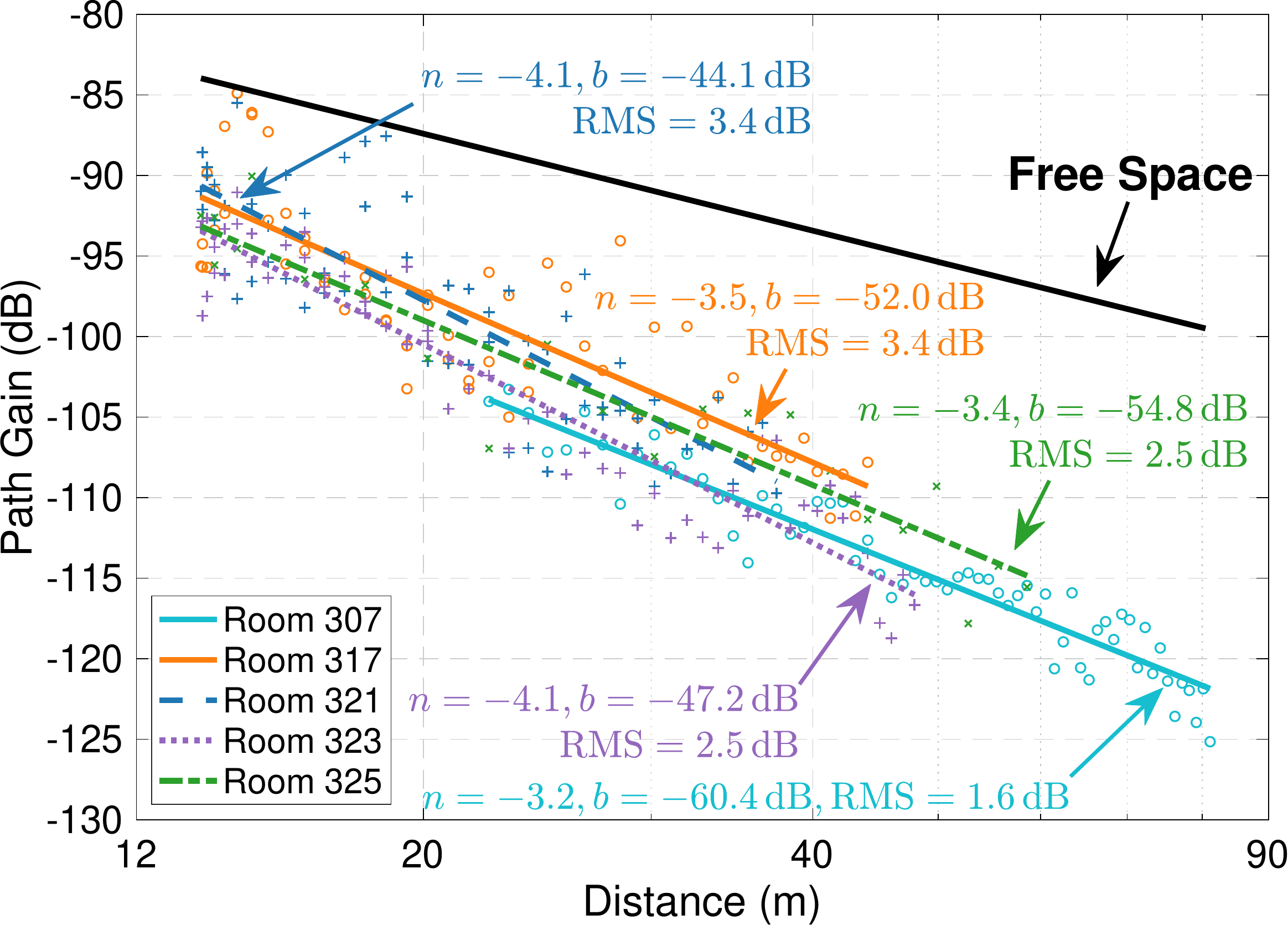}
\label{fig:classrooms-parking}}
\hspace{5pt}
\subfloat[]{
\includegraphics[width=0.33\linewidth]{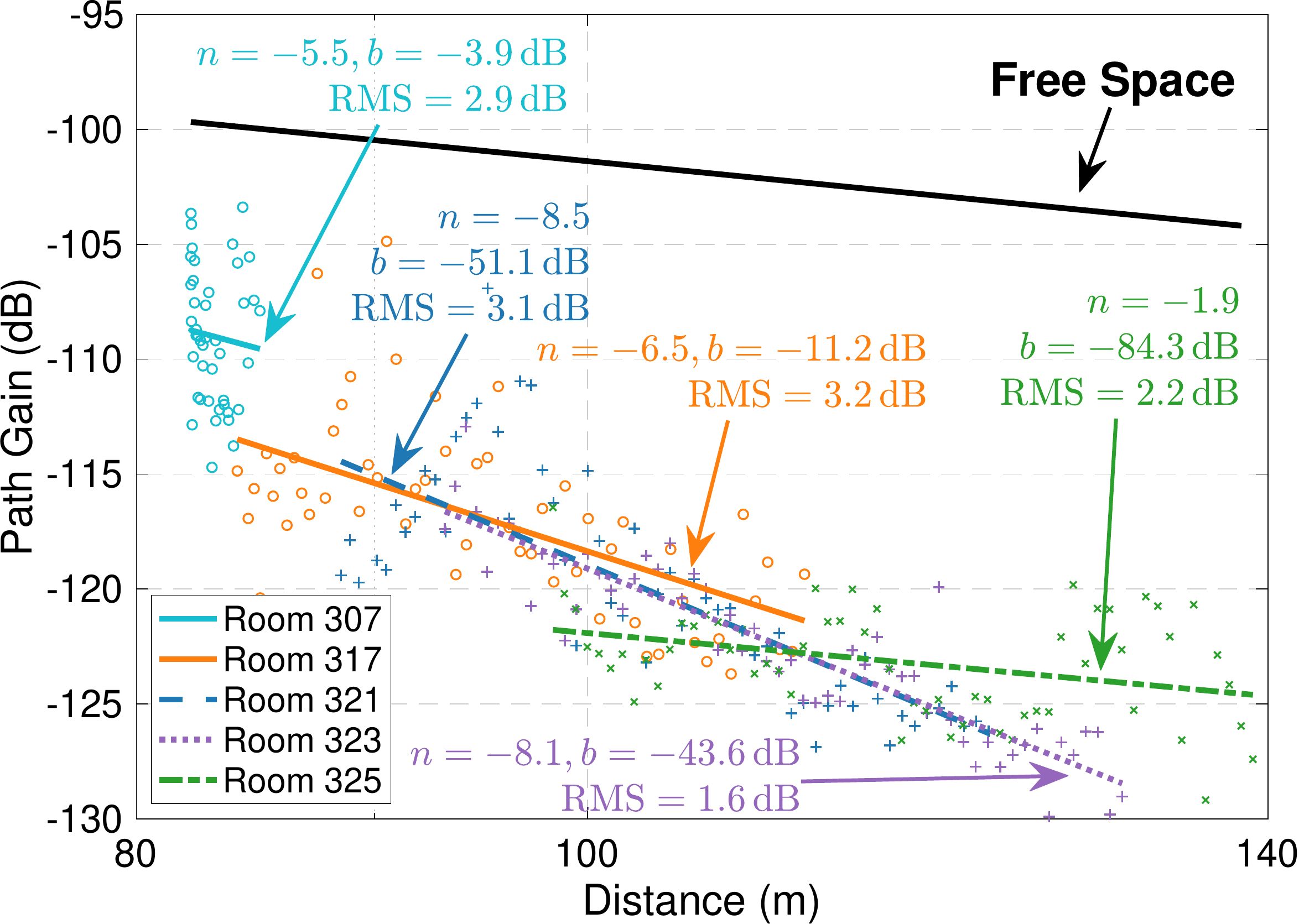}
\label{fig:classrooms-basketball}}
\vspace{-1\baselineskip}
\caption{Summary of path gain measurements taken at HMS. (a) Map of measurement locations. Maroon lines represent paths along which the channel sounder Tx was moved for each classroom Rx location, which correspond to entries in Table~\ref{T:measurements}. (b), (c) Per-classroom path gain models with the Tx placed along (b) the parking lot directly outside the classrooms and (c) the basketball court. Distances represent the three-dimensional Euclidean distance between Tx and Rx.}
\label{fig:school-measurements}
\vspace{-0.5\baselineskip}
\end{figure*}

\begin{figure*}[t]
\subfloat[]{
\includegraphics[width=0.18\linewidth]{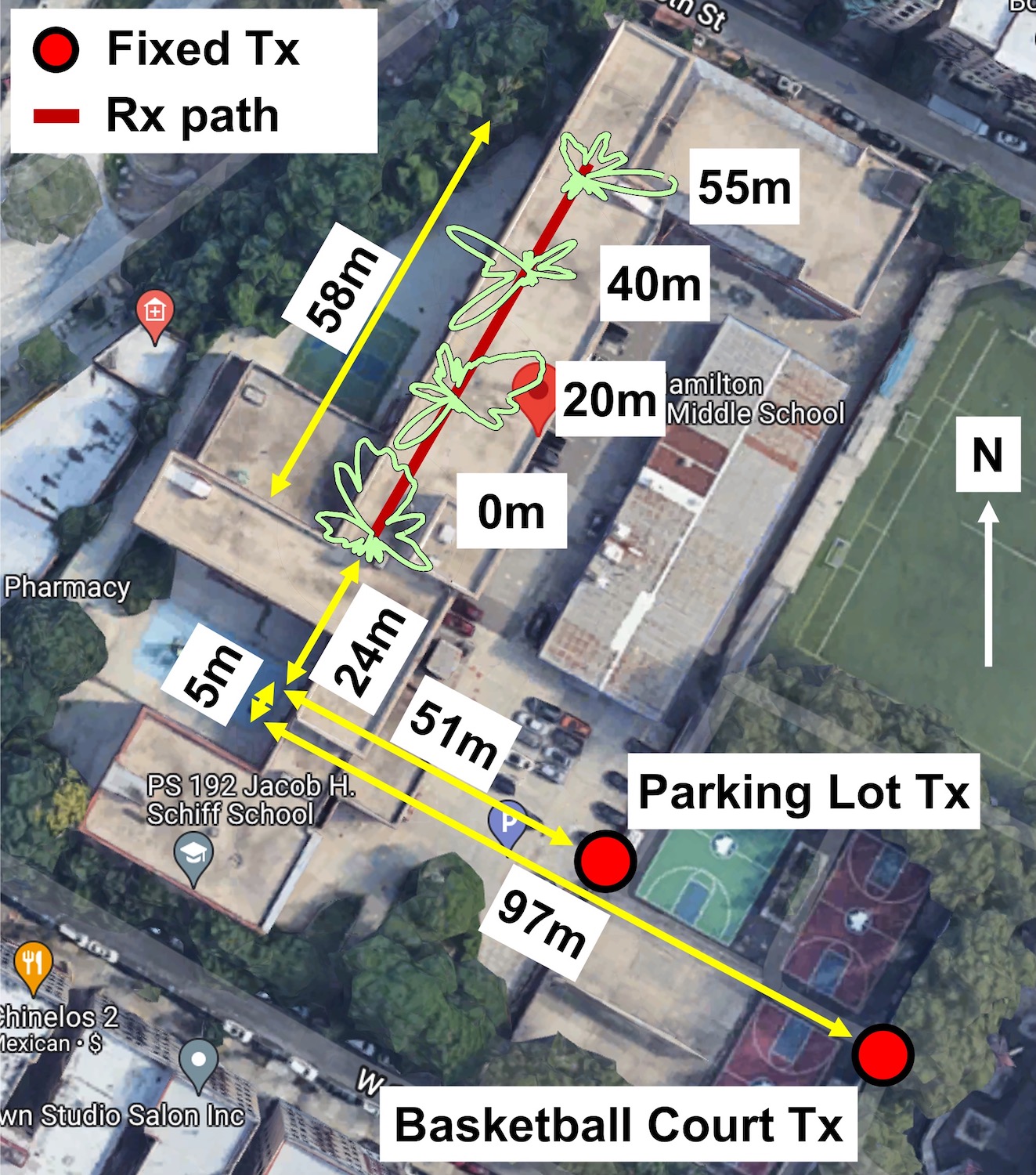}
\label{fig:hgms_hallway_map}}
\hspace{3pt}
\subfloat[]{
\includegraphics[width=0.31\linewidth]{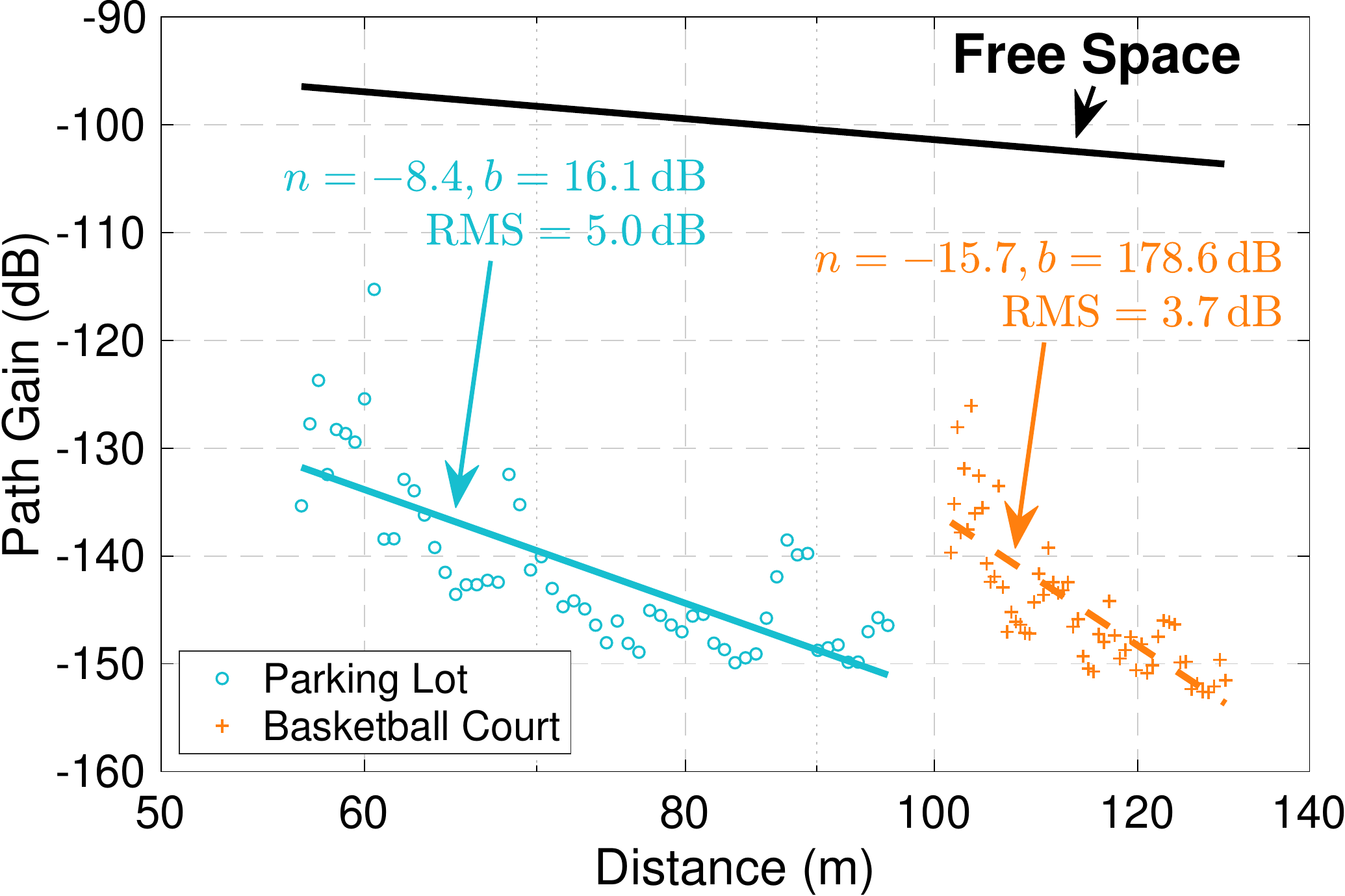}
\label{fig:hallway-compare-pg}}
\hspace{3pt}
\subfloat[]{
\includegraphics[width=0.215\linewidth]{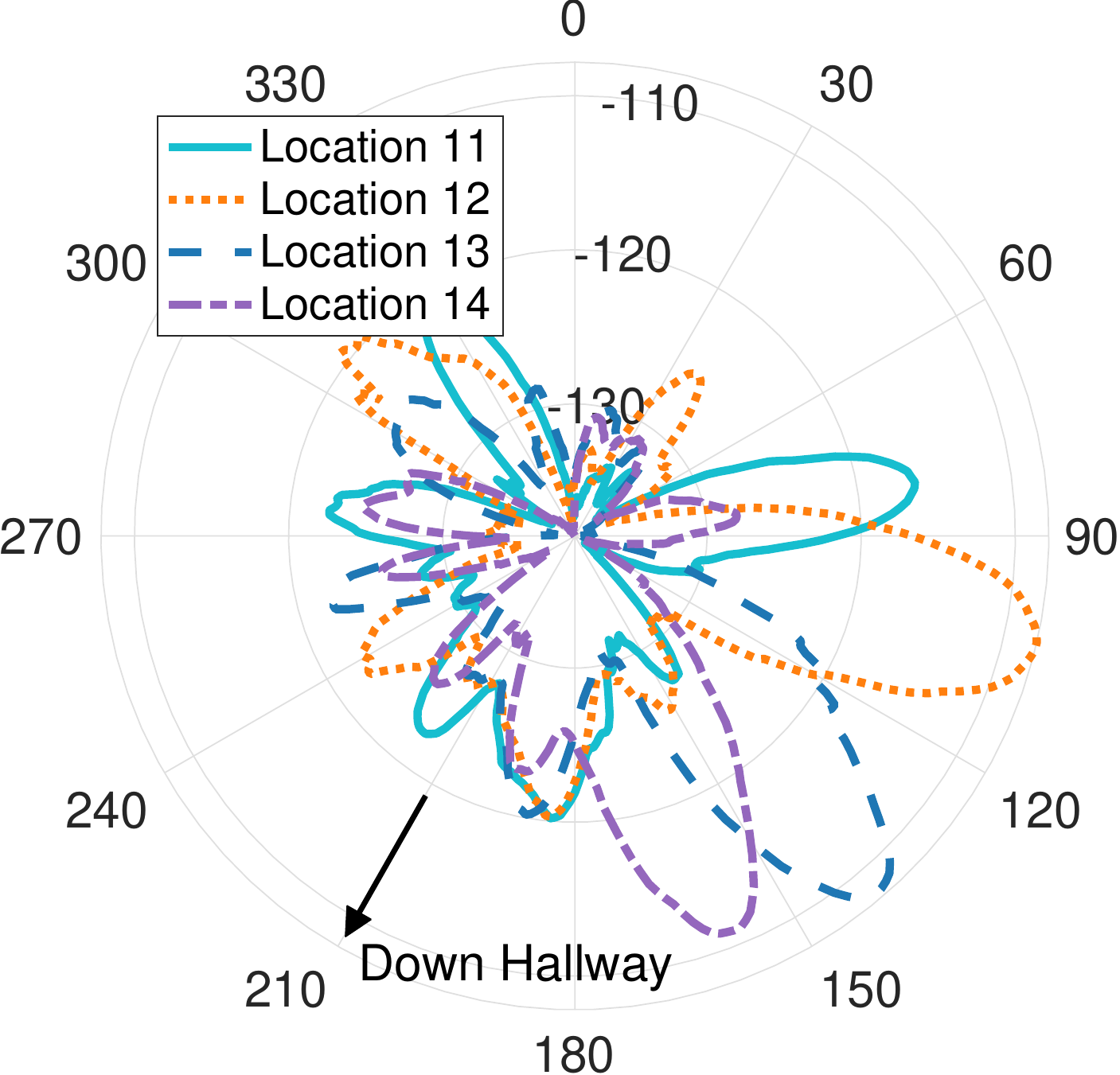}
\label{fig:hallway-compare-pas}}
\hspace{3pt}
\subfloat[]{
\includegraphics[width=0.215\linewidth]{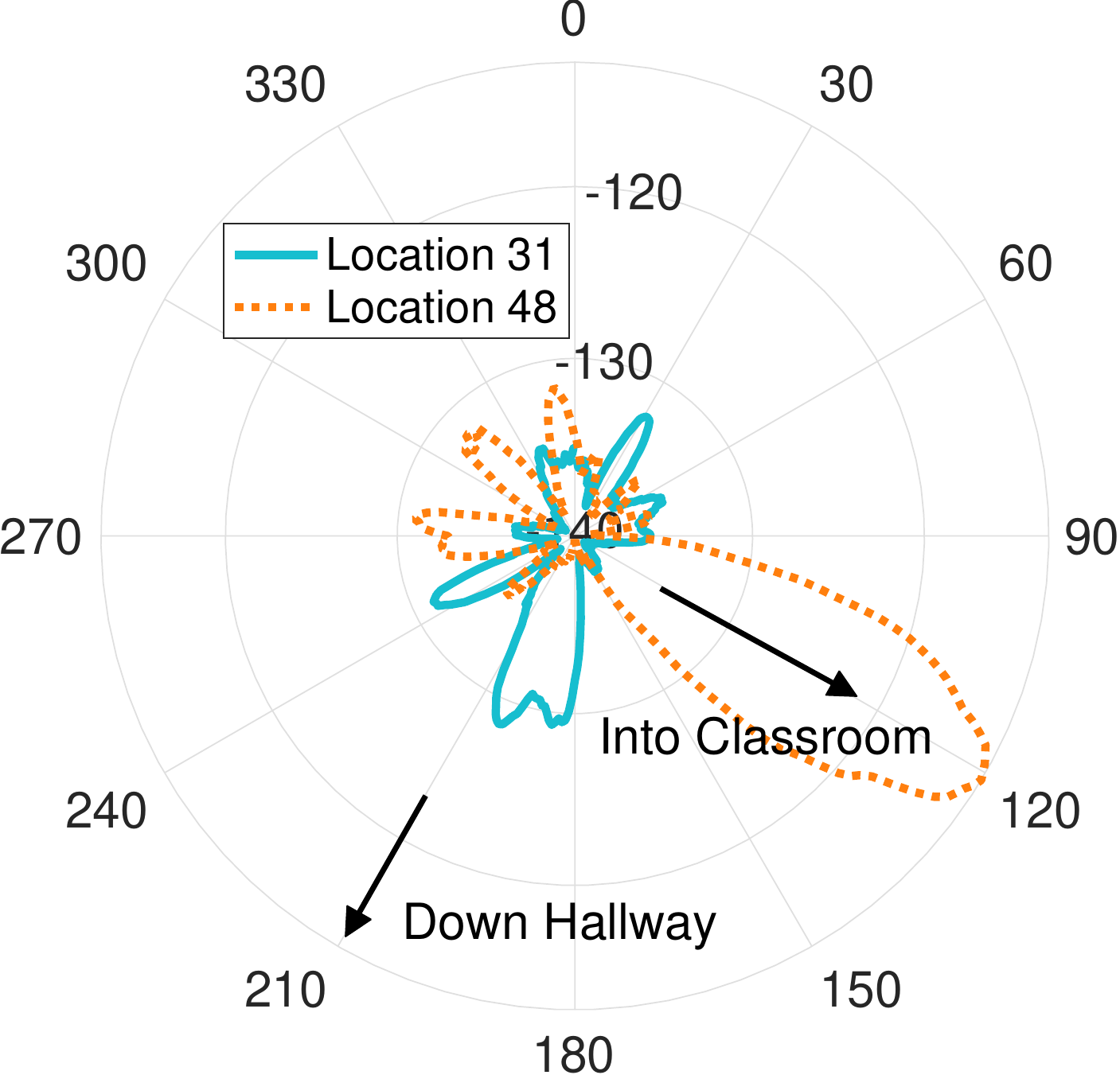}
\label{fig:hallway-compare-pas-paths}}
\vspace{-1\baselineskip}
\caption{Hallway measurements at HMS taken from two different Tx locations. (a) Map of the hallway measurements and example power angular spectra for the Tx located in the parking lot. (b) Path gain models for both Tx locations.  (c) Successive angular spectra showing the peak angle rotating as the Rx moves past a classroom door, demonstrating the presence of a strong propagation path via the nearest classroom. (d) Two other angular spectra showing two different \addedMK{dominant} propagation paths; one through a classroom, and one down the hallway.}
\label{fig:hallway-compare}
\end{figure*}

\section{Case Study: A Public School}
\label{sec:school}
As \textbf{HMS} uses traditional glass for its windows, it experiences a significantly lower path loss compared to the other measured locations. Furthermore, \textbf{HMS} is a representative example of a public school building located within an NYC neighborhood with comparatively low Internet access. These two characteristics make \textbf{HMS} a location of particular interest. We analyze the measurements at \textbf{HMS} in classrooms which are mapped in Figure~\ref{fig:school-measurements}\subref{fig:hgms-map} and enumerated in Table~\ref{T:measurements}, and a hallway mapped in Figure~\ref{fig:hallway-compare}\subref{fig:hgms_hallway_map}. We use these measurements to compare path gain models for the individual classrooms and study how the mmWave signal propagates into the indoor hallway. Access to \textbf{HMS} was facilitated by the COSMOS RET/REM program.


\subsection{Classroom Measurements}
\label{sec:school-classrooms}
Measurements at \textbf{HMS} were taken with the Rx located in five classrooms along the third floor of the school building. We note that the classrooms are all very regular in dimension as well as interior layout. The Tx was moved along two paths, one along the school parking lot located directly outside the classrooms, and the other along the basketball courts located at a greater distance. A map of the school and measurement locations, along with path gain results for the two Tx paths are shown in Figure~\ref{fig:school-measurements}. 

The fitted models for the measurements with the Tx located in the parking lot in Figure~\ref{fig:school-measurements}\subref{fig:classrooms-parking} show a high degree of similarity, with similar fitted slopes close to $n=4$, in line with the theoretical model developed in~\cite{chizhik2021universal} for outdoor-to-indoor propagation at oblique incidence angles. The measured path gain values from different classrooms are largely overlapping, with no clear dependence on the particular classroom, which is an understandable result given the uniformity of the five classrooms considered. The relatively low 10-20\thinspace{dB} excess loss above free space in Figure~\ref{fig:glass_compare}\subref{fig:glass_pg} indicates a strong potential for OtI coverage.

Similar results with the Tx located in the basketball court are shown in Figure~\ref{fig:school-measurements}\subref{fig:classrooms-basketball}. Unlike the measurements with the Tx in the parking lot, there is some dependence on the classroom being measured. In particular, Room 307 has a noticeably higher path gain compared to the other classrooms. One possible reason is the row of trees visible near the middle of the map in Figure~\ref{fig:school-measurements}\subref{fig:hgms-map}. As seen from ground level at the basketball court, these trees did partially block the view of the windows for classrooms 317 to 325, which likely accounts for the higher loss experienced by these classrooms.



\subsection{Hallway Measurements}
\label{sec:school-hallway}
We also conducted measurements by moving the Rx along an interior hallway located behind the row of classrooms indicated in Figure~\ref{fig:hallway-compare}\subref{fig:hgms_hallway_map}. The Tx was kept in two fixed positions, one in the parking lot and the other at the basketball court, noted by the ``Fixed Tx'' locations in the same figure. The Rx was moved along the hallway in an identical manner for both Tx locations\addedMK{, leading to a total of 116 measurements taken of the interior hallway}. The path gain and azimuth beamforming gain measurements are shown in Figure~\ref{fig:hallway-compare}.

The path gain results in Figure~\ref{fig:hallway-compare}\subref{fig:hallway-compare-pg} show a 5\thinspace{dB} difference between the two models, with a large value for $n$ indicating a fast drop-off in received power as the Rx moves down the hallway. \addedMK{The plotted distance in Figure~\ref{fig:hallway-compare}\subref{fig:hallway-compare-pg} is the three-dimensional Euclidean distance between Tx and Rx; the Rx was moved along a 58\thinspace{m} linear distance down the hallway in both measurements. This distance is compressed within the three-dimensional Euclidean distance, creating the particularly steep slopes.} We note that there was no direct line-of-sight path from Tx to Rx. Indoor locations far from a window typically have several candidate propagation paths~\cite{shakya2022dense}. In the case of these hallway measurements, we consider two likely methods~\cite{chizhik2021universal}: (i) via the room most normal to the Tx, and (ii) via the room closest to the location of the Rx. We study the propagation mechanism by investigating the angular spectra $\bar{S}(d, \phi)$ measured at several Tx-Rx links.

Figure~\ref{fig:hallway-compare}\subref{fig:hgms_hallway_map} shows how the angular spectra evolve as the Rx moves down the interior hallway (i.e., away from Room 307 which is the room most normal to the Tx). There is no clear trend in the direction of the peak angles, lacking a persistent dominant direction along the hallway which would be characteristic of propagation method (i). 
Figure~\ref{fig:hallway-compare}\subref{fig:hallway-compare-pas} shows how the peak angle rotates as the Rx moves  past the doorway of Room 317, where locations $\{11, 12\}$ and $\{13, 14\}$ are locations on either side of the doorway. For such locations, it is clear that the dominant method is (ii) and the Rx is receiving a signal through the doorway of the closest classroom along the hallway.
 
The angular spectra in Figure~\ref{fig:hallway-compare}\subref{fig:hallway-compare-pas-paths} show that there are some Rx locations which receive a signal peak from the direction down the hallway, towards Room 307. This represents propagation method (i), and so we find that both propagation methods are active in this OtI scenario, though method (ii) seems to be dominant.

\section{Multi-User Support Potential}
\label{sec:reverse_tc}
The support of multiple users with multiple beams is an important consideration in mmWave MIMO systems. Therefore, in addition to the OtI scenarios in Table~\ref{T:measurements}, we considered an OtI scenario whereby the locations of the Tx and Rx were reversed. This allows for the Rx to emulate a BS and measure from which directions the signal is received from the indoor UE. By moving the Tx indoors between different classrooms at \textbf{TEA} and placing the Rx on an open outdoor balcony 15\thinspace{m} above street level, we can evaluate the \addedMK{feasibility of simultaneous support of multiple users on different beams. This is highly dependent on beam overlap; users are best served on spatially disjoint beams to avoid IUI.} 


We measured 17 different Tx-Rx links across five different classrooms. The power angular spectra $\bar{S}(d, \phi)$ for each classroom were converted to the linear scale and normalized. For all combinations of classroom pairs, without repetition, we computed the cross correlation of the normalized amplitudes \addedMK{as a measure of beam overlap.}

\begin{figure}[t!]
\subfloat[]{
\includegraphics[width=0.56\columnwidth]{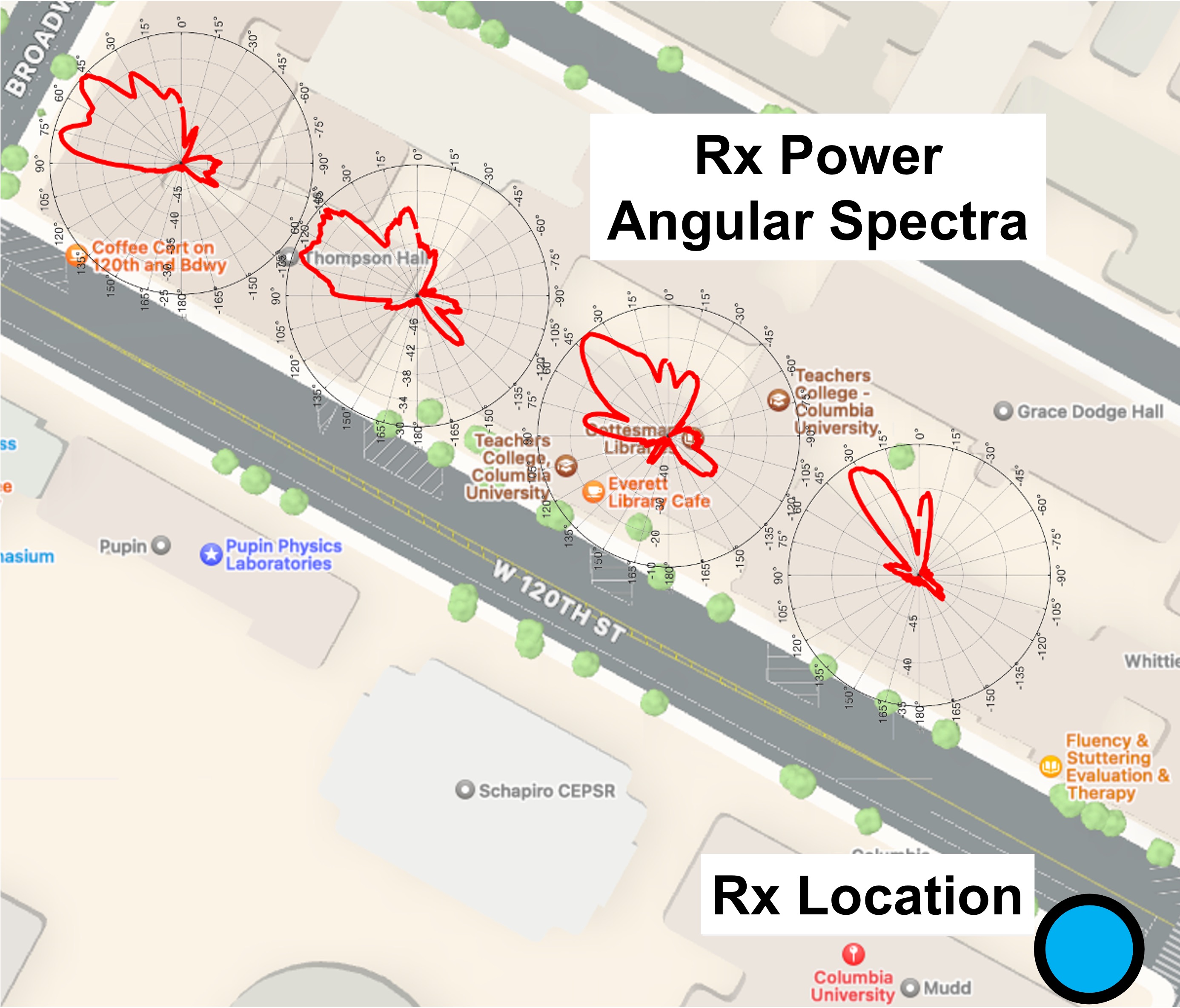}
\label{fig:tc_reverse_beam}}
\hspace{2pt}
\subfloat[]{
\includegraphics[width=0.33\columnwidth]{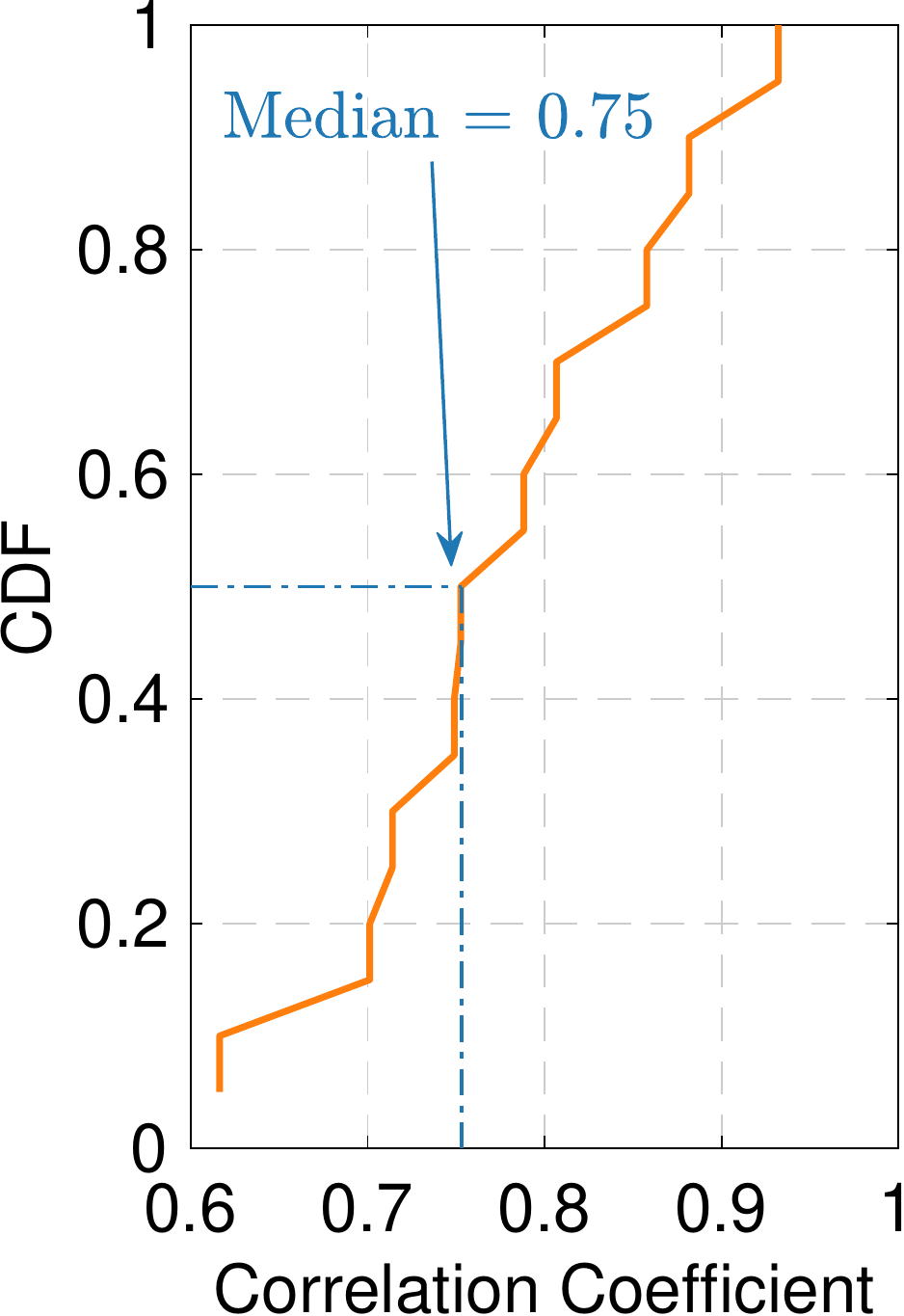}
\label{fig:reverse_tc_cdf}}
\vspace{-1\baselineskip}
\caption{Measurements at TEA with Tx and Rx reversed. (a) Map of Rx location on the balcony and select angular spectra overlaid on the corresponding Tx location. (b) CDF of correlation coefficient between  power angular spectra.}
\vspace{-1\baselineskip}
\label{fig:reverse_tc}
\end{figure}

Figure~\ref{fig:reverse_tc}\subref{fig:tc_reverse_beam} shows that the dominant beam direction of the received signal does rotate towards the transmitter as it moves between classrooms. However, as seen in Figure~\ref{fig:reverse_tc}\subref{fig:reverse_tc_cdf}, cross correlation between received beams produces a high median correlation coefficient of 0.75, likely caused by the similar oblique incidence angles for links further down the street. These high coefficients indicate a high level of inter-user-interference. Therefore, a 28 GHz BS deployment near the street intersection of \textbf{TEA} may have limited capability to spatially multiplex users with multiple beams. Street intersections are a popular location for BS deployments, but these results indicate that the BS would likely need to be placed across from the center of the building to use beamforming and beam steering capabilities most effectively. 

\section{Glass-Dependent OtI Data Rates}
\label{sec:OtI-coverage}
The models in Section~\ref{sec:results-glass} are now used to develop a measure of \addedMK{link} rate coverage for OtI scenarios with ``traditional'' or Low-e glass. Table~\ref{T:parameters} defines typical parameters for the 28\thinspace{GHz} BS and UE representative of recent advances in state-of-the art mmWave hardware~\cite{sadhu201728, deng2021dual, hwang2021dualband, ershadi2021gate, chang2020linear}. We select conservative values for these parameters to reduce the possibility of overestimating data rates, and we include an additional 5\thinspace{dB} of losses in $NF$. The resulting Rx noise floor is $N = -174+10\log_{10}B+NF = -76$\thinspace{dBm}. In this analysis, we assume that the BS and UE are able to efficiently align their transmit and receive beams~\cite{zia2022effects}.


\addedMK{As the signal-to-noise (SNR) will determine the achievable data rate, we present a relevant measure of data rate coverage by considering the 10\textsuperscript{th} percentile $SNR(d)$, $SNR_{10}(d)$, which defines the SNR that 90\% of users will exceed. The SNR in dB may be computed as
$SNR(d) = P_{Tx} + G_{Tx} + G_{LNA} + G_{Rx} - G_{deg}(d) + G_{path}(d) - N$, where $G_{path}(d)$ is computed from our path gain model and $G_{deg}(d)$ is computed from the median azimuth beamforming gain.}

\addedMK{By using the empirical models, the SNR will end up as a normally distributed random variable $SNR\sim \mu(M)+\sigma(M)\cdot\mathcal{N}(0,1)$, 
where $M$ is the path gain model being considered. As the model given in $M$ is itself a function of $d$, this will result in a normally distributed SNR variable for every distance $d$.}

\addedMK{We consider three SNR boundaries: 25\thinspace{dB}, 14\thinspace{dB}, and 4\thinspace{dB}. These represent SNRs at which 256QAM 4/5, 16QAM 1/2, and QPSK 3/10 modulation and coding schemes (MCS) may be received with goodput factors of 0.7, 0.7, and 1.0 respectively~\cite{3gpp256qamstudy,3gppperformance,peralta20185g}. We can estimate link rates using an impaired Shannon capacity $\hat{D} = \rho\beta B\log_2(1+10^{(SNR-C)/10})$, where $\rho=0.6$ is the overhead factor, $\beta$ is the goodput factor, and $C=3$\thinspace{dB} is implementation loss. This leads to link rates of 2.5, 1.2, and 0.5\thinspace{Gbps} for 256QAM, 16QAM, and QPSK respectively, close to values in 3GPP reference material~\cite{3gpp256qamstudy,3gppperformance}.}

\subsection{\addedMK{Buildings with Traditional Glass}}
As shown in Sections~\ref{sec:results-glass} and~\ref{sec:school}, buildings with ``traditional'' glass experience  lower path loss, suggesting a strong potential for OtI coverage at 28\thinspace{GHz}. From Figure~\ref{fig:glass_compare}\subref{fig:glass_pg}, we set the slope $n=3$, intercept $b=-59.8$\thinspace{dB}, and $\sigma=4.3$\thinspace{dB}. \addedMK{For each outdoor link distance $d\in\{10, 11,...,200\}$, we compute the 10\textsuperscript{th} percentile path gain given by the log-normal distribution 
$G_{path,i}(d) = b+n\cdot10\log_{10}d+\sigma \cdot\mathcal{N}(0,1)$. We also compute the median azimuth beamforming gain degradation $G_{deg,i}(d) = 14.5-G_{az,median}$ where $G_{az,median}$ is the median of the azimuth beamforming gain distribution for traditional glass in Figure~\ref{fig:glass_compare}\subref{fig:glass_bfg}.}

The top curve in Figure~\ref{fig:OtI-coverage} demonstrates \addedMK{$SNR_{10}(d) > 25$ for $d \leq 68$, meaning} that 256-QAM modulation can be supported for up to 90\% of indoor users at a link distance of up to 68\thinspace{m}. This corresponds to \addedMK{$\hat{D} > 2.5$\thinspace{Gbps}}. 16-QAM 1/2 MCS for 90\% of indoor users can be supported at link distances up to 175\thinspace{m}, corresponding to \addedMK{$\hat{D} > 1.2$\thinspace{Gbps}}. These measurements covered a variety of Tx and Rx locations at \textbf{HMS} as shown in Figure~\ref{fig:school-measurements}\subref{fig:hgms-map}, with many links occluded by foliage and with a large variation in the AoI. Thus, we believe these results to be representative of building constructions with traditional glass in typical urban environments. A 68\thinspace{m} link distance subtends 81$^\circ$ with a typical 10\thinspace{m} building standoff, which is within the beamsteering capability of phased array antennas that would be suitable for outdoor BSes~\cite{sadhu201728}.

\subsection{\addedMK{Buildings with Low-e Glass}}
\begin{figure}[t!]
\includegraphics[width=0.83\columnwidth]{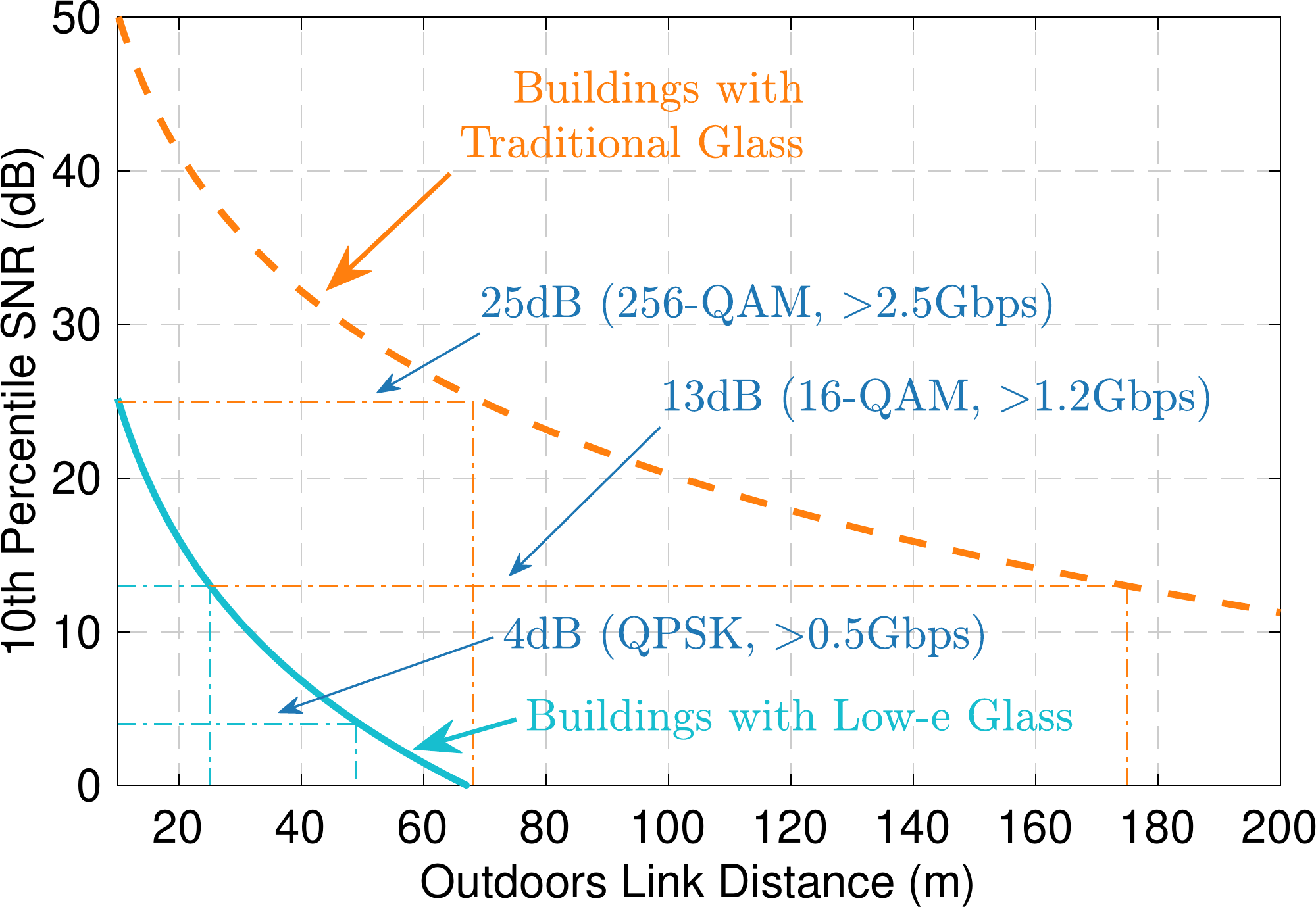}
\vspace{-1\baselineskip}
\caption{\addedMK{10\textsuperscript{th} percentile SNR predictions for buildings with \mbox{Low-e} or traditional glass windows, with coverage ranges $d^*$ labelled for various MCS.}}
\label{fig:OtI-coverage}
\end{figure}

We repeat the \addedMK{SNR calculations} using the Low-e model in Figure~\ref{fig:glass_compare}\subref{fig:glass_pg}, setting $n=3$, $b=-79.6$\thinspace{dB}, and $\sigma=8.4$\thinspace{dB}, producing the the lower curve of Figure~\ref{fig:OtI-coverage}. The results show that 256-QAM coverage cannot be supported for at least 90\% of users even at the shortest realistic Tx-Rx link distance. Instead, 16-QAM 1/2 (\addedMK{$\hat{D} > 1.2$\thinspace{Gbps}}) and QPSK  3/10 (\addedMK{$\hat{D} > 0.5$\thinspace{Gbps}}) MCS may be supported up to 25\thinspace{m} and 49\thinspace{m}, respectively. As the Low-e glass model was computed with measurements from six distinct buildings, we believe this result is representative of buildings with Low-e glass.

\begin{table}[t!]
\caption{Typical device parameters for a 28\thinspace{GHz} Tx (BS) and Rx (UE)}
\vspace{-\baselineskip}
\footnotesize
\begin{tabular}{|c|c|c|c|}
\hline
\textbf{Quantity}        & \textbf{Symbol} & \textbf{Value}   & \textbf{Ref.} \\ \hline
Tx Power        & $P_{Tx}$    & $+28$\thinspace{dBm} & \cite{sadhu201728}     \\ \hline
Tx Antenna Gain         & $G_{Tx}$    & 23\thinspace{dBi} &  \cite{sadhu201728}    \\ \hline
Rx LNA Gain    & $G_{LNA}$     & 13\thinspace{dB} & \cite{deng2021dual} \\ \hline
Rx Antenna Gain         & $G_{Rx}$    & 9\thinspace{dBi}  &  \cite{hwang2021dualband}  \\ \hline
Rx Noise Figure & $NF$     & $4+5$\thinspace{dB} & \cite{ershadi2021gate}  \\ \hline
Bandwidth       & $B$      & 800\thinspace{MHz} & \cite{sadhu201728, chang2020linear} \\ \hline
\end{tabular}
\vspace{-\baselineskip}
\label{T:parameters}
\end{table}

The coverage experienced by an indoor UE has a large variation depending heavily on the window material. Indoor coverage potential is significantly higher for buildings with older, thinner glass. However, coverage at gigabit data rates is still feasible even in buildings with modern construction if BS is nearby ($\sim$20\thinspace{m}).




\section{Conclusion}
\label{sec:conclusion}
We addressed the lack of extensive OtI mmWave measurements by conducting a large-scale measurement campaign consisting of over \addedMK{2,200} Tx-Rx links across seven building sites in West Harlem, NYC. We used the measurements to develop models for OtI path gain under various conditions. Among other things, these models show that data rates in excess of \addedMK{2.5\thinspace{Gbps} are achievable for at least 90\% of indoor users in typical public school buildings with lightpole BS deployments at distances up to 68\thinspace{m} away. Rates in excess of 1.2\thinspace{Gbps} may be achieved even with distant BS placements up to 175\thinspace{m} away. Similar lightpole deployments up to 49\thinspace{m} range are capable of providing data rates in excess of \addedMK{500}\thinspace{Mbps} for users in buildings that use modern Low-e glass.} We expect the results to inform the deployment of mmWave networks in urban areas with low Internet access, thereby helping to improve connectivity and bridging the digital divide.

While we show that high data rates in OtI scenarios are achievable, we also show that OtI multi-user support by a mmWave BS is challenging, with potentially high inter-user-interference. This illustrates the need for careful design of beamforming algorithms which take OtI scenarios into account. This is one of the subjects of our future research, which will be supported  by further measurement to ensure model accuracy. We will also use the 28\thinspace{GHz} phased array antenna modules integrated in the COSMOS Testbed~\cite{chen2021programmable} to implement and test the designed algorithms as well as make wideband channel measurements, which can produce other important results, including the delay spread and channel coherence time. 

\section{Acknowledgements}
This work was supported by NSF grants CNS-1827923, OAC-2029295, and AST-2037845, NSF-BSF grant CNS-1910757\addedMK{, and ANID PIA/ APOYO grant AFB180002}. We thank Angel Daniel Estigarribia and Zixiang Zheng for their help with the measurements. We thank Basil Masood, Taylor Riccio, and Jennifer Govan for their support during the measurement campaigns at Hamilton Grange Middle School, Miller Theatre, and Teachers' College. We thank Tingjun Chen for his helpful comments and suggestions. 

\printbibliography
\end{document}